\documentclass[11pt]{article}
\usepackage{amsmath,amssymb,bm}
\usepackage{graphicx}
\usepackage{enumerate}
\usepackage{natbib}
\usepackage{url} % not crucial - just used below for the URL 
\usepackage[dvipsnames]{xcolor}
\usepackage{xr-hyper}
\usepackage{hyperref}
\usepackage[margin=2.5cm]{geometry}
\hypersetup{colorlinks=true,linkcolor=OliveGreen,citecolor=MidnightBlue,urlcolor=black}

\newcommand{\edit}[1]{\textcolor{black}{#1}}
\newcommand{\editm}[1]{\textcolor{black}{#1}}

\begin{document}

\def\spacingset#1{\renewcommand{\baselinestretch}%
{#1}\small\normalsize} \spacingset{1}

%%%%%%%%%%%%%%%%%%%%%%%%%%%%%%%%%%%%%%%%%%%%%%%%%%%%%%%%%%%%%%%%%%%%%%%%%%%%%%

  \title{\bf Statistical inference for multivariate extremes via a geometric approach}
  \author{Jennifer Wadsworth \hspace{.2cm}\\
    Department of Mathematics and Statistics, Lancaster University\\
    and \\
    Ryan Campbell \\
    Department of Mathematics and Statistics, Lancaster University}
  \maketitle

%\if1\blind
%{
%	\title{\bf Statistical inference for multivariate extremes via a geometric approach}
%	\author{Jennifer Wadsworth \hspace{.2cm}\\
%		Department of Mathematics and Statistics, Lancaster University\\
%		and \\
%		Ryan Campbell \\
%		Department of Mathematics and Statistics, Lancaster University}
%	\maketitle
%} \fi
%
%\if0\blind
%{
%	\bigskip
%	\bigskip
%	\bigskip
%	\begin{center}
%		{\LARGE\bf Statistical inference for multivariate extremes via a geometric approach}
%	\end{center}
%	\medskip
%} \fi

\bigskip
\begin{abstract}
A geometric representation for multivariate extremes, based on the shapes of scaled sample clouds in light-tailed margins and their so-called limit sets, has recently been shown to connect several existing extremal dependence concepts. However, these results are purely probabilistic, and the geometric approach itself has not been fully exploited for statistical inference. We outline a method for parametric estimation of the limit set shape, which includes a useful non/semi-parametric estimate as a pre-processing step. More fundamentally, our approach provides a new class of asymptotically-motivated statistical models for the tails of multivariate distributions, and such models can accommodate any combination of simultaneous or non-simultaneous extremes through appropriate parametric forms for the limit set shape. Extrapolation further into the tail of the distribution is possible via simulation from the fitted model. A simulation study confirms that our methodology is very competitive with existing approaches, and can successfully allow estimation of small probabilities in regions where other methods struggle. We apply the methodology to two environmental datasets, with diagnostics demonstrating a good fit.

\end{abstract}

\noindent%
\vfill

\newpage
\spacingset{1.4} % DON'T change the spacing!
\section{Introduction}
\label{sec:intro}

\subsection{Multivariate extreme value theory}

%--------
% Addition to improve initial intro 

Multivariate extreme value theory provides the basis for estimation of rare event probabilities that involve the effect of more than one variable. Applications are diverse and include estimating flood risk \citep{Keefetal13b,EngelkeHitz20}, extreme air pollution levels \citep{HeffernanTawn04,Vettorietal19}, structural design \citep{ColesTawn94}, dietary risk assessment \citep{Chautru15} and financial risk assessment \citep{ZhangHuang06,Hilaletal14}. 
%--------

The study of multivariate extremes primarily began in the 1970s and 80s, with the theoretical study of multivariate regular variation \citep{deHaan1970,deHaanResnick77,Resnick1987}. \edit{Multivariate regular variation} is intrinsically tied up with the componentwise block maximum method for multivariate extremes. Suppose we have $n$ independent replicates of a random vector $\bm{Y}_i \in \mathbb{R}^d$, $i=1,\ldots,n$; the componentwise maximum vector is 
\begin{align*}
\bm{M}_n = (M_{n,1},\ldots,M_{n,d}) = \left(\max_{1 \leq i \leq n} Y_{1,i},\ldots, \max_{1 \leq i \leq n} Y_{d,i}\right).
\end{align*}
Univariate extreme value theory tells us that if, for each $j=1,\ldots,d$, there exists $a_{n,j}>0, b_{n,j}$ such that $\{M_{n,j} - b_{n,j}\}/a_{n,j}$ converges to a non-degenerate random variable, then the distribution of this limiting variable is generalized extreme value \citep{FisherTippett28, Gnedenko43}, which is the only univariate \emph{max-stable} distribution. \edit{A distribution is max-stable if it is invariant to the operation of taking (componentwise) block maxima, up to marginal location and scale changes.} The additional condition for joint convergence of the entire vector $\{\bm{M}_n -\bm{b}_n\}/\bm{a}_n$ to a multivariate max-stable distribution is \edit{multivariate regular variation}. Since this represents an assumption on the dependence structure it can be expressed in standardized margins: a common choice is to set $X_{P,j} = 1/\{1-F_j(Y_j)\}$, where $X_{P,j}$ follows a standard Pareto distribution if $Y_j \sim F_j$ has a continuous distribution, else it is asymptotically Pareto. A common way to express the \edit{multivariate regular variation} assumption is
\begin{align}
\lim_{t \to \infty} \Pr(\bm{X}_P / \|\bm{X}_P\| \in B,\|\bm{X}_P\|>ts~\mid~\|\bm{X}_P\|>t) = s^{-1}H(B), \label{eq:mrv}
\end{align}
where $B \subset \mathcal{S}_{d-1}=\{\bm{v}\in[0,1]^d: \edit{\|\bm{v}\|} = 1\}$ is a measurable set with $H(\partial B) = 0$. Assumption~\eqref{eq:mrv} shows that large values of the ``radial'' component $\|\bm{X}_P\|$ become independent of the ``angular'' component $\bm{X}_P / \|\bm{X}_P\|$, which follows some probability distribution $H$ on $\mathcal{S}_{d-1}$, commonly referred to as the \emph{spectral measure}. \edit{The choice of norm $\|\cdot\|$ is arbitrary, see e.g.\ \citet[][Chap. 8]{Beirlantetal04}, but the most common choice is the $L_1$ norm $\|\cdot\|_1$, so that $\mathcal{S}_{d-1}=\{\bm{v}\in[0,1]^d: \sum_{j=1}^d v_j = 1\}$.}

Statistical methodology for multivariate extremes followed shortly after this theoretical study, and focused initially on inference for data arising as componentwise block maxima through parametrized forms of multivariate max-stable distributions \citep{Tawn90}. This was soon followed by more direct exploitation of the \edit{multivariate regular variation} assumption~\eqref{eq:mrv}, \editm{whereby parametric models were proposed for the spectral measure $H$, and inference performed on these \citep{ColesTawn91}}.

\begin{figure}[h]
	\centering
	\includegraphics[width=0.4\textwidth]{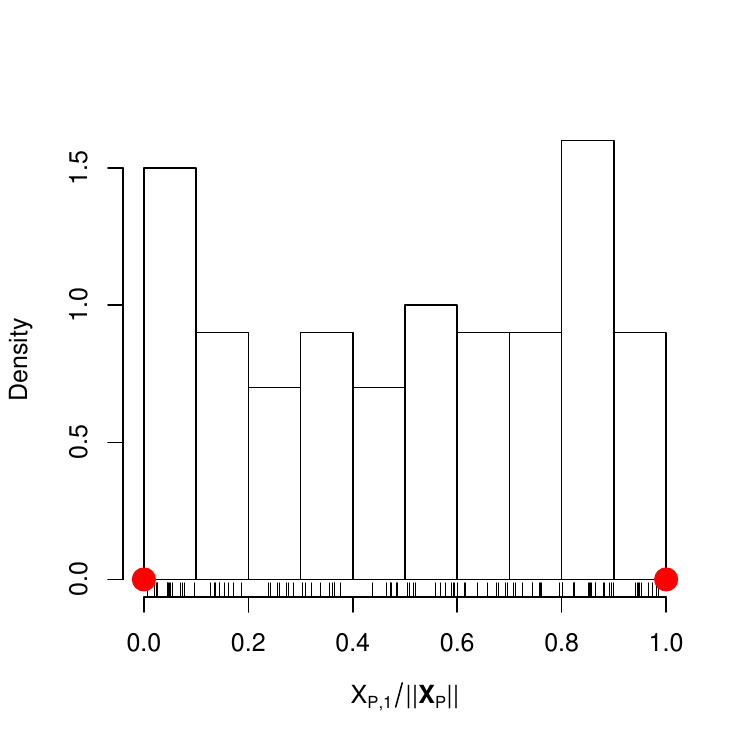}
	\includegraphics[width=0.4\textwidth]{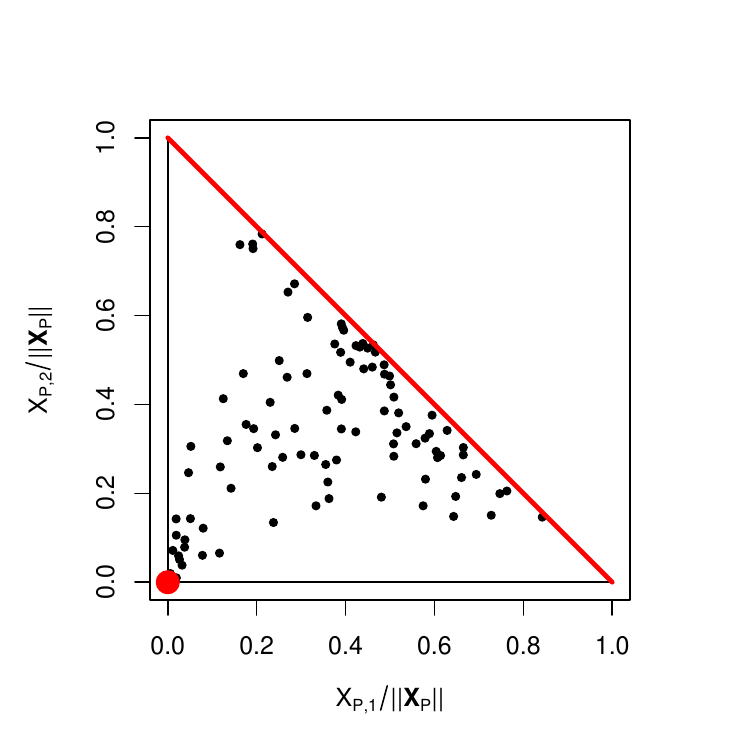}
	\caption{\edit{Illustration of the distribution of $\bm{X}_P/\|\bm{X}_P\|_1$, given that $\|\bm{X}_P\|_1$ is large, for two examples. Left panel: $\bm{X}_P$ has a bivariate Gaussian dependence structure with correlation parameter $\rho=0.8$. Right panel: $(X_{P,1},X_{P,2})$ have an inverted Clayton dependence structure while $X_{P,3}$ is simulated conditional upon the value of $X_{P,2}$ so that these variables have an inverted logistic dependence structure. The position of mass of the theoretical limiting $H$ is illustrated by red dots / lines.}}
	\label{fig:HMRV}
\end{figure}

The study of componentwise maxima is a natural multivariate extension of the univariate block maximum approach, and the associated \edit{multivariate regular variation} dependence condition~\eqref{eq:mrv} widely applicable. However, it has been known for a long time that while examples not satisfying~\eqref{eq:mrv} are rare, the number of examples for which this assumption forms a \emph{useful} basis for statistical inference is very much smaller. This is because, for many dependence structures, mass of the spectral measure $H$ accumulates on one or more regions of the form
\begin{align}
\mathbb{B}_C = \{\bm{v} \in \mathbb{S}_{d-1}: v_j > 0, j \in C; v_k = 0, k \not\in C\}, \qquad C \subset \{1,\ldots, d\}. \label{eq:BC}
\end{align}
When this is the case, joint extremes of the random vector $\bm{Y}$ (or equivalently $\bm{X}_P$) may not always occur simultaneously; see, e.g., \citet{Goixetal17} or \citet{Simpsonetal20} for a more detailed explanation. In practice, however, we never observe mass on such sets $\mathbb{B}_C$ at finite levels. \edit{This is illustrated in Figure~\ref{fig:HMRV}, which displays the distribution of $\bm{X}_P/\|\bm{X}_P\|_1$ when the associated radial variable $\|\bm{X}_P\|_1$ exceeds its 0.98 quantile for two examples. In the left panel, the true limiting spectral measure $H$ places mass only on the points $\mathbb{B}_{\{1\}}=\{(1,0)\}$ and $\mathbb{B}_{\{2\}}=\{(0,1)\}$, yet at observable levels, the distribution of angles is relatively evenly spread over $\mathbb{B}_{\{1,2\}}$, represented by the interval $(0,1)$. In the right panel, the limiting spectral measure places mass only on $\mathbb{B}_{\{3\}}$ and $\mathbb{B}_{\{1,2\}}$, but once again, at observable levels we see all values in $\mathbb{B}_{\{1,2,3\}}$. A consequence of this mismatch between finite-sample and limiting distribution is a common modelling assumption that $H$ places all mass on $\mathbb{B}_{\{1,\ldots,d\}} = \{\bm{v} \in \mathbb{S}_{d-1}: v_j > 0, j \in \{1,\ldots,d\}\}$, leading to overestimation of the probability of joint extremes. Moreover, even if one successfully detects the location of mass of the limiting object $H$, this does not lead to a practical strategy for performing extrapolation beyond the observed values; to achieve this, more detailed information on the behaviour of $\bm{X}_P$ before the limiting regime is required.}

\subsection{Geometric approach to multivariate extremes}
The early study of \edit{multivariate regular variation} was followed by a smaller body of work that examined the convergence of light-tailed multivariate sample clouds onto so-called limit sets \citep{Davisetal88,KinoshitaResnick91}. These ideas did not have a clear link with multivariate max-stable models and did not lead to the same proliferation of statistical methodology. More recently, several papers have revisited this geometric approach from a theoretical perspective \citep{BEN2010, BalkemaNolde10, BalkemaNolde12, Nolde14,NoldeWadsworth21} and in some cases shown how the shape of the limit set links to whether joint extremes of certain variables can occur.

\begin{figure}
	\centering
	\includegraphics[width=0.3\textwidth]{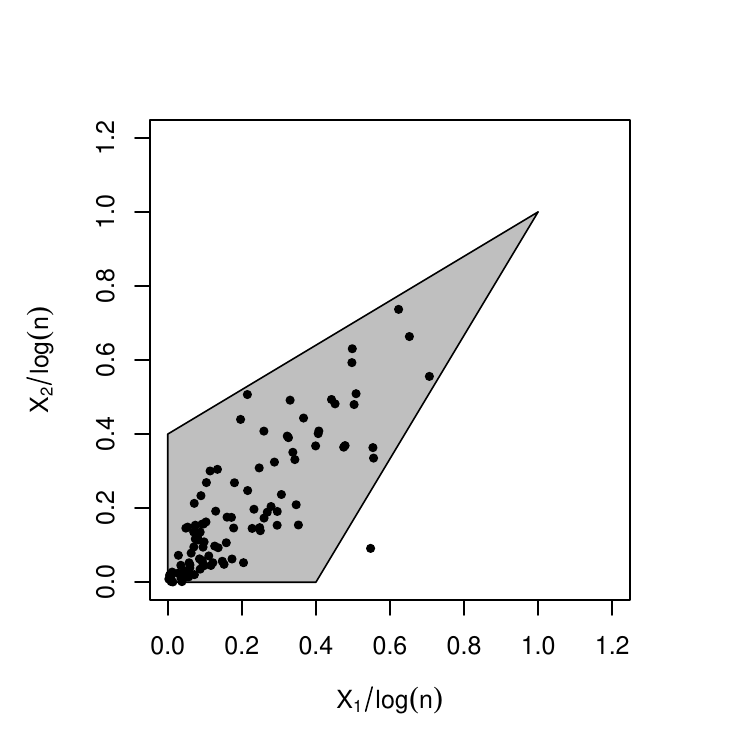}
	\includegraphics[width=0.3\textwidth]{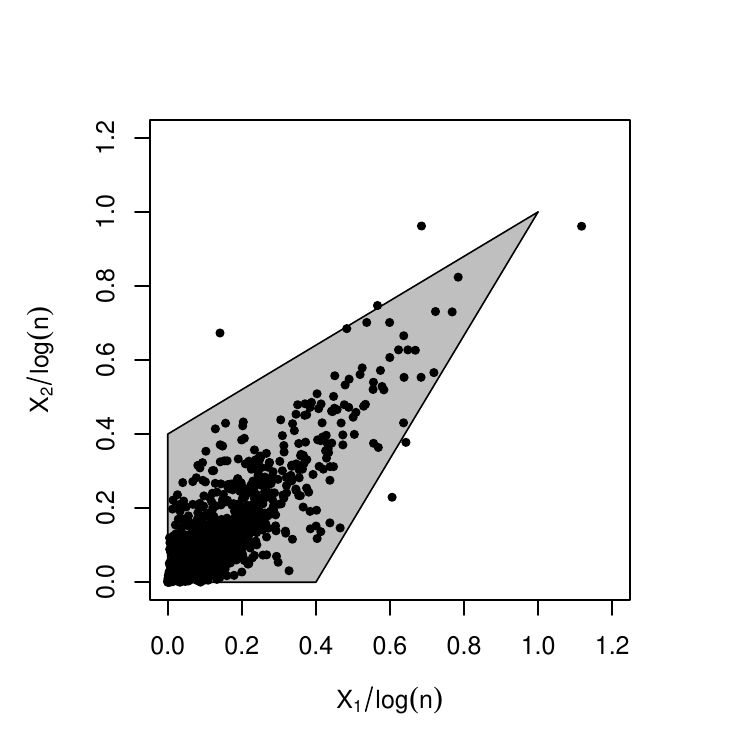}
	\includegraphics[width=0.3\textwidth]{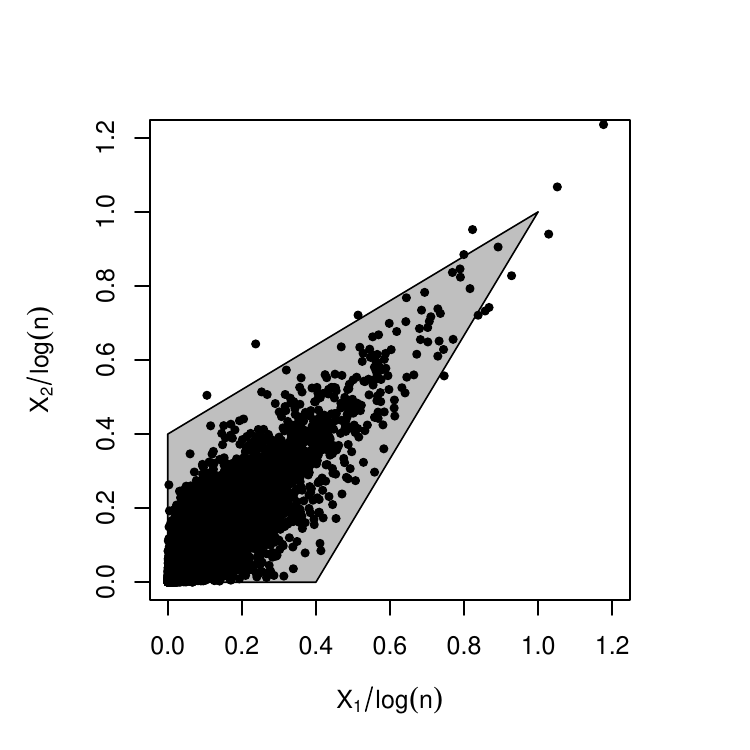}
	\caption{\edit{Illustration of the convergence of the scaled sample cloud $N_n$ onto a limit set. From left to right, sample sizes are $n=100,1000,10000$. The limit set $G$ is depicted by the grey polygon.}}
	\label{fig:convillus}
\end{figure}

To make ideas more concrete consider $n$ independent copies of a random vector $\bm{X}_{i}$, $i=1,\ldots,n$, with standard exponential margins; in practice, this will typically involve marginal transformation of the original vectors $\bm{Y}_i$. The \emph{scaled $n$-point sample cloud} is defined as 
 \begin{align*}
 N_n = \{\boldsymbol{X}_{1}/\log n, \ldots, \boldsymbol{X}_{n}/\log n\},
 \end{align*}
and we assume that this converges onto a \emph{limit set} $G=\{\bm{x}\in \mathbb{R}^d_+: g(\bm{x}) \leq 1\}$, where $g$ is the 1-homogeneous \emph{gauge function} of the limit set. \edit{This convergence is illustrated in Figure \ref{fig:convillus} for data with a logistic dependence structure, where the shape of the limit set can be seen to emerge in the scaled sample cloud as $n$ becomes large.} The precise sense of \edit{convergence of $N_n$ onto $G$}, and necessary and sufficient conditions for it, can be found in \citet{BEN2010}. \edit{Loosely, these conditions say that the expected number of points from $N_n$ lying in sets that intersect with the limit set tends to infinity, whereas the expected number of points lying in sets that are disjoint from the limit set converges to zero.} However, these \edit{specific} conditions are rather unintuitive \edit{and make it difficult to determine the form of $G$ for a given distribution}, which led \citet{Nolde14} and \citet{NoldeWadsworth21} to consider \edit{alternative} conditions in terms of the joint Lebesgue density of $\bm{X}$, \edit{when} it exists. Denoting this joint density by $f_{\bm{X}}$, a sufficient condition for convergence of $N_n$ onto $G$ is
\begin{align}
\lim_{t \to \infty}-\log f_{\bm{X}}(t \bm{x})/t  = g(\bm{x}) , \qquad \bm{x} \in [0,\infty)^d, \label{eq:logdens}
\end{align}
for a continuous gauge function $g$. Given that many statistical models have tractable joint densities and continuous gauge functions, equation~\eqref{eq:logdens} provides a simple way to determine the form of $g$, and hence $G$, in several examples \citep{NoldeWadsworth21}.  \edit{Further illustrations} of \edit{limit sets} $G$ are given for $d=2,3$ in Section~\ref{sec:examplelimitsets} of the supplement.

The shape of $G$ is important as a description of the extremal dependence of the underlying random vector. Limit sets exist for a much more general class of light-tailed marginal distributions than exponential, but we \editm{specialize} to this case so that there is a clear correspondence between the shape of $G$ and the dependence structure. \edit{In this case, the coordinatewise supremum of the limit set $G$ is the point $(1,\ldots,1)$, since for independent copies of an exponential variable $X_i$, the random variable $\max_{1 \leq i \leq n} X_i/\log n$ converges in probability to 1. Gauge functions, and therefore limit sets, of lower dimensional margins indexed by $J\subset\{1,\ldots,d\}$ can be found through the following minimization operation \citep{NoldeWadsworth21}:
\begin{align*}
g_J(\bm{x}_J) = \min_{x_k \geq 0, k \not\in J} g(\bm{x}), 
\end{align*}
where $\bm{x}_J = (x_j)_{j \in J}$. Exponential margins implies that for singleton sets $J=\{j\}$, $g_{\{j\}}(x_j) = x_j$.}

 \citet{NoldeWadsworth21} showed how $G$ can be used to determine an array of extremal dependence measures which generally relate to representations of multivariate extremes that are more useful than \edit{multivariate regular variation} when the spectral measure $H$ places mass on one or more sets $\mathbb{B}_C$ as in equation~\eqref{eq:BC}. These include expressions for determining the residual tail dependence coefficient \citep{LedfordTawn97}, key elements of the conditional extremes model \citep{HeffernanTawn04}, the angular dependence function \citep{WadsworthTawn13}, and the dependence coefficients of \citet{Simpsonetal20}, which can be used to help determine the sets $\mathbb{B}_C$ on which $H$ places mass.

Given the importance of the shape of $G$, a natural question that arises is how to estimate this from a sample of data. To date this question has been studied very little indeed; \citet{JacobMasse96} study estimation from a theoretical perspective but with no implementation. Very recently, \citet{SimpsonTawn22} outlined an estimation approach in the bivariate case. 
 
In this paper, we consider estimation of $G$ as part of a wider new approach to the statistical analysis of extreme values. \editm{While} $G$ is an object of interest in \editm{itself}, we direct our methodology more broadly at the question of statistical modelling and extrapolation for multivariate extreme values rather than focusing only on the descriptive aspects of extremal dependence that come from estimation of $G$. Our modelling approach allows in principle for any combination of joint extremes of sub-vectors of $\bm{Y}$ (equivalently, $H$ may place mass on any valid combination of sets $\mathbb{B}_C$), and permits extrapolation in all directions -- i.e., into the joint tail where all variables are large, or into other regions of the multivariate tail where only some variables are large. Existing alternatives to methodology based on \edit{multivariate regular variation} do not capture these possibilities in a coherent manner. 

\edit{To illustrate the potential importance of being able to capture complex structure in extremes, consider the dataset of river flow measurements from \citet{Simpsonetal20} that will also be analyzed in Section~\ref{sec:river}. Their analysis showed that there are some events where all four rivers were extreme simultaneously, but that there were also extreme episodes involving single rivers, or groups of two or three rivers without the others. This might be explained physically by the weather patterns causing the extremes, and the relationships between catchments. While \citet{Simpsonetal20} introduced and estimated coefficients to help determine this structure, they did not provide any modelling approach that could account for it, as we do here.}

Section~\ref{sec:Model} outlines our statistical model and assumptions. Section~\ref{sec:Examples} details theoretical examples that demonstrate applicability of the method. We focus on  details of statistical inference in Section~\ref{sec:inference}, and use simulation to show that our approach is very competitive for estimation of extreme set probabilities in a wide range of scenarios in Section~\ref{sec:simstudy}. Section~\ref{sec:data} contains applications to oceanographic and fluvial datasets, and we conclude in Section~\ref{sec:discussion}.

\section{Model and assumptions}
\label{sec:Model}

Here and throughout the rest of the paper, we assume that we have a random vector $\bm{X}$ with standard exponential margins and joint Lebesgue density denoted by $f_{\bm{X}}$. Marginal transformation can be applied as a standard step via estimation of each marginal distribution function. The assumption of a joint density is very common for statistical analysis, as it is required for most likelihood-based inference, for example. We further assume that the scaled sample cloud $N_n$ converges onto a limit set $G$ whose shape can either be described by a continuous gauge function $g$, or that we are only interested in the continuous part.

 Assumption~\eqref{eq:logdens}, which yields a sufficient condition for convergence of $N_n$ onto $G$, can equivalently be expressed $f_{\bm{X}}(t\boldsymbol{x})=\exp\{-tg(\boldsymbol{x})[1+o(1)]\}$ for $g(\bm{x})>0$ as $t \to \infty$. The homogeneity of $g$ suggests making the radial-angular transformation $R=\sum_{j=1}^d X_{j}, \boldsymbol{W} = \boldsymbol{X}/R$; such transformations are \editm{common} in multivariate extremes, but normally on Pareto, rather than exponential, margins. The Jacobian of this transformation is $r^{d-1}$, which leads to joint density of $(R,\bm{W})$: $ f_{R,\bm{W}}(r,\boldsymbol{w}) = r^{d-1} f_{\bm{X}}(r\bm{w}) = r^{d-1}\exp\{-rg(\boldsymbol{w})[1+o(1)]\}$, as $r \to \infty$. This in turn means that the conditional density of $R|\boldsymbol{W}=\boldsymbol{w}$ satisfies 
 \begin{align}
 f_{R|\bm{W}}(r|\boldsymbol{w}) \propto r^{d-1} \exp\{-rg(\boldsymbol{w})[1+o(1)]\}, \qquad r \to \infty. \label{eq:RgivenW}
 \end{align}
 We recognize the form of a gamma kernel in the non-asymptotic terms of~\eqref{eq:RgivenW}, suggesting that when $R|\bm{W}=\bm{w}$ is large, its distribution could potentially be well approximated by a gamma distribution. Indeed, if the $o(1)$ term in the exponent is negligible, this suggests a truncated-and-renormalized gamma approximation above a high threshold $r_0(\bm{w})$ of the conditional distribution $R|\bm{W}=\bm{w}$.
 
 A valid concern is whether the $o(1)$ term in the exponent of~\eqref{eq:RgivenW} is really negligible. In Section~\ref{sec:Examples} we detail several examples which in fact have the more helpful asymptotic form $f_{R|\bm{W}}(r|\boldsymbol{w}) \propto r^{d-1} \exp\{-rg(\boldsymbol{w})\}[1+o(1)]$, i.e., with the $o(1)$ outside of the exponent, and give explicit rates for this term. Based on this \edit{latter} asymptotic representation, we focus in this paper on the model
 \begin{align}
 R|\{\bm{W}=\bm{w},R>r_0(\bm{w})\} \overset{.}{\sim}\mbox{truncGamma}(\alpha, g(\boldsymbol{w})), \label{eq:rwgamma}
 \end{align}
 where $\alpha>0$ is the gamma shape, and $g(\bm{w})$ is the gamma rate parameter. In most examples, the theoretical shape parameter is $\alpha=d$, but for modelling purposes the flexibility of an estimated shape is desirable. By parametrizing flexible forms for the gauge function $g(\bm{w}) = g(\bm{w}; \bm{\theta})$, we can use \edit{approximation}~\eqref{eq:rwgamma} to estimate these parameters. Full details of our approach are given in Section~\ref{sec:inference}, including diagnostic plots for assessing \edit{approximation}~\eqref{eq:rwgamma}.

\section{Examples}
\label{sec:Examples}

\edit{In this section we consider a variety of examples. The convergence onto a limit set $G$ holds very broadly, and in many examples the gauge function for this limit set in exponential margins can be recovered fully or partly from convergence~\eqref{eq:logdens}. The form of the gauge function and limit set for several examples, including multivariate $t_\nu$, light-tailed elliptical, skew-normal, generalized hyperbolic, certain mixture distributions and multivariate generalized Pareto forms has been derived in \citet{BEN2010}, \citet{Nolde14}, \citet{NoldeWadsworth21}, and \citet{Zhangetal22}, for example, although not always in exponential margins.}

The validity and quality of the truncated gamma approximation in~\eqref{eq:rwgamma} to the conditional density in~\eqref{eq:RgivenW} depends on the $o(1)$ term. Since this lies in the exponent, it is not always guaranteed to be negligible. In this section, we explicitly calculate the density of $R|\bm{W}=\bm{w}$ for various theoretical examples, showing that most in fact have the form $f_{R|\bm{W}}(r|\boldsymbol{w}) \propto r^{d-1} \exp\{-rg(\boldsymbol{w})\} [1+o(1)]$, as $r \to \infty$. The exception to this is the Gaussian dependence structure, for which we find $f_{R|\bm{W}}(r|\boldsymbol{w}) \propto r^{\alpha(\bm{w})-1} \exp\{-rg(\boldsymbol{w})\} [1+o(1)]$, as $r \to \infty$, i.e., the conditional gamma form is still applicable, but the shape parameter depends on the value of $\bm{w}$. Nonetheless, further investigations, described briefly below and in more detail in Section~\ref{sec:gauss} of the supplement, show the assumption of a common shape in model~\eqref{eq:rwgamma} does not appear problematic in practice. This is also supported by our simulation study in Section~\ref{sec:simstudy}. More generally, we will incorporate model checking of assumption~\eqref{eq:rwgamma} into our statistical analysis.

For each distribution, we \editm{provide} the overall form of $f_{R|\bm{W}}(r|\bm{w})$, with further calculations given in Section~\ref{sec:rwderivations} of the supplement. We denote the ordered values of the vector $\bm{w}$ (and similarly $\bm{x}$) by $w_{(1)}\geq w_{(2)} \geq \cdots \geq w_{(d)}>0$, assuming the minimum to be positive. In the convergence rates given below, we assume a strict ordering  $w_{(1)} > w_{(2)} >\cdots > w_{(d)}>0$; where this is not the case, following the derivations in the supplement, one usually observes improved rates, e.g., $O(e^{-r(w_{(d-2)}-w_{(d)})})$ replacing $O(e^{-r(w_{(d-1)}-w_{(d)})})$ if $w_{(d-2)}>w_{(d-1)} = w_{(d)}$.

\paragraph{Multivariate max-stable and generalized Pareto distributions}
\edit{Multivariate max-stable distributions are most readily expressed by their distribution functions. In exponential margins,
\begin{align*}
\Pr(\bm{X} \leq \bm{x}) = \exp\left\{-V\left([-\log(1-e^{-\bm{x}})]^{-1}\right)\right\},
\end{align*}
where $V: \mathbb{R}^d_+ \to \mathbb{R}_+$ is the homogeneous of order $-1$ \emph{exponent function}, and operations are applied componentwise.} The general asymptotic form of the density for a max-stable distribution in exponential margins is therefore
\begin{align*}
f_{\bm{X}}(t\bm{x}) = \exp\{-V(e^{t\bm{x}}+1/2+O(e^{-t{\bm{x}}}))\}\sum_{\pi \in \Pi}\prod_{s \in \pi} V_s(e^{t\bm{x}}+1/2+O(e^{-t\bm{x}})) \times e^{t \sum_{j=1}^d x_j}[1+O(e^{-2tx_{(d)}})],
\end{align*}
$t\to \infty$, where $\Pi$ is the set of all partitions of $\{1,\ldots,d\}$, and $V_s(\bm{z}) = \partial^{|s|}V(\bm{z})/\prod_{j \in s}\partial z_j$.

We focus on the $d$-dimensional logistic distribution, for which $V(\bm{z}) = \left(\sum_{j=1}^d z_j^{-1/\gamma}\right)^\gamma$ with parameter $\gamma \in (0,1]$. This distribution has \editm{gauge function} $g(\bm{x}) = \sum_{j=1}^d x_j/\gamma + (1-d/\gamma)x_{(d)}$, and 
\begin{align*}
f_{R|\bm{W}}(r|\bm{w}) \propto r^{d-1} e^{-rg(\bm{w})}[1+O(e^{-r(w_{(d-1)}-w_{(d)})/\gamma})+O(e^{-rw_{(d)}})], \qquad r \to \infty.
\end{align*}

The simpler form of the densities make calculations more straightforward for corresponding multivariate generalized Pareto distributions (MGPDs), which are related to max-stable distributions \citep{RootzenTajvidi06,Rootzenetal18}. The support of MGPDs whose margins have unit scale and zero shape is contained in $\{\bm{x}\in \mathbb{R}^d: x_{(1)}>0\}$.  Densities for several models for which the spectral measure $H$ places mass only on $\mathbb{B}_{\{1,\ldots,d\}}$ are given in \citet{Kiriliouketal18}; in such cases the dependence structure can be determined by focusing on large values of $\bm{x}>\bm{0}$. Further details are in the supplement, \edit{Section}~\ref{sec:rwderivations}.

For the MGPD associated to the negative logistic max-stable distribution \citep{Galambos75,Dombryetal16}, $g(\bm{x}) = (1+d\gamma) x_{(1)} - \sum_{j=1}^d x_j\gamma$, $\gamma>1$ and
\begin{align*}
f_{R|\bm{W}}(r|\bm{w}) \propto r^{d-1} e^{-rg(\bm{w})}\left[1+O\left(e^{r(w_{(2)}-w_{(1)})\gamma}\right)\right], \qquad r \to \infty.
\end{align*}

For the MGPD associated to the Dirichlet max-stable distribution \citep{ColesTawn91}, $g(\bm{x}) = (1+\sum_{j=1}^d \theta_j) x_{(1)} - \sum_{j=1}^d \theta_j x_j$, for all $\theta_j>0$, and
\begin{align*}
f_{R|\bm{W}}(r|\bm{w}) \propto r^{d-1} e^{-rg(\bm{w})}\left[1+O\left(e^{r(w_{(2)}-w_{(1)})}\right)\right], \qquad r \to \infty.
\end{align*}

\paragraph{Inverted max-stable distributions}
Inverted max-stable distributions \edit{are derived by translating the joint lower tail of max-stable distributions to be the joint upper tail. This is achieved by applying a monotonically decreasing marginal transformation to a max-stable random vector. In exponential margins they} have density
\begin{align*}
f_{\bm{X}}(\bm{x}) = \exp\{-l(\bm{x})\}\sum_{\pi \in \Pi}\prod_{s \in \pi} l_s(\bm{x}),
\end{align*}
where $l$ is the \emph{stable tail dependence function} of the corresponding max-stable distribution, obtained via $l(\bm{x}) = V(1/\bm{x})$, and  $l_s(\bm{x}) = \partial^{|s|}l(\bm{x})/\prod_{j \in s}\partial x_j$. The gauge function is always $g(\bm{x}) = l(\bm{x})$. Owing to the fact that $l_s(\bm{x})$ is homogeneous of order $1-|s|$, we obtain
\begin{align*}
f_{R|\bm{W}}(r|\bm{w}) \propto r^{d-1} \exp\{-rg(\bm{w})\}[1+O(r^{-1})], \qquad r \to \infty.
\end{align*}

\paragraph{Multivariate Gaussian distribution}
We consider the multivariate Gaussian dependence structure with correlation matrix $\Sigma$. When one or more correlation parameters is negative then the continuous convergence $-\log f_{\bm{X}}(t\bm{x})/t \to g(\bm{x})$ fails when components of $\bm{x}$ are zero \citep{NoldeWadsworth21}. Since we are considering $w_{(d)}>0$ this is not an issue here, but we note that to fully capture negative association it is ideal to reformulate ideas in terms of Laplace, rather than exponential, margins; see \citet{NoldeWadsworth21} and Section~\ref{sec:discussion}. 
For $\Sigma$ with non-negative entries, $g(\bm{x}) = (\bm{x}^{1/2})^{\top}\Sigma^{-1}\bm{x}^{1/2}$, \edit{where $\bm{x}^{1/2} = (x_1^{1/2},\ldots,x_d^{1/2})^\top$}, and 
\begin{align*}
f_{R|\bm{W}}(r|\bm{w}) \propto r^{\alpha(\bm{w})-1}\exp\{-rg(\bm{w})\}\left[1+O\left(\frac{(\log r)^2}{r}\right)\right], \qquad \alpha(\bm{w}) = \frac{d}{2} + \frac{(\bm{w}^{1/2})^{\top}\Sigma^{-1}\bm{w}^{-1/2}}{2},
\end{align*} 
 $r \to \infty$. In this case, the gamma shape parameter therefore depends on $\bm{w}$, and the region on which $\alpha(\bm{w})>0$ depends on the entries of $\Sigma$. We investigate this further in Section~\ref{sec:gauss} of the supplement, showing that local estimates of $\alpha$ do not vary strongly with $\bm{w}$ and may reasonably be assumed constant. We also show that results from our model are useful even in the (typically small) regions where $\alpha(\bm{w}) \leq 0$. 

\paragraph{Multivariate $t_{\nu}$ distribution}

We consider the multivariate $t$ distribution with $\nu$ degrees of freedom, focusing only on positive dependence; see the supplement for further comment. The gauge function is $g(\bm{x}) = (1+d/\nu)x_{(1)}-\sum_{j=1}^d x_j/\nu$, and
\begin{align*}
f_{R|\bm{W}}(r|\bm{w}) \propto r^{d-1} e^{-rg(\bm{w})}\left[1+O(e^{r(w_{(2)}-w_{(1)})/\nu}) +O(e^{-2 r w_{(d)}/\nu})\right], \qquad r \to \infty.
\end{align*}

\paragraph{Clayton and inverted Clayton copulas}

We consider the Clayton and inverted Clayton copulas with parameter $\gamma>0$. The Clayton copula has $g(\bm{x}) = \sum_{j=1}^d x_j$, and
\begin{align*}
f_{R|\bm{W}}(r|\bm{w}) \propto r^{d-1} e^{-rg(\bm{w})}[1+O(e^{-r w_{(d)}})], \qquad r \to \infty.
\end{align*}
The inverted Clayton copula has $g(\bm{x}) = (1+d\gamma)x_{(1)}-\sum_{j=1}^d x_j \gamma$, and
\begin{align*}
f_{R|\bm{W}}(r|\bm{w}) \propto r^{d-1} e^{-rg(\bm{w})}[1+O(e^{-r(w_{(2)}-w_{(1)})})], \qquad r \to \infty.
\end{align*}

\noindent
In the supplement, we also calculate $f_{R|\bm{W}}(r|\bm{w})$ for a trivariate vine copula example.

\section{Statistical inference}
\label{sec:inference}
\subsection{Calculating the threshold \texorpdfstring{$r_0(\bm{w})$}{r0w}}
\label{sec:threshold}

\edit{To implement model~\eqref{eq:rwgamma}, we firstly need to calculate $r_0(\bm{w})$, which represents a high threshold of the conditional distribution $R|\bm{W} = \bm{w}$.}
A natural approach to calculating \edit{this threshold} is quantile regression, treating $\bm{W}$ as the covariate. A similar approach has been taken in the context of establishing covariate-dependent thresholds in univariate extreme value analysis \citep{NorthropJonathan11}. When data are bivariate, so that $\bm{W} \in \mathcal{S}_1$ is equivalent to $W \in [0,1]$, this approach is straightforward. However, standard parametric quantile regression requires a high degree of manual tuning to ensure that the model form captures the relation between $R$ and $\bm{W}$ well. We therefore suggest using additive quantile regression \citep{Fasioloetal21} via the corresponding R \editm{package} \texttt{qgam}.

\begin{figure}
	\centering
	\includegraphics[width=0.3\textwidth]{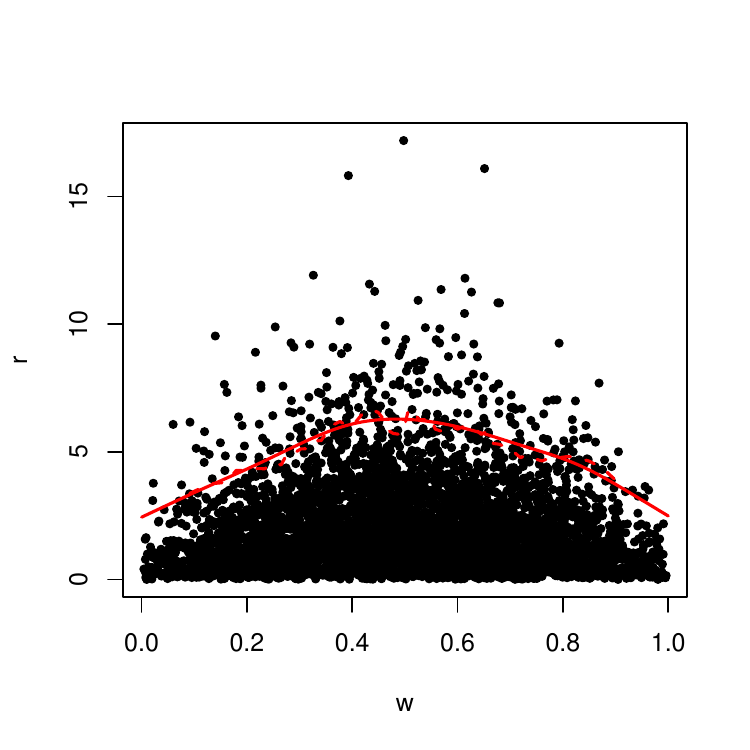}
	\includegraphics[width=0.3\textwidth]{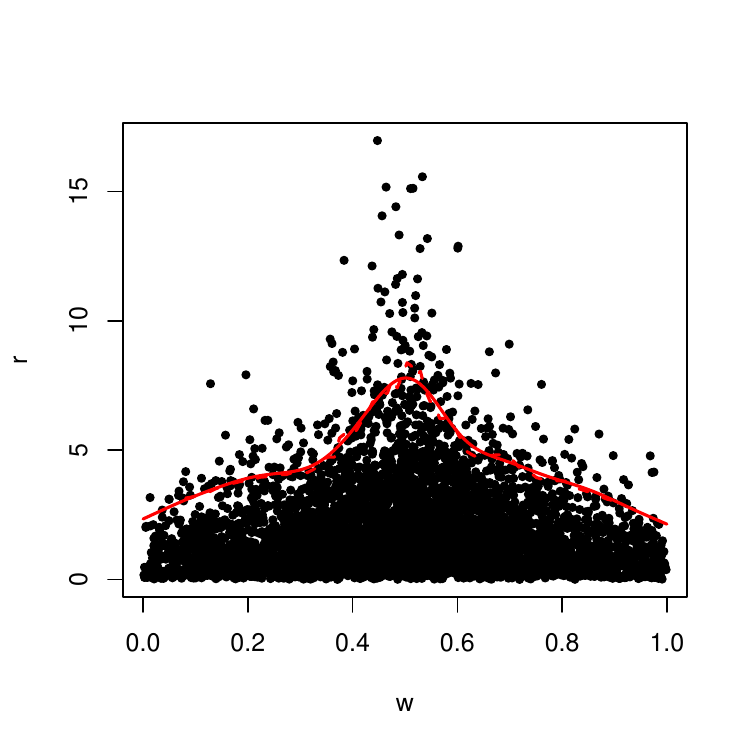}
	\includegraphics[width=0.3\textwidth]{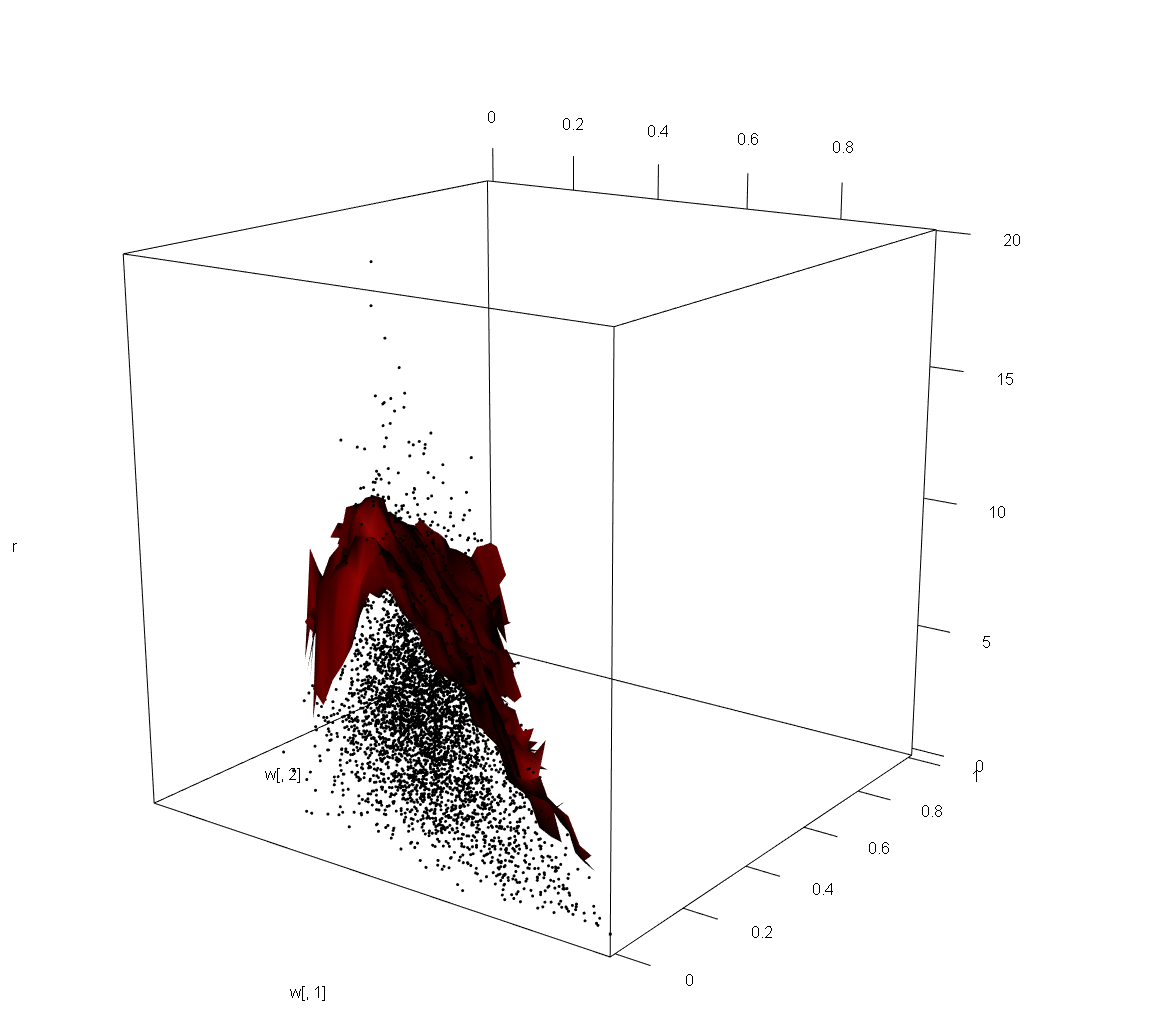}\\
	\includegraphics[width=0.3\textwidth]{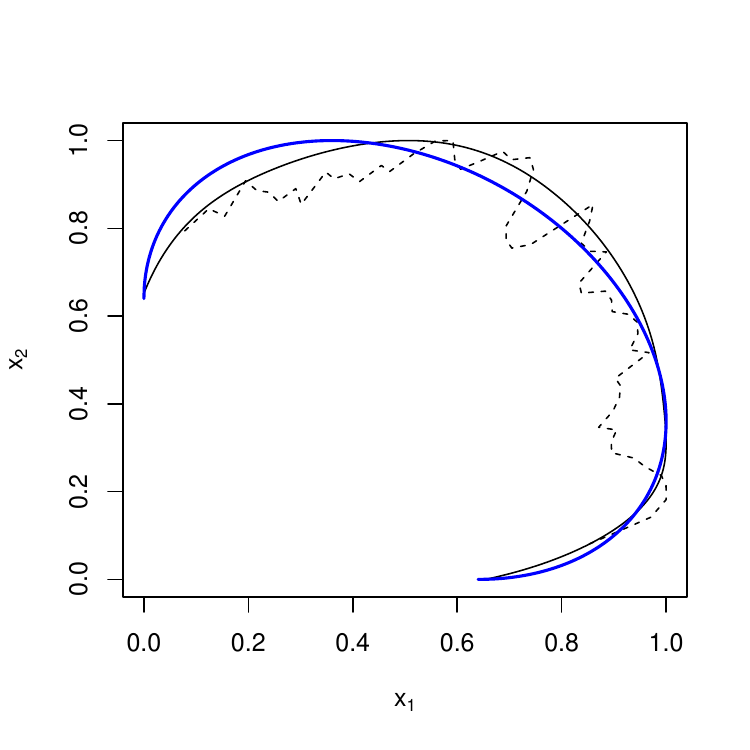}
	\includegraphics[width=0.3\textwidth]{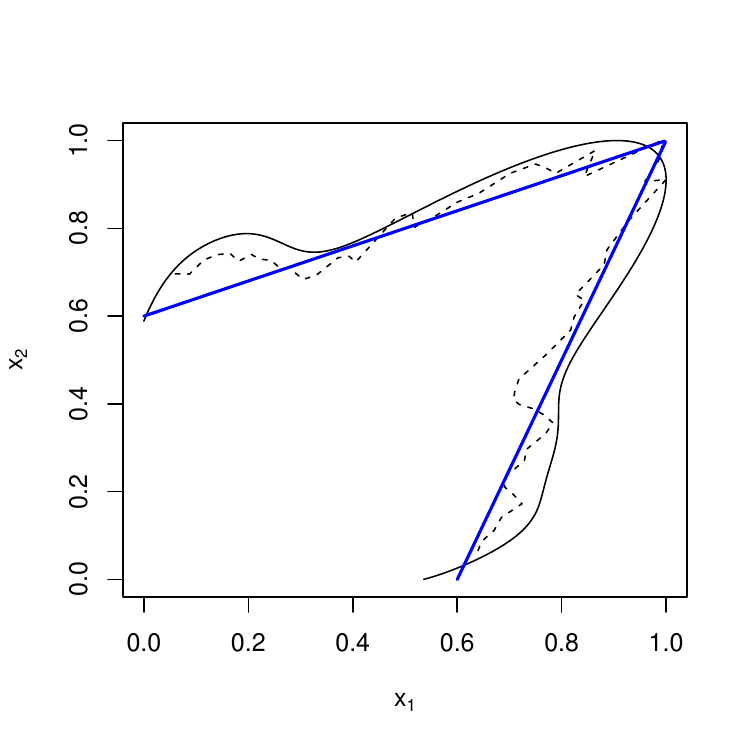}
	\includegraphics[width=0.3\textwidth]{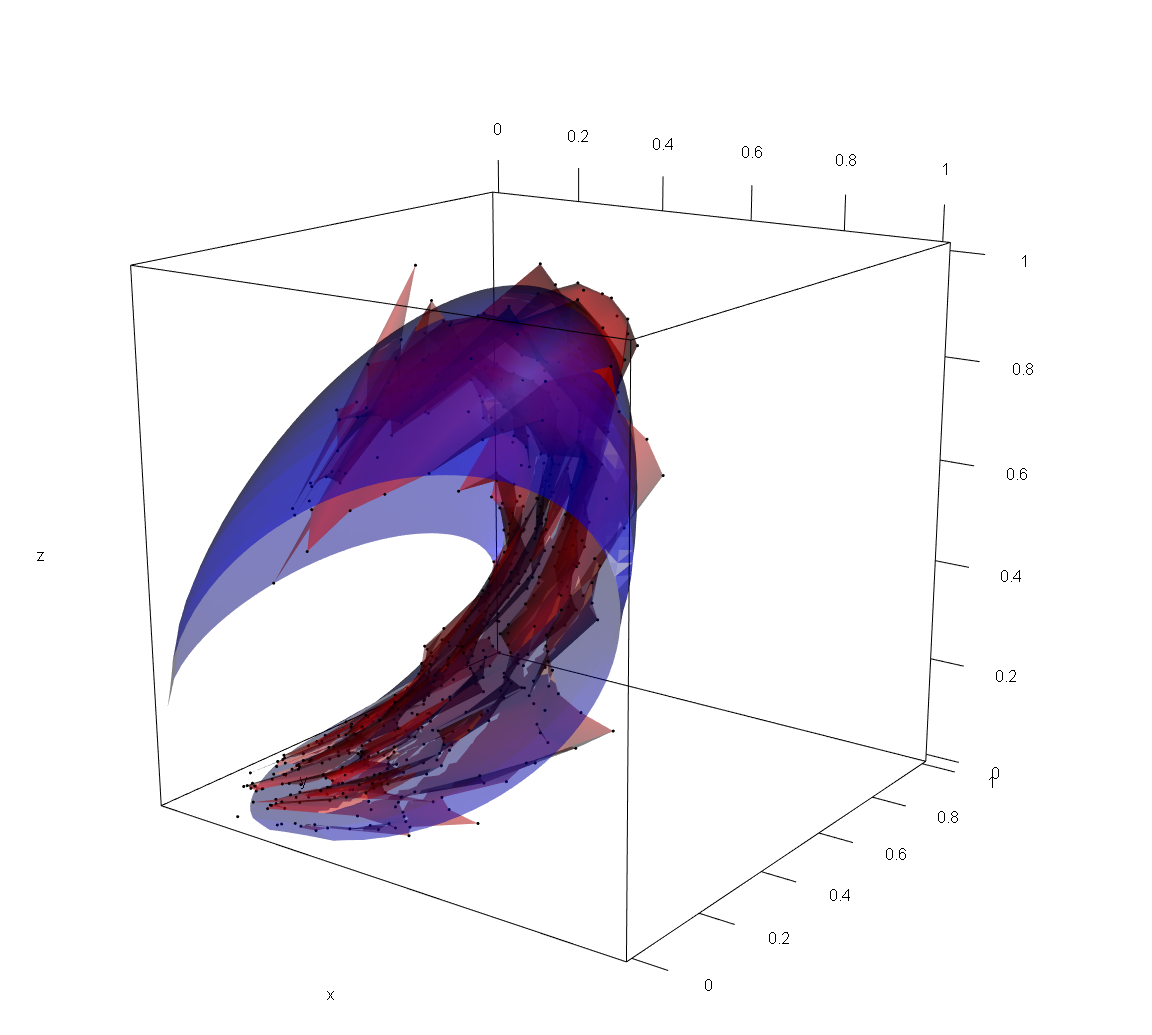}
	\caption{Top row: $R$ against $\bm{W}$, with the estimated 0.95 quantile of $R|\bm{W}=\bm{w}$ in red. In the left and centre plots, solid lines represent the output from \texttt{qgam}, and dashed lines from rolling-windows quantiles. In the right plot, the surface is calculated through a rolling-windows technique. Bottom row: Plots of \edit{$\bm{x} = \bm{v}r_0(\bm{v})$}, rescaled to lie in $[0,1]^d$ \edit{as per equation \eqref{eq:empiricalgauge}}.  In the left and centre plots, solid \edit{black} lines represent the output from \texttt{qgam}, and dashed lines from rolling-windows quantiles. The blue lines are the unit level sets of $g(\bm{x})$, with $g$ the true gauge function. In the right plot, the red surface comes from the rolling-windows technique, and the blue surface is the unit level set of the true gauge function.}
	\label{fig:r0w}
\end{figure}

When $\bm{W} \in \mathcal{S}_{d-1}$, $d>2$, then both parametric and additive quantile regression become more difficult due to the specific support of $\bm{W}$ on the simplex. A simple alternative is to calculate quantiles of $R|\bm{W}=\bm{w}$ from overlapping blocks of $\bm{W}$ values, which is feasible for relatively low dimensions, but becomes more laborious as $d$ grows. \edit{The top row of} Figure~\ref{fig:r0w} illustrates the concepts for $d=2,3$. In each case, $r_0(\bm{w})$ is calculated as the 0.95 quantile of $R|\bm{W}=\bm{w}$.

%--------
%-------- 

In the second row of Figure~\ref{fig:r0w} we demonstrate that the threshold $r_0(\bm{w})$, suitably rescaled, can be viewed as a non-/semi-parametric estimate of $g$. The reason for this can roughly be explained by considering the case where the gamma approximation is exact. \edit{Let $\bar{F}(r|\bm{w})$ be the (gamma) survival function of $R|\bm{W}=\bm{w}$, then for quantile regression at level $\tau \in (0,1)$, $\bar{F}(r_0(\bm{w})|\bm{w}) = 1-\tau$. We have
\begin{align}
\bar{F}(r_0(\bm{w})|\bm{w}) & = 1- \int_0^{r_0(\bm{w})} \frac{g(\bm{w})^\alpha}{\Gamma(\alpha)}v^{\alpha-1}e^{-v g(\bm{w})} \,\mbox{d}v \notag \\
& =  1- \int_0^{r_0(\bm{w})g(\bm{w})} \frac{s^{\alpha-1}}{\Gamma(\alpha)}e^{-s} \,\mbox{d}s = 1-\tau, \label{eq:r0wg}
\end{align}
using the change of variables $s=g(\bm{w})v$. Equation~\eqref{eq:r0wg} is solved by taking $r_0(\bm{w}) = C_\tau / g(\bm{w})$, with $C_\tau$ the solution to the equation $\int_0^{C_\tau} \frac{s^{\alpha-1}}{\Gamma(\alpha)}e^{-s} \,\mbox{d}s = \tau$. Since the gamma approximation is only asymptotically valid, we have in practice that $r_0(\bm{w}) \approx C_\tau / g(\bm{w})$ for $\tau$ close to 1.} \edit{To plot unit level sets of the gauge function $g$, we plot points $\bm{x} = \bm{v}/g(\bm{v})$, where $\bm{v}$ is a sequence of points covering the simplex $\mathcal{S}_{d-1}$. Consequently we can compare $r_0(\bm{w})$ to $g$ by plotting points $\bm{x} = \bm{v}r_0(\bm{v})/C_\tau$. However, since the gamma approximation is not exact, we instead scale each margin so that the coordinatewise supremum exactly equals one, by plotting  
\begin{align}
\bm{x} = \left(v_1 r_0(\bm{v})/\max\{v_1 r_0(\bm{v})\},\ldots, v_d r_0(\bm{v})/\max\{v_d r_0(\bm{v})\}\right). \label{eq:empiricalgauge}
\end{align}
}
 We will use \edit{the observation that links $r_0(\bm{w})$ and $g(\bm{w})$} later to assist with model checking, but note that, combined with extension of additive quantile regression to higher dimensions, \edit{this} presents a very interesting avenue for future work.

\subsection{Likelihood}

In order to fit model~\eqref{eq:rwgamma}, we use likelihood-based inference. For $n_0$ independent observations of $R_i|\{\bm{W}_i = \bm{w}_i, R_i>r_0(\bm{w}_i)\}$, $i=1,\ldots, n_0$, the likelihood that we maximize is
\begin{align}
L(\bm{\psi}) = \prod_{i=1}^{n_0} \frac{g(\bm{w}_i; \bm{\theta})^{\alpha}}{\Gamma(\alpha)} \frac{r_i^{\alpha - 1} e^{-r_ig(\bm{w}_i;\bm{\theta})}}{\bar{F}(r_0(\bm{w}_i); \alpha, g(\bm{w}_i;\bm{\theta}))}, \label{eq:lik}
\end{align}
where $\bm{\psi} = (\alpha,\bm{\theta})^\top$ and $\bar{F}(\cdot; \alpha, g(\bm{w};\bm{\theta}))$ represents the gamma survival function with shape parameter $\alpha$, and rate parameter $g(\bm{w};\bm{\theta})$. Estimates of uncertainty in the maximum likelihood estimators may be obtained through the inverse Hessian matrix, subject to model validity and independence checks, or via the bootstrap. In practice, many datasets exhibit weak-to-moderate temporal dependence, so that while likelihood~\eqref{eq:lik} may be used for parameter estimation \citep[e.g.][]{ChandlerBate07}, block-bootstrap techniques will be preferable for estimation of uncertainty.

\subsection{Gauge functions and model selection}

\subsubsection{Gauge functions from specific distributions}
\label{sec:gaugespecific}
Key to a successful fit of model~\eqref{eq:rwgamma} via likelihood~\eqref{eq:lik} are flexible parametrized forms of $g$ that are able to capture a wide variety of limit set shapes. In Section~\ref{sec:Examples}, we detail various forms of gauge function that come from different underlying distributions, some of which are illustrated in Section~\ref{sec:examplelimitsets} of the supplement. Further forms can also be found in \citet{NoldeWadsworth21}. Any of these parametric forms could be fitted as a candidate model, and standard model-selection techniques, such as information criteria, used to establish a best choice; we will demonstrate this in our simulation study of Section~\ref{sec:simstudy}. 

A key attraction of our new approach to inference for multivariate extremes is the ability to be able to capture the complex dependence structures that arise when different sub-groups of variables can potentially be co-extreme while the others are small. Under \edit{multivariate regular variation}, this corresponds to the spectral measure \edit{$H$} placing mass on sets $\mathbb{B}_C$ as described in Section~\ref{sec:intro}. In order to capture these scenarios, we consider the gauge function corresponding to the asymmetric logistic distribution \citep{Tawn90}, which can place mass on any valid combination of sets $\mathbb{B}_C$. The full expression for this involves minimization over several components, and is given in Section~\ref{sec:asylog} of the supplement. Figure~\ref{fig:AL} depicts some of the potential limit sets arising from this structure when $d=3$. 

\begin{figure}
	\centering
	\includegraphics[width=0.3\textwidth]{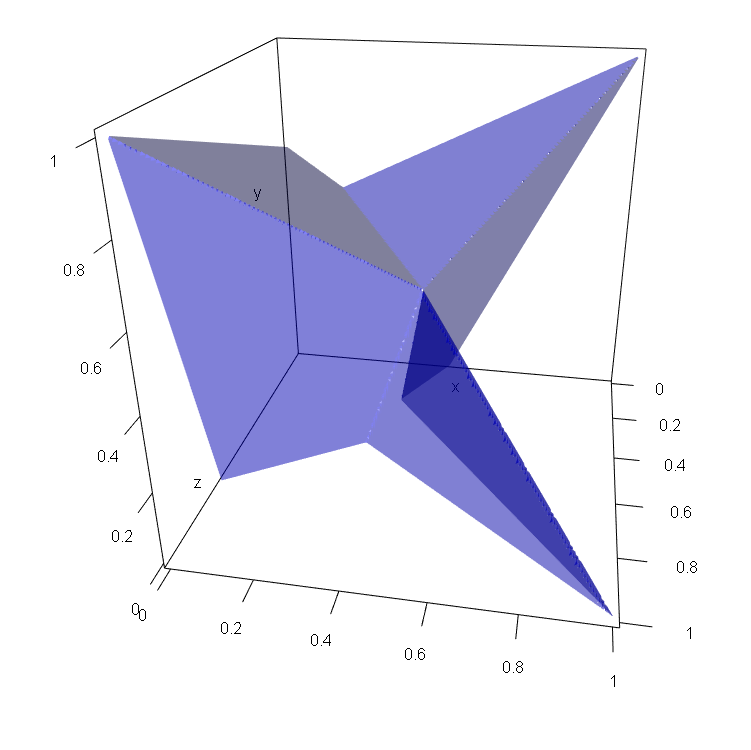}
	\includegraphics[width=0.3\textwidth]{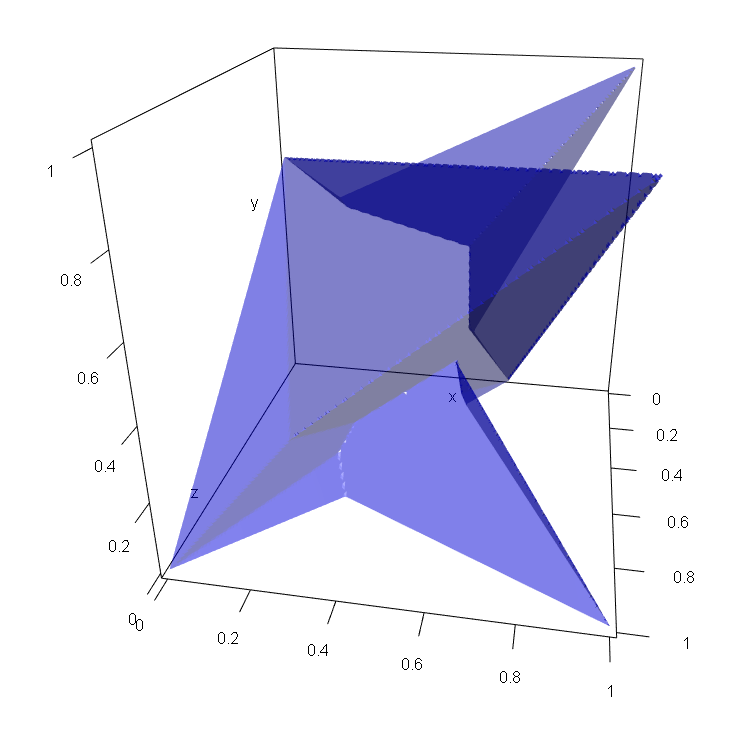}
	\includegraphics[width=0.3\textwidth]{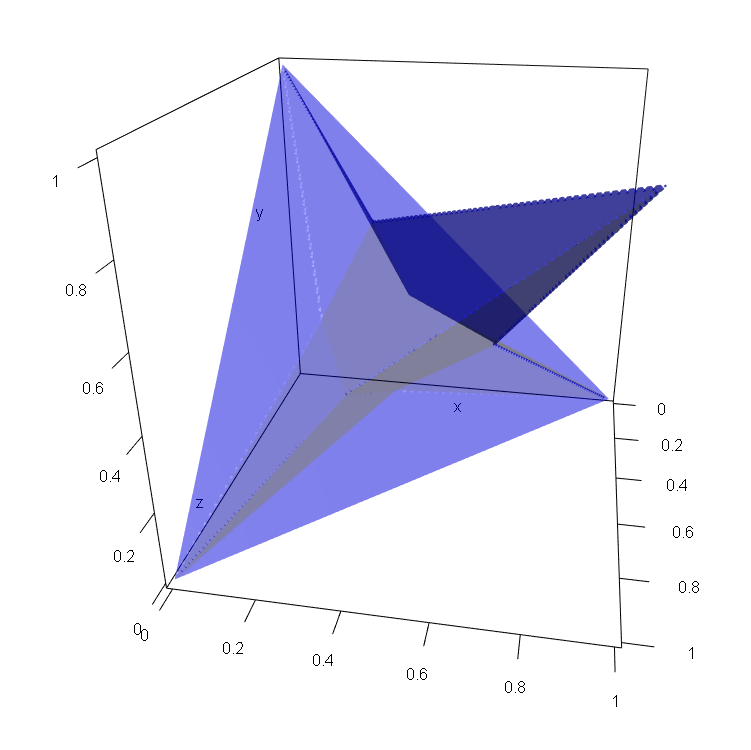}
	\caption{Example limit sets \edit{$G$} (area between surface and planes $x_j=0$) from the asymmetric logistic distribution. Left: mass of $H$ on $\mathbb{B}_{\{1,2\}}$, $\mathbb{B}_{\{1,3\}}$, $\mathbb{B}_{\{2,3\}}$ with parameters $\gamma_{\{1,2\}}=0.5$, $\gamma_{\{1,3\}}=0.2$, $\gamma_{\{2,3\}}=0.7$. Centre: mass on $\mathbb{B}_{\{3\}}$, $\mathbb{B}_{\{1,2\}}$, $\mathbb{B}_{\{1,3\}}$, $\mathbb{B}_{\{1,2,3\}}$ with $\gamma_{\{1,2\}}=0.5$, $\gamma_{\{1,3\}}=0.5$, $\gamma_{\{1,2,3\}}=0.7$. Right: mass on $\mathbb{B}_{\{1\}}$, $\mathbb{B}_{\{2\}}$, $\mathbb{B}_{\{3\}}$, $\mathbb{B}_{\{1,2,3\}}$ with $\gamma_{\{1,2,3\}}=0.5$.}
	\label{fig:AL}
\end{figure}

\subsubsection{Additively mixing gauge functions}
\label{sec:additivemixing}

The gauge functions described in Section~\ref{sec:Examples} provide a starting point for inference on model~\eqref{eq:rwgamma}, but may not always be flexible enough to capture the structures of observed data. We now consider how to mix gauge functions to generate more flexible models. 
\edit{As mentioned in Section \ref{sec:intro}, the} limit sets $G$ for data with exponential margins have coordinatewise supremum equal to $(1,\ldots,1)$\edit{; equivalently}, the one-dimensional marginal gauge functions are \edit{$g_{\{j\}}(x_j) = x_j$}. Each form of $g$ given in Section~\ref{sec:Examples}
 satisfies this constraint, and we \edit{require} any scheme for mixing gauge functions to \editm{also} satisfy this, since they will be applied to data in exponential margins.

\begin{figure}
	\centering
	\includegraphics[width=0.3\textwidth]{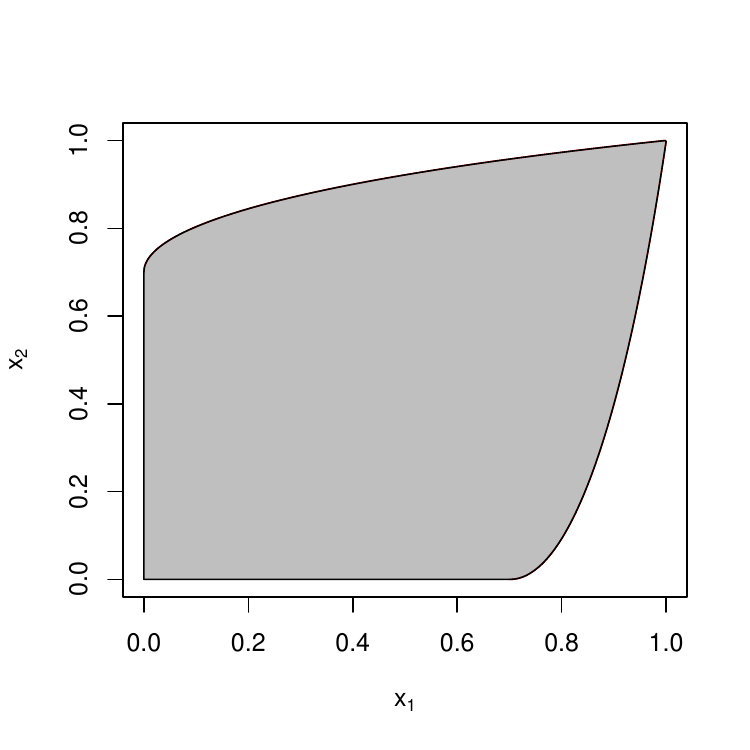}
	\includegraphics[width=0.3\textwidth]{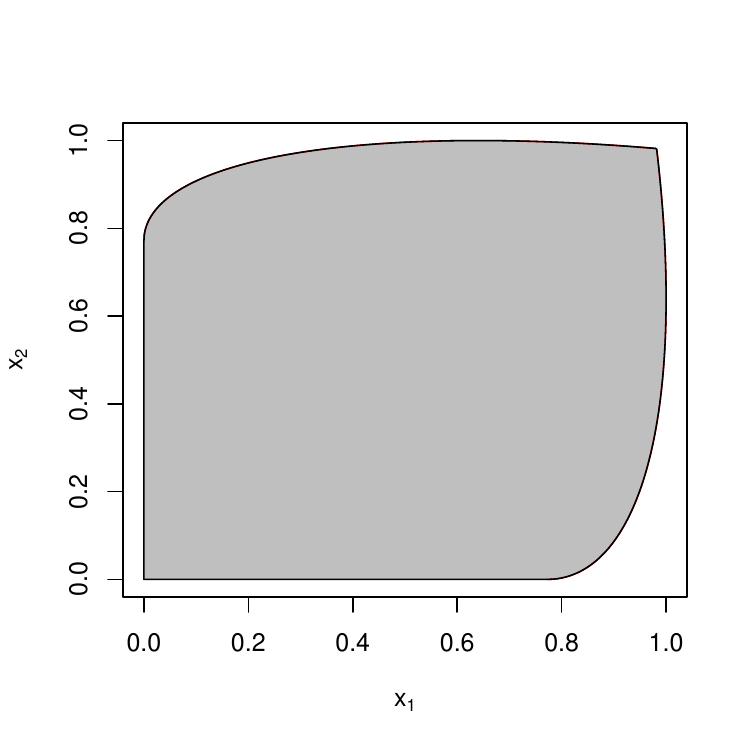}
	\includegraphics[width=0.3\textwidth]{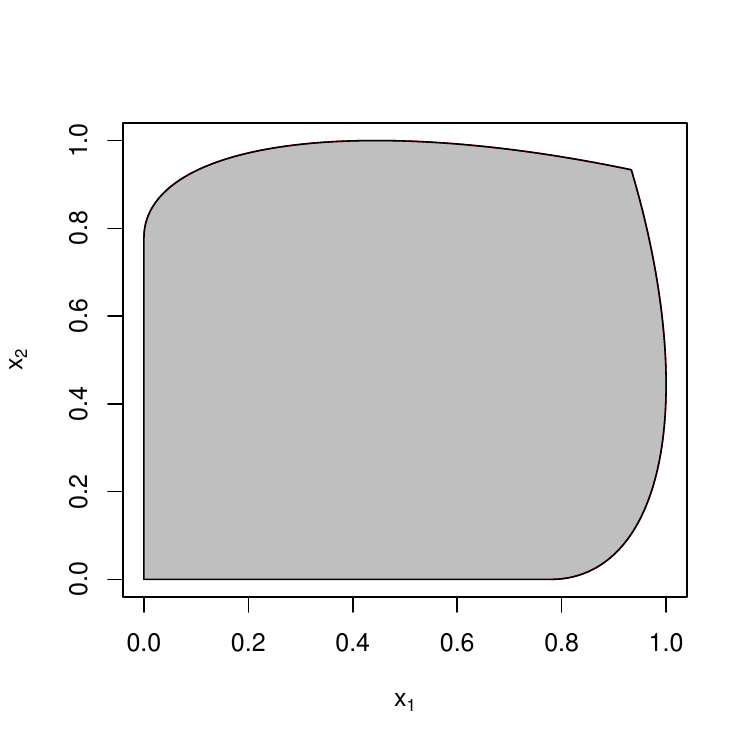}
	\caption{Examples of \edit{limit sets (grey shaded region) from} additively mixed gauge functions. In each case the component gauge functions are $g^{[1]}(x_1,x_2) = (x_1+x_2-2\rho(x_1 x_2)^{1/2})/(1-\rho^2)$ with $\rho =0.5$ and $g^{[2]}(x_1,x_2) = (x_1+x_2)/\gamma+(1-2/\gamma)\min(x_1,x_2)$ with $\gamma =0.5$. From left to right, the weights are $a_1=1,2,3$.} 
	\label{fig:additivemixing}
\end{figure}

 A simple way to mix that retains \edit{the marginal condition $g_{\{j\}}(x_j) = x_j$} is via minimization: $g(\bm{x}) = \min\{g^{[1]}(\bm{x}),\ldots,g^{[m]}(\bm{x})\}$, for $g^{[1]},\ldots,g^{[m]}$ each satisfying \edit{this marginal condition}. The resulting gauge function is the one that would correspond to a mixture density $f_{\bm{X}}(\bm{x}) = \sum_{k=1}^m \pi_k f_{\bm{X}}^{[k]}(\bm{x})$ with $\sum_{k=1}^m \pi_k =1$ and $\pi_k \in (0,1)$ for each $k$. However, such an approach has the effect of retaining the most protruding part of each limit set and may not yield the most realistic shapes; some examples are given in Section~\ref{sec:mixinggauge} of the supplement. Instead we focus on additive mixing, defining
 \begin{align*}
 \tilde{g}(\bm{x}) =  a_{1} g^{[1]}(\bm{x}) + \cdots + a_{m-1} g^{[m-1]}(\bm{x}) + g^{[m]}(\bm{x}), \qquad a_1,\ldots,a_{m-1} >0.
 \end{align*}
The resulting function is denoted by $\tilde{g}$ as in general it will not satisfy \edit{the marginal condition}, and will need to be rescaled to do so. Suppose that the coordinatewise supremum of the set $\tilde{G} = \{\bm{x}:\tilde{g}(\bm{x}) \leq 1\}$ is $\tilde{\bm{c}}=(\tilde{c}_1,\ldots,\tilde{c}_m)$. Then the rescaled gauge function $g(\bm{x}) = \tilde{g}(\tilde{c}_1 x_1, \ldots, \tilde{c}_d x_d)$ satisfies \edit{$g_{\{j\}}(x_j) = x_j$}. Some examples of limit sets from additively mixed functions are depicted in Figure~\ref{fig:additivemixing}. Interestingly, we observe for $d=2$ that this process is able to interpolate between limit sets for which $g(1,1)=1$ and have a ``pointy'' shape, to those with $g(1,1)<1$ and are described by \citet{BalkemaNolde12} as ``blunt''. The former arise for dependence structures representing joint extremes ($H$ places mass only on $\mathbb{B}_{\{1,2\}}$), while the latter arise for those representing separate extremes ($H$ places mass only on $\mathbb{B}_{\{1\}}$ and $\mathbb{B}_{\{2\}}$). Figures in Section~\ref{sec:mixinggauge} of the the supplement also show that for $d=3$ we retain the ability to move between ``pointy'' \edit{limit set} shapes representing joint occurrence of extremes in some components and ``blunt'' shapes representing separate extremes. We focus in the figures only on the case $m=2$, and leave theoretical study of this phenomenon for any $m$ to future work. 

Note that when using additive mixing, the component gauge functions $g^{[k]}(\bm{x})$ need not satisfy \edit{the marginal condition $g_{\{j\}}(x_j) = x_j$} due to the rescaling. This allows, for example, one to include the Gaussian gauge function $g(\bm{x}) = (\bm{x}^{1/2})^{\top} \Sigma^{-1}\bm{x}^{1/2}$ when $\Sigma$ has negative entries, and increases the flexibility of this approach. In practice, we use numerical methods to \editm{find the vector $\tilde{\bm{c}}$ for rescaling}.

\subsection{Model checking}
\label{sec:modelchecking}

We propose checking the fitted model from likelihood~\eqref{eq:lik} via probability-probability (PP) plots. The fitted distribution function (df) of the truncated gamma model is
\begin{align*}
\widehat{F}_{\mbox{tg}}(r|\bm{w},r_0(\bm{w})):=\Pr(R \leq r |\bm{W}=\bm{w},R>r_0(\bm{w})) = 1- \frac{\bar{F}(r; \widehat{\alpha}, g(\bm{w}; \widehat{\bm{\theta}}))}{\bar{F}(r_0(\bm{w}); \widehat{\alpha}, g(\bm{w}; \widehat{\bm{\theta}}))},
\end{align*}
with $\bar{F}$ as in likelihood~\eqref{eq:lik}, and $\widehat{\alpha}, \widehat{\bm{\theta}}$ representing the maximum likelihood estimates of the parameters. The PP plot for \editm{$n_0$} observations with $R_i>r_{0}(\bm{w}_i)$ is the set of points: \edit{$\{(i/(n_0+1), u_{(n_0-i+1)})\}$, where $u_i = \widehat{F}_{\mbox{tg}}(r_{i}; \bm{w}_{i}, r_0(\bm{w}_{i})))$, and $u_{(1)} \geq u_{(2)} \geq \cdots \geq u_{(n_0)}$ represent the ordered sample of $u_i$.} This diagnostic will be demonstrated in Section~\ref{sec:data}.

Comparison of the ``empirical'' estimate of the gauge function $\hat{g}(\bm{w}) \approx \hat{C}/r_0(\bm{w})$, as outlined in Section~\ref{sec:threshold}, provides another check on the form of the fitted model. As was seen in Section~\ref{sec:threshold}, while we do not expect perfect correspondence between $\hat{g}(\bm{w})$ and $g(\bm{w}; \widehat{\bm{\theta}})$, we can expect to see broad similarities in shape. Again we use this in Section~\ref{sec:data}.

\subsection{Prediction}
\label{sec:Prediction}

A key aspect of our proposed geometric framework for statistical inference is that we can use simulation from the fitted model to estimate probabilities of lying in extreme regions, enabling extrapolation outside the range of the observed data. Up to this point, we have focused on the conditional distribution of $R|\{\bm{W}=\bm{w},R>r_0(\bm{w})\}$. In order to perform extrapolation and estimate multivariate tail probabilities, we need realizations of the distribution of $\bm{X}$ in some suitably extreme region. Notationally it is helpful to introduce an alternative radial variable, $R' = R / r_0(\bm{W})$, so that $\bm{X} = R\bm{W} = R' r_0(\bm{W})\bm{W}$. Given a particular value of $\bm{W}=\bm{w}$, our extreme region to date has been $\{R>r_0(\bm{w})\}$. Now considering our extreme region across all $(R,\bm{W})$ values, this corresponds to $\{R'>1\}$. In Figure~\ref{fig:r0w}, all points above the red line / surface in the top row are those with $\{R'>1\}$.

We focus initially on simulating an arbitrary number of points satisfying the conditioning event $\{R'>1\}$, and discuss below adaptations for simulating above higher thresholds. To get draws from the distribution of $\bm{X}|R'>1$, we \editm{multiply simulations from two components:}
\begin{enumerate}[(i)]
	\item the distribution of $\bm{W}|R'>1$;
	\item the distribution of $R|\{\bm{W}=\bm{w},R>r_0(\bm{w})\}$.
\end{enumerate}
 
\begin{figure}
	\centering
	\includegraphics[width=0.4\textwidth]{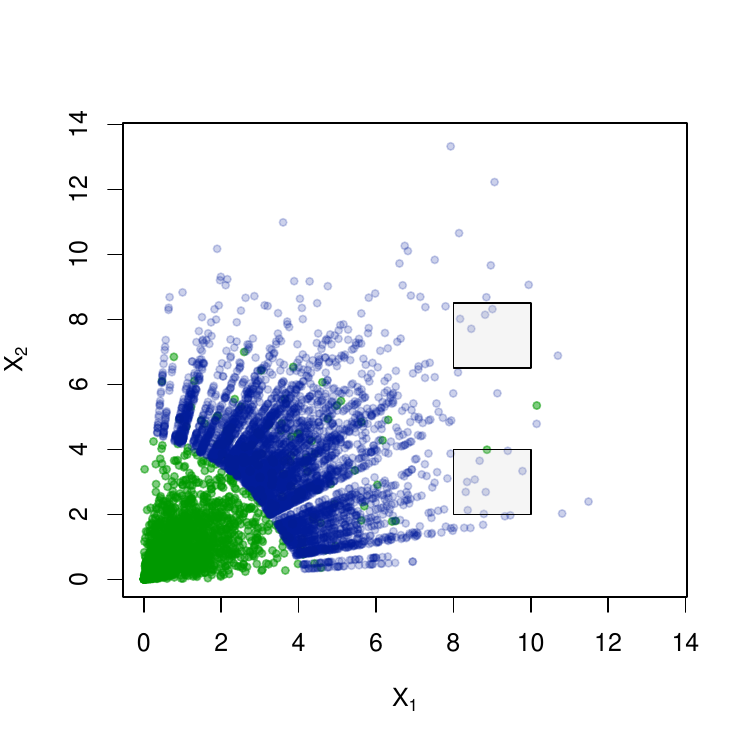}
	\includegraphics[width=0.4\textwidth]{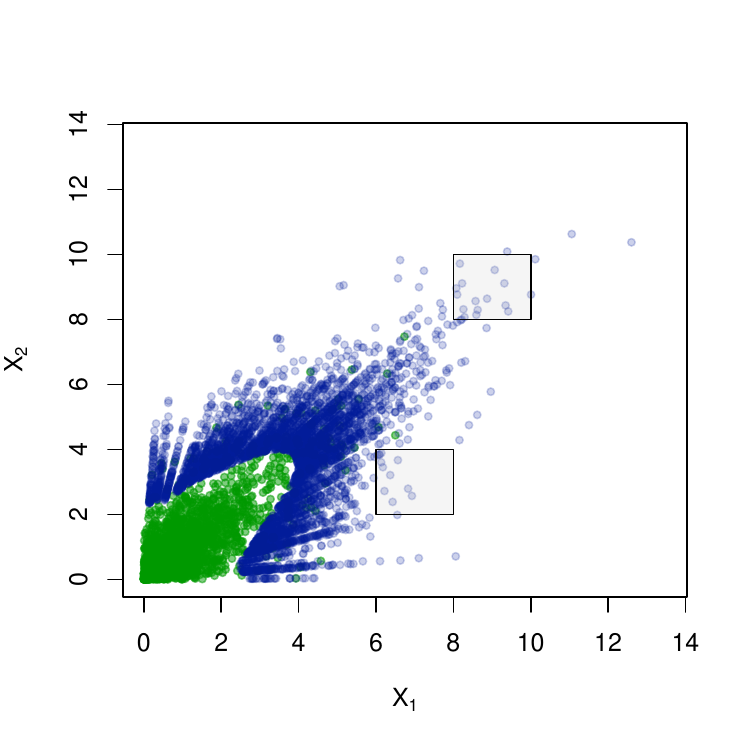}
	\caption{Example of 5000 points simulated from $\bm{X}|R'>1$ in blue for the inverted logistic (left) and logistic (right) distributions. Models were fitted to 2500 data points shown in green. Light grey squares represent potential sets $B$ in equation~\eqref{eq:extset}.}
	\label{fig:simulation}
\end{figure}

The second of these is a simple case of simulating from the fitted truncated gamma distribution, which can be done via the \edit{inverse probability integral transform using standard efficient algorithms to calculate the gamma quantile function}. For the first of these, we may either resample from the empirical distribution of $\bm{W}|R'>1$, or we could fit a parametric model to such samples and simulate from this. We opt for the former in this work, and note the latter as a potential line of future investigation. Figure~\ref{fig:simulation} shows 5000 draws simulated from $\bm{X}|R'>1$, based on a model fitted to 2500 data points.

To estimate the probability of lying in extreme sets, we exploit the simple equation
\begin{align}
\Pr(\bm{X} \in B) = \Pr(\bm{X} \in B |R'>1)\Pr(R'>1), \label{eq:extset}
\end{align}
for any set $B$ lying entirely within the region $\{\bm{x} \in \mathbb{R}^d_+ : \sum_{j=1}^d x_j > r_0(\bm{x}/\sum_{j=1}^d x_j)\}$; some examples are given in Figure~\ref{fig:simulation}. The first probability on the right-hand side of~\eqref{eq:extset} can be estimated empirically from the simulated draws. The second probability may be estimated from the dataset as the proportion of points $R'$ exceeding 1. When quantile regression at level $\tau$ has been used to find $r_0(\bm{w})$, we expect the proportion of points \editm{above the threshold} to be near $1-\tau$.

The fact we can simulate an arbitrary number of points from our model with the condition $\{R'>1\}$ means that in principle we can extrapolate quite a way beyond the observed data. Nonetheless, such an approach may be computationally demanding for very extreme sets that require a large number of simulations. We consider now how to simulate given the condition $\{R'>k\}$, with $k>1$; results will be illustrated in Section~\ref{sec:simstudy}.

Simulation from the distribution of $R|\{\bm{W}=\bm{w},R>k r_0(\bm{w})\}$ is again straightforward from the fitted truncated gamma distribution. Simulation from the distribution of $\bm{W}|R'>k$ is more challenging if $k$ is sufficiently high that there are few or no empirical samples available. However, we have the relation
\begin{align}
f_{\bm{W}}(\bm{w}|R'>k) &= \frac{\int_{k}^\infty f_{R',\bm{W}}(r',\bm{w}|R'>1)\,\mathrm{d}r' }{\int_{\mathcal{S}_{d-1}}\int_{k}^\infty f_{R',\bm{W}}(r',\bm{v}|R'>1)\,\mathrm{d}r' \,\mathrm{d}\bm{v}} \notag\\
&=\frac{f_{\bm{W}}(\bm{w}|R'>1)\int_{k}^\infty f_{R'|\bm{W}}(r'|\bm{w},R'>1)  \,\mathrm{d}r' }{\int_{\mathcal{S}_{d-1}}\int_{k}^\infty f_{R'|\bm{W}}(r'|\bm{v},R'>1) f_{\bm{W}}(\bm{v}|R'>1)\,\mathrm{d}r' \,\mathrm{d}\bm{v}}, \label{eq:fwgivenrpgk}
\end{align}
where $f_{\bm{U}}(\cdot|V>v)$ denotes the density of a random vector $\bm{U}|V>v$. Note that
\begin{align*}
\int_{k}^\infty f_{R'|\bm{W}}(r'|\bm{w},R'>1)  \,\mathrm{d}r' = \int_{k r_0(\bm{w})}^\infty f_{R|\bm{W}}(r|\bm{w},R>r_0(\bm{w}))  \,\mathrm{d}r,
\end{align*}
so that under the truncated gamma \edit{approximation}~\eqref{eq:rwgamma} for $R|\{\bm{W}=\bm{w},R>r_0(\bm{w})\}$, we have the \edit{proportionality} \editm{statement}
\begin{align}
f_{\bm{W}}(\bm{w}|R'>k) \propto f_{\bm{W}}(\bm{w}|R'>1) \frac{\bar{F}(k r_0(\bm{w}); \alpha, g(\bm{w}))}{\bar{F}(r_0(\bm{w}); \alpha, g(\bm{w}))}. \label{eq:impweights}
\end{align}
The ratio of gamma survival functions in~\eqref{eq:impweights} can therefore be used as importance weights to derive an approximate sample from the distribution of $\bm{W}|R'>k$, using a sample from the distribution of $\bm{W}|R'>1$.

Finally, to estimate $\Pr(R'>k)$, so that we can calculate extreme probabilities as in equation~\eqref{eq:extset}, note that the constant of proportionality in~\eqref{eq:impweights} is $\Pr(R'>k|R'>1)$, from the denominator of equation~\eqref{eq:fwgivenrpgk}. An estimate of this is therefore
\begin{align*}
\widehat{\Pr}(R'>k|R'>1) = \frac{1}{n_0} \sum_{i=1}^{n_0} \frac{\bar{F}(k r_0(\bm{w}_i); \alpha, g(\bm{w}_i))}{\bar{F}(r_0(\bm{w}_i); \alpha, g(\bm{w}_i))},
\end{align*}
where $\bm{w}_i, i=1,\ldots,n_0$ are the angles corresponding to the values for which $R'>1$. Lastly, $\widehat{\Pr}(R'>k) = \widehat{\Pr}(R'>k|R'>1)\widehat{\Pr}(R'>1)$, where $\widehat{\Pr}(R'>1)$ is estimated empirically, as previously. We note that another alternative to this procedure is to fit the generalized Pareto distribution to $R'|R'>1$ and use this fitted model to estimate $\Pr(R'>k|R'>1)$. Our investigation into this found that both options perform similarly for relatively small $k$, but the \edit{generalized Pareto} model introduces extra uncertainty for larger $k$, and so we stick to the first approach in Section~\ref{sec:simstudy}.

\edit{In our experience we have found that estimates of $\Pr(\bm{X} \in B)$ are relatively insensitive to the precise choice of $k$, provided both that $k$ is large enough to ensure that several sample points lie in $B$, and that $B \subset \{\bm{x} \in \mathbb{R}^d_+ : \sum_{j=1}^d x_j > kr_0(\bm{x}/\sum_{j=1}^d x_j)\}$, as is required for the analogue of equation~\eqref{eq:extset} to hold. The simplicity of checking this latter condition depends on the shape of $B$ and of $r_0(\bm{w})$, but it is easy to check visually for $d=2$, and it may crudely be checked by ensuring that $k<\sum_{j=1}^d \tilde{x}_{l,j} / r_0(\tilde{\bm{x}}_l/\sum_{j=1}^d \tilde{x}_{l,j})$ for a sample of points $\tilde{\bm{x}}_l, l=1,\ldots,m$ along the boundary of $B$. We recommend taking an intermediate $k$ that is slightly smaller than the maximum for which this series of $m$ inequalities holds, to safeguard against the crudeness of this check.}

\subsection{\edit{Summary of inference and prediction procedures}}
\edit{For convenience, here we briefly summarize the procedures for inference and prediction via the geometric framework.}

\edit{\begin{enumerate}
		\item Determine a high threshold $r_0(\bm{w})$ of the distribution of $R|\bm{W}=\bm{w}$ for all $\bm{w} \in \mathcal{S}_{d-1}$ using either additive quantile regression or a rolling-windows approach.
		\item Select a set of candidate parametric gauge functions $g(\cdot;\bm{\theta})$ and for each one fit the truncated gamma likelihood~\eqref{eq:lik} to the $n_0$ values of $R|\{\bm{W}=\bm{w},R>r_0(\bm{w})\}$.
		\item Compare model fits using selection criteria such as the Akaike or Bayesian information criterion.
		\item Use diagnostics such as the PP plot and comparison with the empirically-estimated gauge function to confirm acceptable fit of the best model(s).
		\item Letting $R'=R/r_0(\bm{W})$, simulate new realizations from the distribution of $\bm{X}|R'>1$ by drawing from the empirical distribution of $\bm{W}|R'>1$ and multiplying by draws of $R|\{\bm{W}=\bm{w},R>r_0(\bm{w})\}$ from the fitted truncated gamma distribution. If required, adapt these steps to simulate from the distribution of $\bm{X}|R'>k$ with $k>1$.
		\item Estimate $\Pr(\bm{X}\in B)$ using equation~\eqref{eq:extset}, or suitable adaptation if $R'>k$.
\end{enumerate}}

\section{Simulation study}
\label{sec:simstudy}

We now demonstrate the performance of our methods against existing approaches for analyzing multivariate extremes. Our focus lies on estimation of probabilities $\Pr(\bm{X} \in B)$ for three sets $B$ that lie in different parts of the region where $\bm{X}$ may be considered extreme.

 We begin with the bivariate case, which is well-established and understood, demonstrating that our methodology gives estimates with low bias in each situation, performing competitively with other methods across a range of scenarios. Specifically, we compare with estimation methodology based on multivariate regular variation (MRV), hidden regular variation \citep{LedfordTawn97} (HRV) and the conditional extreme value model (CE) of \citet{HeffernanTawn04}. 
 \edit{The simplest approach to implementing MRV methodology is to use the approximation $\Pr(\bm{X} \in v+B') \approx e^{-v} \Pr(\bm{X} \in B')$, where we take as the set of interest $B=v+B'$, and $B'$ is extreme, but in the range of the data so can be estimated empirically. This is a nonparametric implementation, but parametric assumptions are possible as well. Specifically we can also assume that equation~\eqref{eq:mrv} holds at finite levels and choose a parametric form for the angular measure $H$. We adopt both techniques below.} HRV is a refinement of MRV that allows for situations where the spectral measure $H$ places no mass on $\mathbb{B}_{\{1,\ldots,d\}}$. \edit{Implementation of this methodology relies on exploiting the relation $\Pr(\bm{X} \in v+B') \approx e^{-v/\eta}\Pr(\bm{X} \in B')$, where $\eta \in (0,1]$ is the residual tail dependence coefficient; this is estimated using the Hill estimator \citep{LedfordTawn97}. Parametric models based on HRV exist \citep{RamosLedford09}, but are generally poorly-justified since the so-called ``hidden angular measure'' is often not a finite measure over the unit simplex; we therefore do not consider these here.} Like MRV however, the asymptotics of HRV are suited only to extrapolating into regions where all variables are large simultaneously. \edit{Implementation of the CE methodology follows the original approach suggested in \citet{HeffernanTawn04}, adapted to exponential margins.}
 \edit{Following the bivariate case}, we move on to the more difficult case of $d=3$, and show that we can substantially outperform the CE model in this setting, which is the only other \edit{viable approach for providing} an estimate of the probabilities of interest.

\subsection{Dimension $d=2$}
\label{sec:simd2}
For the bivariate case, we perform estimation based on 5000 datapoints simulated from \edit{four} different dependence structures: (I) logistic distribution with parameter $\gamma=0.4$; (II) Gaussian distribution with $\rho=0.8$; (III) inverted logistic distribution with $\gamma=0.7$; \edit{(IV) logistic distribution with $\gamma=0.8$. Distributions~(I) and~(IV) represent moderately strong and weak logistic dependence structures, respectively}. In Section~\ref{sec:simstudyextra} of the supplement we show examples of the \edit{four} datasets, and three sets of interest $B_1 = (10,12) \times (10,12)$, $B_2=(10,12) \times(6,8)$, and $B_3 = (10,12)\times(2,4)$.

In each case we fit model~\eqref{eq:rwgamma} to the data using four different gauge functions: those corresponding to \edit{the unique} distributions (I)--(III), where the parameter is to be estimated from the data, and the function $g(\bm{x}; \theta) = \max\{(x_1 - x_2)/\theta, (x_2 - x_1)/\theta, (x_1 + x_2)/(2 - \theta)\}$. We select the model that yields the lowest value of the Akaike information criterion (AIC) for the prediction step, thereby avoiding using knowledge of the true data-generating process. Recall that before fitting model~\eqref{eq:rwgamma}, we need to calculate a high threshold $r_0(w)$. In Section~\ref{sec:threshold}, we described using either additive quantile regression or a rolling-windows quantile calculation for this. We used both techniques in the simulation study, setting $\tau=0.95$, finding relatively little difference in the performance of the resulting inference, particularly in comparison to differences across extreme-value methodologies. Therefore, to keep presentation focused, we detail only the results where $r_0(w)$ was found using the simpler rolling-windows quantile method.  Although our focus is on extreme probability estimation, we also display (non-)parametric estimates of $G$, obtained via $\hat{g}(w)$ and $g(w;\widehat{\bm{\theta}})$, in Section~\ref{sec:simstudyextra}. 

\edit{For the parametric MRV approach, we employ a similar strategy to our geometric approach. After transforming to radial-angular coordinates $\|\bm{X}_P\|$ and $\bm{X}_P/\|\bm{X}_P\|_1$ from Pareto margins, we take all angles for which the corresponding radius exceeds the 0.95 quantile of radii, and fit a parametric form for the density of $H$ via maximum likelihood. We choose between five parametric models for $H$ using AIC, where the true logistic density for distributions (I) and (IV) is among the choices. Probabilities are estimated using numerical integration over $B$ using the fitted model for angles, combined with Pareto density for radii.}

\begin{figure}[htpb]
	\centering
	\includegraphics[width=0.3\textwidth]{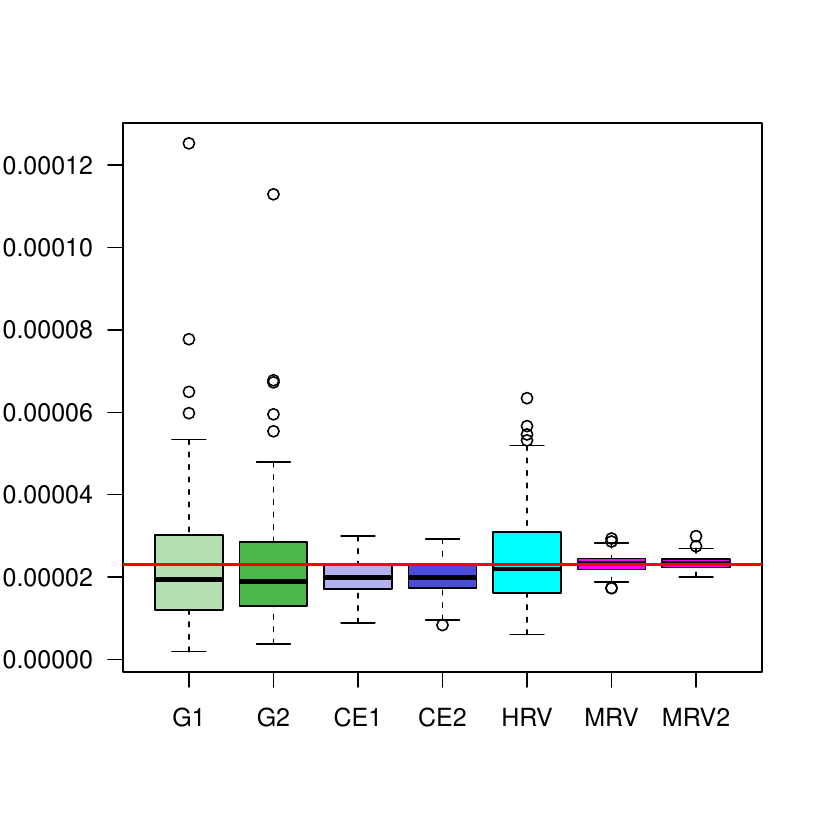}
	\includegraphics[width=0.3\textwidth]{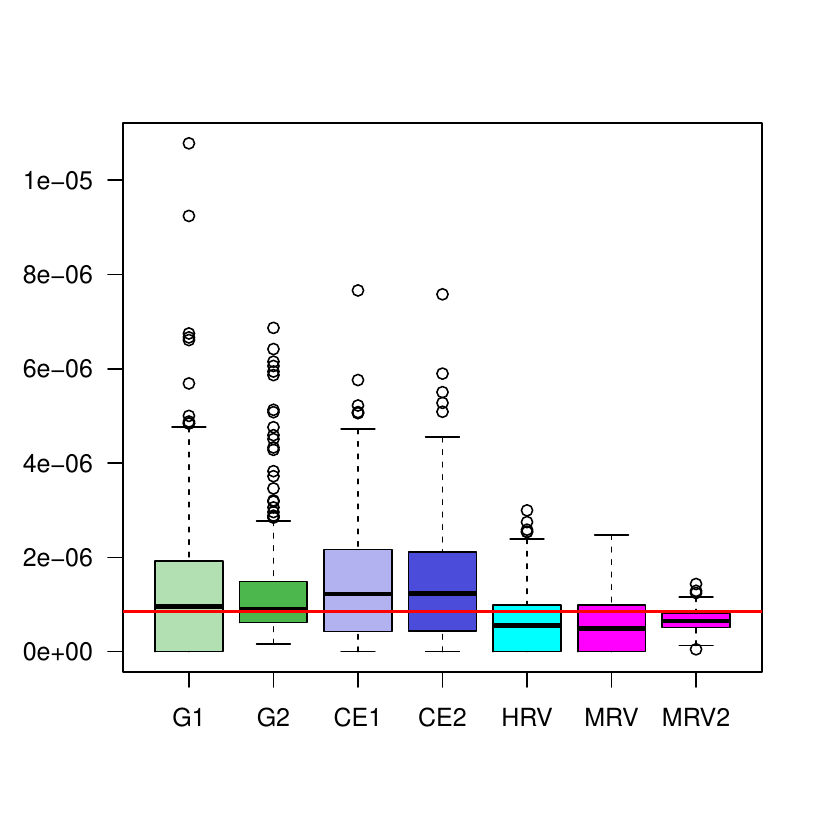}
	\includegraphics[width=0.3\textwidth]{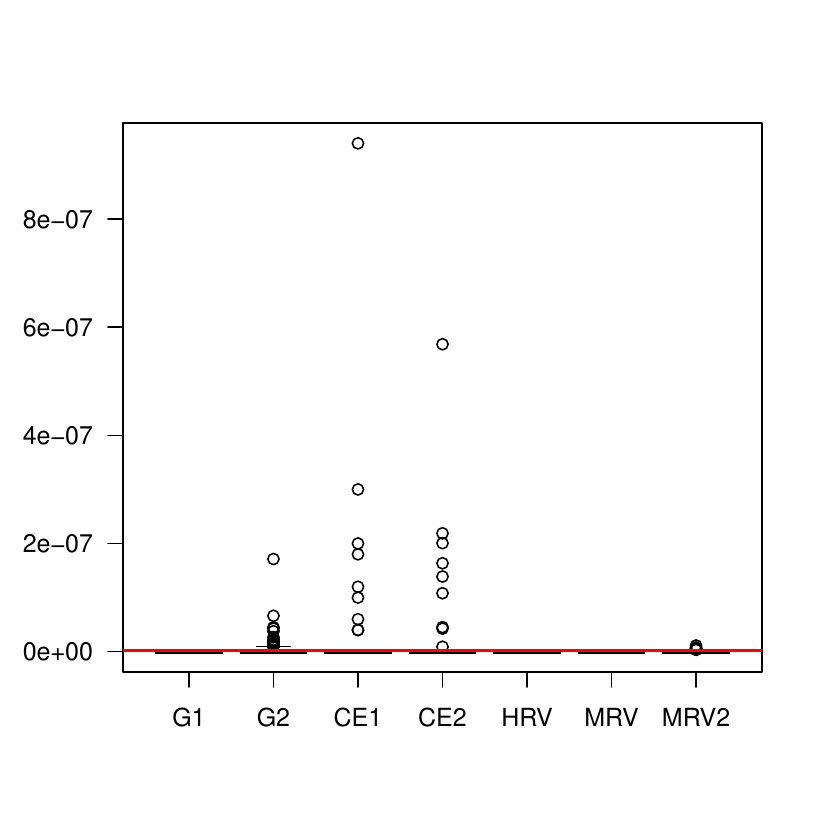}\\
	\includegraphics[width=0.3\textwidth]{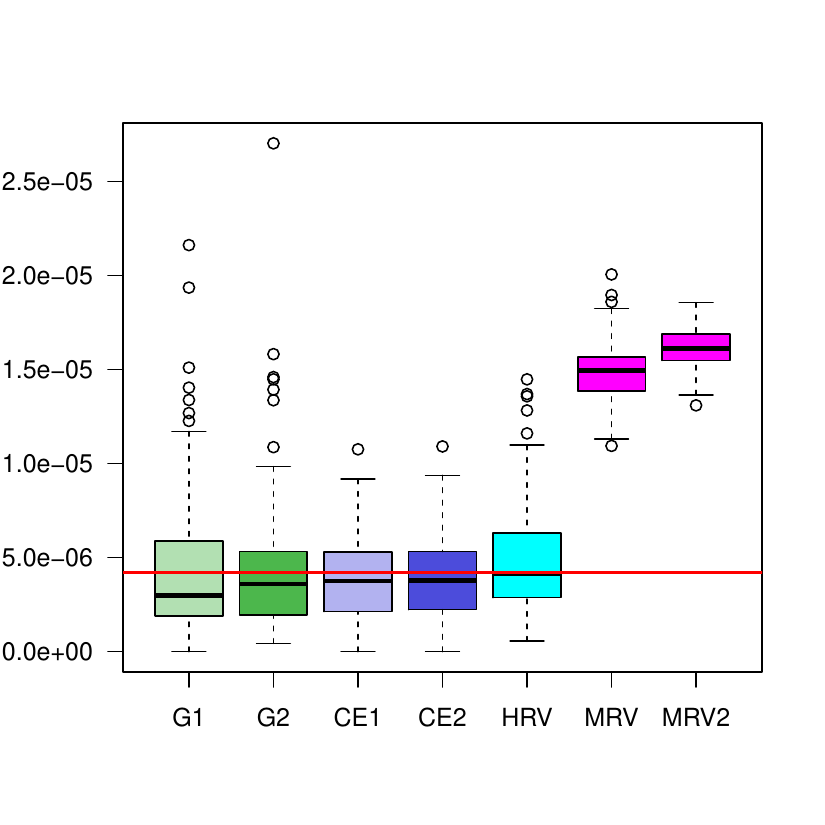}
	\includegraphics[width=0.3\textwidth]{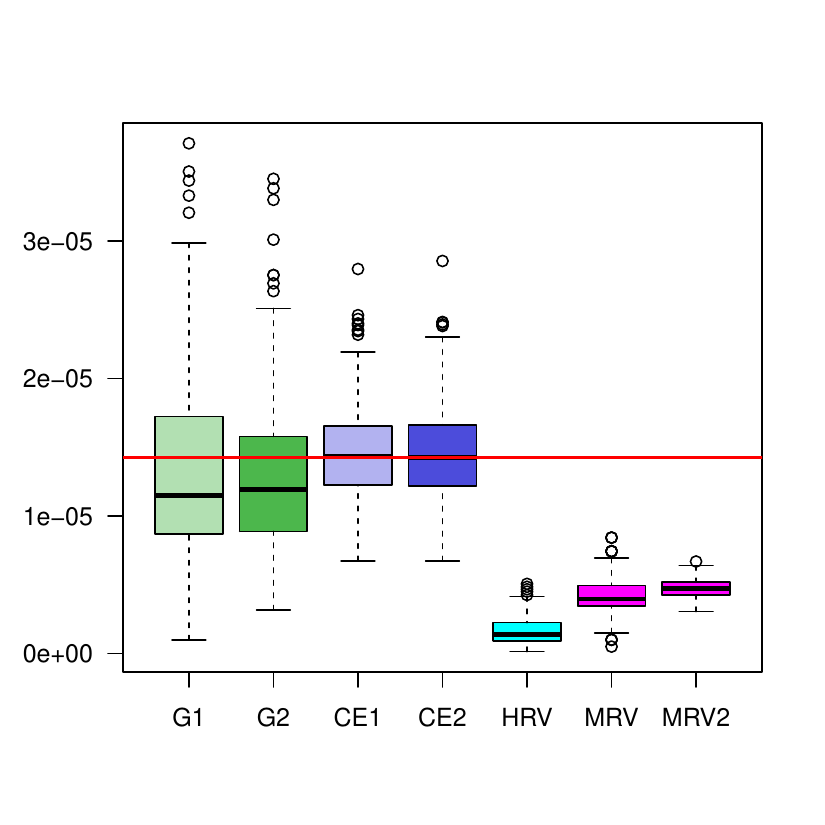}
	\includegraphics[width=0.3\textwidth]{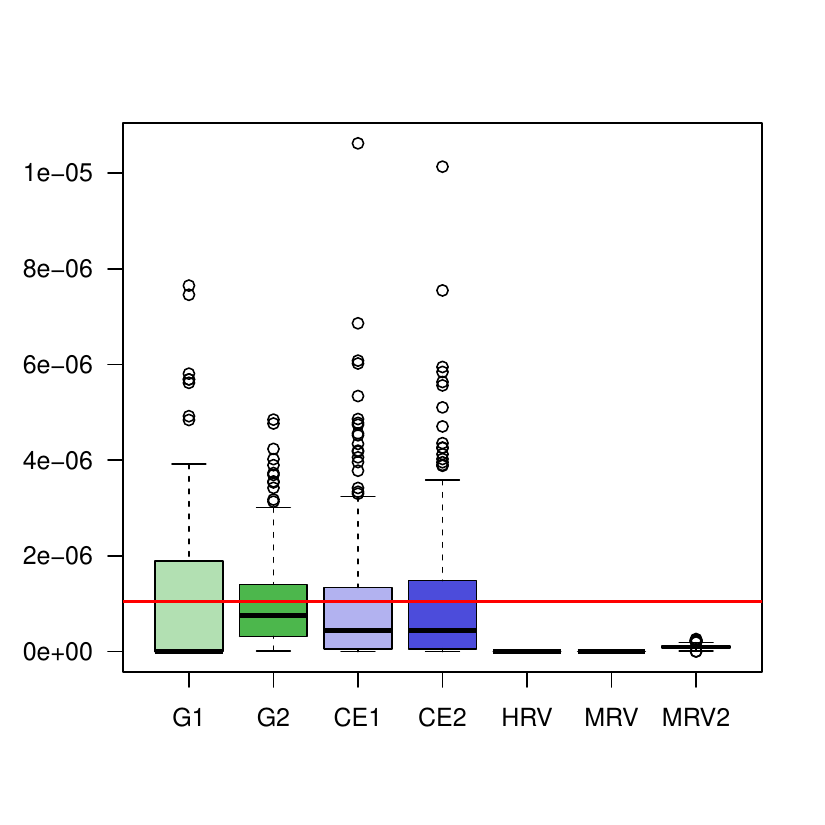}\\
	\includegraphics[width=0.3\textwidth]{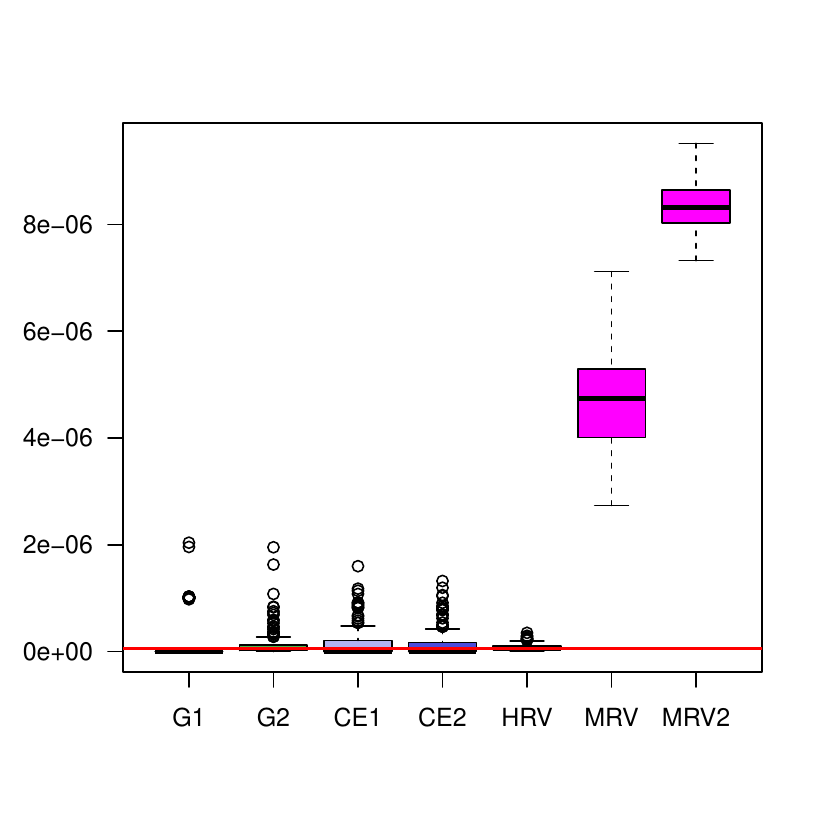}
	\includegraphics[width=0.3\textwidth]{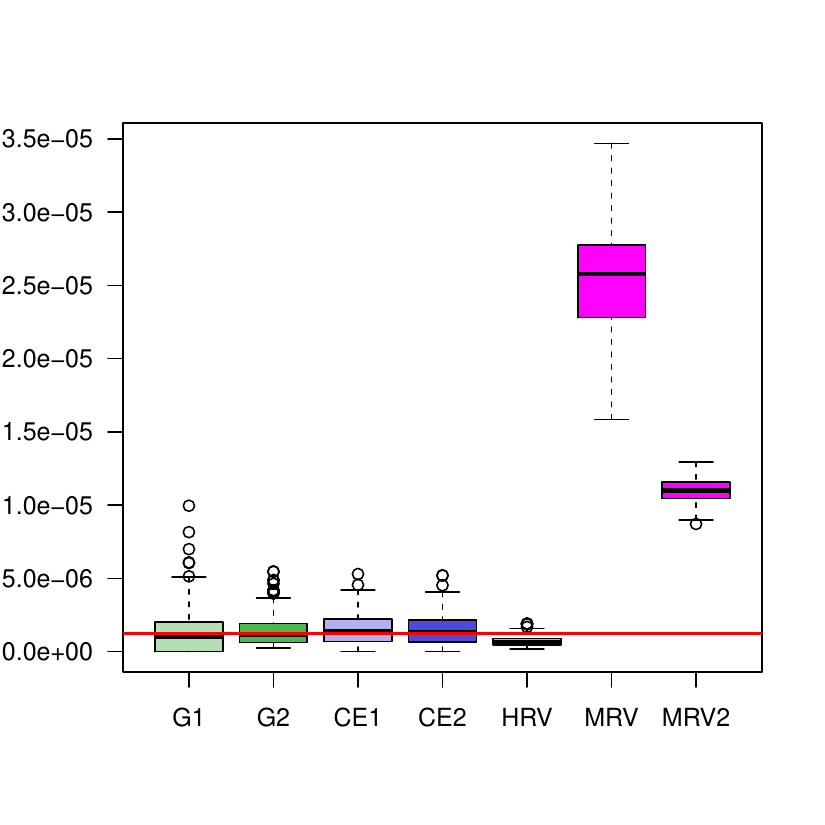}
	\includegraphics[width=0.3\textwidth]{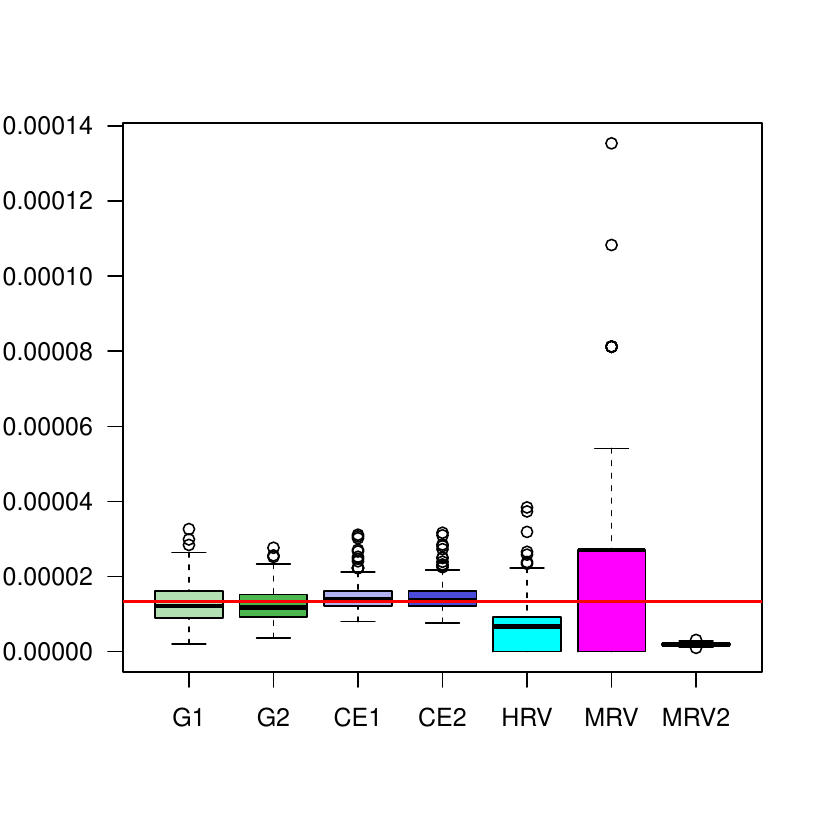}
	\includegraphics[width=0.3\textwidth]{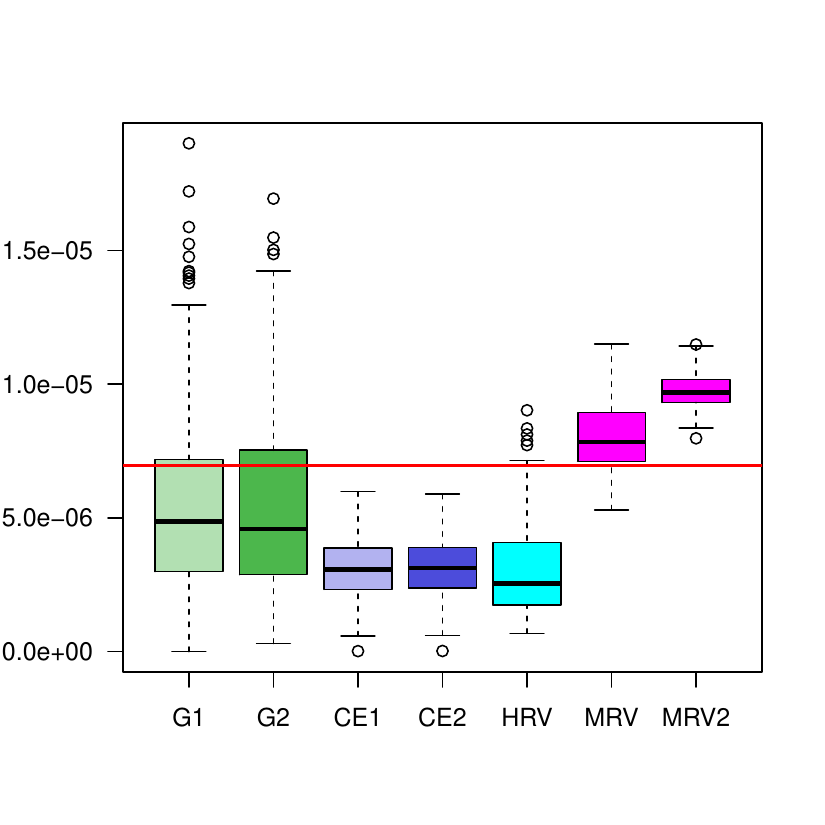}
	\includegraphics[width=0.3\textwidth]{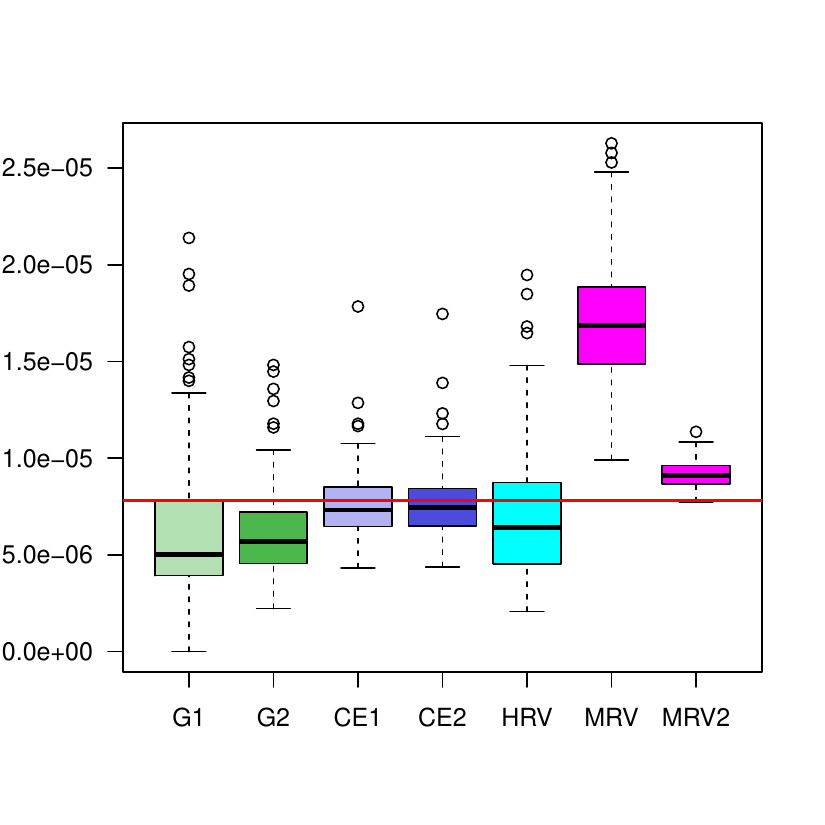}
	\includegraphics[width=0.3\textwidth]{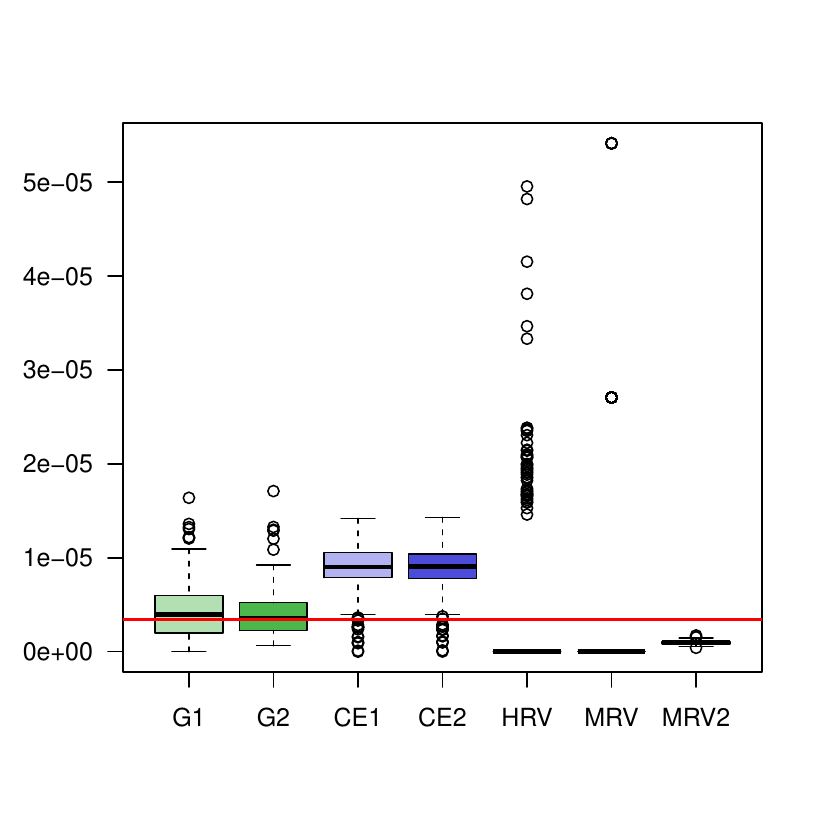}
	\caption{\footnotesize{Boxplots of the estimated probabilities for $d=2$. From left to right columns represent $\Pr(\bm{X} \in B_1),\Pr(\bm{X} \in B_2),\Pr(\bm{X} \in B_3)$, respectively. From top to bottom, datasets are (I), (II), (III), \edit{(IV)} respectively. Green boxplots, labelled G1, G2, give results from our geometric approach: G1 is calculated from $\bm{X}|R'>1$; G2 is calculated from $\bm{X}|R'>k$, where $k$ is determined as the maximum value such that all sets $B_1, B_2, B_3$ lie in the region $\{\bm{x}:x_1+x_2>k r_0(x_1/(x_1+x_2))\}$. Dark blue boxplots, labelled CE1, CE2 give results from the conditional extremes model: CE1 is calculated from $\bm{X}|X_1>6.9$; CE2 is calculated from $\bm{X}|X_1>10$. \edit{Turquoise boxplots give results from hidden regular variation (HRV) methodology; purple boxplots represent nonparametric and parametric multivariate regular variation (labelled MRV, MRV2, respectively)}. True values of the probabilities are indicated by horizontal red lines.}}
	\label{fig:2dboxplots}
\end{figure}

\begin{figure}[htpb]
	\centering
	\includegraphics[width=0.3\textwidth]{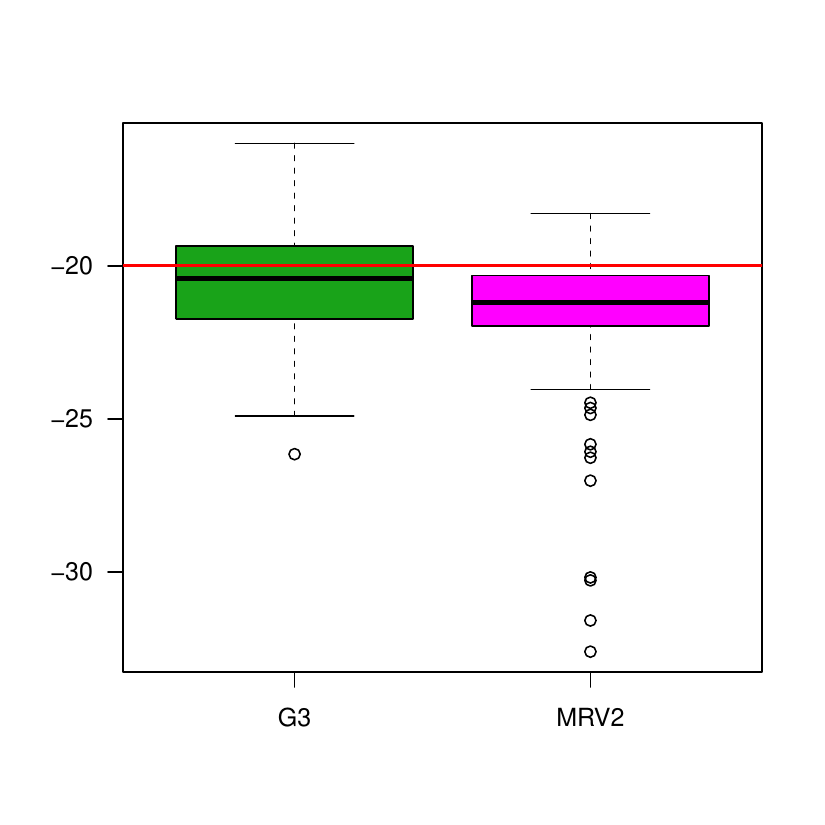}
	\includegraphics[width=0.3\textwidth]{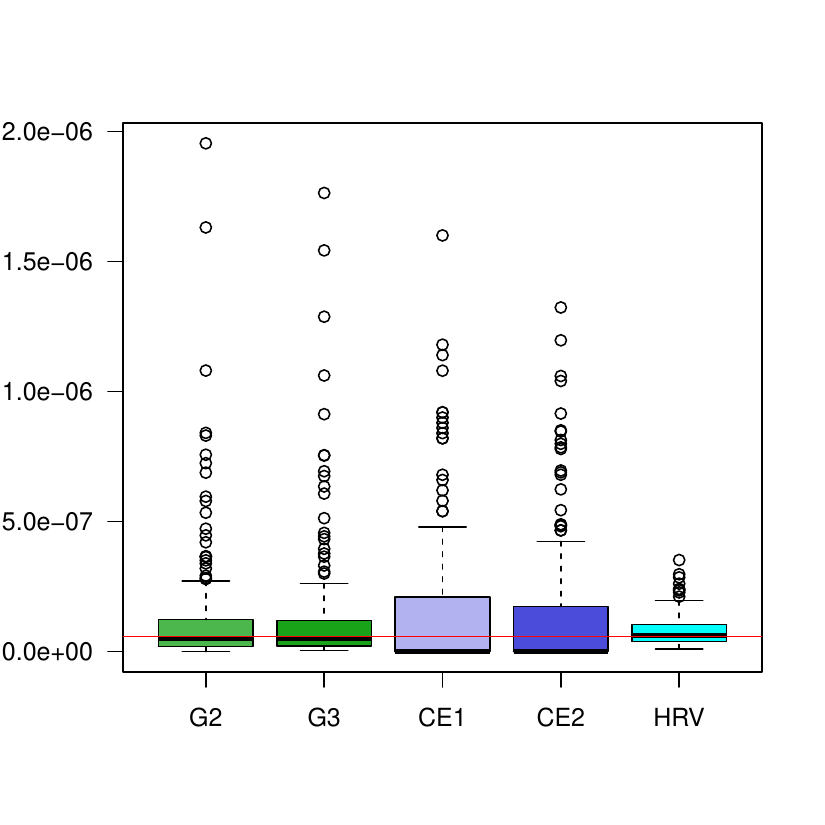}
	\caption{Left: boxplot of log estimates of $\Pr(\bm{X} \in B_3)$ for distribution (I), $d=2$, using the geometric approach at a high threshold as described in the text. Right: estimates of $\Pr(\bm{X} \in B_1)$ for distribution (III), $d=2$, using two different thresholds for the geometric approach (green). Estimates for two different thresholds from the conditional approach are in dark blue, and from HRV in turquoise.}
	\label{fig:extra2dboxplots}
\end{figure}

Figure~\ref{fig:2dboxplots} displays boxplots of the estimated probabilities for 200 repetitions across different methodologies: see the caption for details. For distribution (I), all methods estimate $\Pr(\bm{X} \in B_1)$ with little bias; the smallest variance is attributed to \edit{the MRV approaches}, which is as expected since we are looking at a distribution where $H$ places mass on $\mathbb{B}_{\{1,2\}}$ and estimating a probability in the joint tail. The geometric approach and CE estimate $\Pr(\bm{X} \in B_2)$ relatively well, with the smallest variance attributable to the geometric approach based on $\bm{X}|R'>k$ for suitable $k>1$. HRV and MRV start to exhibit \edit{some} bias because $B_2$ lies outside the joint tail region. For $\Pr(\bm{X} \in B_3)$, all estimates based on \edit{the nonparametric} HRV and MRV \edit{approaches} are equal to zero. For the geometric approach, we are able to estimate this probability well when selecting a suitable $k$. Specifically, in each repetition, we select one of the largest values of $k$ such that $B_3 \subset \{\bm{x}: (x_1+x_2)>k r_0(x_1/(x_1+x_2))\}$. This results in \edit{all} probabilities having a non-zero estimate, compared to 0\% for \edit{nonparametric} HRV/MRV, and 4.5\% for CE (at each of two thresholds). \edit{This probability can be estimated as non-zero by parametric MRV, but with a little bias.} A boxplot of this case is included in the left panel of Figure~\ref{fig:extra2dboxplots}.

\edit{Distribution (IV) also represents the case where $H$ places mass on $\mathbb{B}_{\{1,2\}}$, yet interestingly, MRV gives biased estimates in for all probabilities. This is likely due to the practical rate of convergence to the limiting angular measure $H$ being slower under this weaker dependence scenario. Indeed we see differing estimates from the two MRV approaches, which are based on different effective ``thresholds'' for defining extremes. MRV changes from appreciably over-estimating the probabilities $\Pr(\bm{X} \in B_1)$ and $\Pr(\bm{X} \in B_2)$ to hugely under-estimating $\Pr(\bm{X} \in B_3)$. The geometric approach suggests a small under-estimation of $\Pr(\bm{X} \in B_1)$ and $\Pr(\bm{X} \in B_2)$ and good performance for $\Pr(\bm{X} \in B_3)$. CE moves from large under-estimation to over-estimation moving from $\Pr(\bm{X} \in B_1)$ to $\Pr(\bm{X} \in B_3)$.}

For distributions (II) and (III), the geometric approach and CE exhibit quite similar performance, although CE has a smaller variance for estimates of $\Pr(\bm{X} \in B_2)$ under distribution (II), and of $\Pr(\bm{X} \in B_3)$ under distribution (III). MRV is not an appropriate method for these distributions and always performs badly; HRV is appropriate in the joint tail, where it exhibits similar performance to other methods for (II) and better performance for (III), while it leads to poor estimates in other regions. Additional boxplots in the right panel of Figure~\ref{fig:extra2dboxplots} display more detailed information for the estimates of $\Pr(\bm{X} \in B_1)$ under distribution (III). As described for distribution (I), we also used a suitable $k>1$ for estimating this probability. The geometric approach outperforms CE in this case. This is because, using an appropriate $k$, we are able to simulate points to generate non-zero estimates of the probabilities (\edit{93\%} and 100\% of estimates are positive for the two thresholds shown). In contrast, only \edit{45.5\%} and \edit{46\%} of estimates are positive for CE.

\subsection{Dimension $d=3$}
\label{sec:simd3}
We again perform estimation based on 5000 data points from three different data structures: (I) asymmetric logistic distribution, for which the spectral measure $H$ places mass on $\mathbb{B}_{\{1,2\}},\mathbb{B}_{\{1,3\}},\mathbb{B}_{\{2,3\}}$, with parameters $\gamma_{\{1,2\}}=\gamma_{\{1,3\}}=\gamma_{\{2,3\}}=0.4$; (II) asymmetric logistic distribution where $H$ places mass on $\mathbb{B}_{\{1\}},\mathbb{B}_{\{1,2\}},\mathbb{B}_{\{2,3\}}$ and with parameters $\gamma_{\{1,2\}}=\gamma_{\{2,3\}}=0.4$; (III) distribution constructed by taking an inverted Clayton copula with parameter 2 for $(X_1,X_2)$, with $X_3|X_2=x_2$ drawn from an inverted logistic dependence structure with parameter 0.5. Such a distribution is in the domain of attraction of a spectral measure $H$ placing mass on $\mathbb{B}_{\{1,2\}}, \mathbb{B}_{\{3\}}$. In Section~\ref{sec:simstudyextra} of the supplement we display examples of the three datasets along with sets of interest $B_1 = (8,10)\times(8,10)\times(0.01,3)$, $B_2=(8,10) \times (5,7) \times (0.01,3)$ and $(8,10)\times (2,4) \times (0.01,3)$.

For the $d=3$ case we consider only two methodologies: the geometric approach and CE, \editm{as HRV/MRV only perform well when considering sets $B$ where all variables are of a similar magnitude, and the sets that we are considering are all small in $x_3$}. Moreover for MRV we require mass on $\mathbb{B}_{\{1,2,3\}}$ for good performance of this method. 

For the geometric approach we fit model~\eqref{eq:rwgamma} to the data after identifying potential suitable forms for the gauge function $g$. For this initial step, we calculate the coefficients $\tau_C(\delta)$, and associated estimates of the probability of mass on $\mathbb{B}_C$ as in \citet{Simpsonetal20}, for $\delta=0.4,0.5,0.6$ and $C\subseteq\{1,2,3\}$. These estimates help to identify potential faces of the simplex on which the limiting spectral measure $H$ places mass, and hence a suitable structure for the form of the gauge function.
\edit{Specifically, where they exist, the coefficients $\tau_C(\delta) \in (0,1]$ should be equal to 1 if $H$ places mass on $\mathbb{B}_C$, for all values of the tuning parameter $\delta \in [\delta^\star,1]$ and some $\delta^\star \in [0,1]$. However, because of difficulties in estimating these coefficients precisely in the presence of nuisance parameters, \citet{Simpsonetal20} use them as part of a broader strategy to  estimate of the probability of mass on $\mathbb{B}_C$.}
 If all estimates \edit{for the three values of $\delta$} suggest the same extremal dependence structure \edit{in terms of where $H$ places mass,} then a single model is fitted, where the gauge function corresponds to that of the asymmetric logistic distribution for the identified structure. Otherwise, up to three different models are fitted, and the \editm{model with the lowest AIC is selected}. We note that, for distributions~(I) and~(II), this means that we have the potential to fit the correct model form to the data, subject to its identification via the \citet{Simpsonetal20} methodology, although for distribution (III), we always have a misspecified model.

\begin{figure}[htpb]
	\centering
	\includegraphics[width=0.3\textwidth]{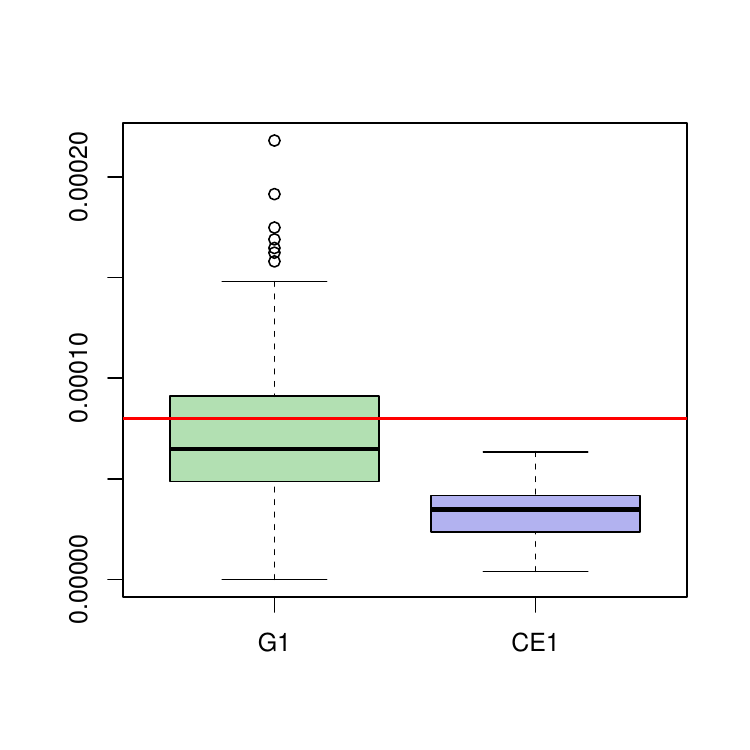}
	\includegraphics[width=0.3\textwidth]{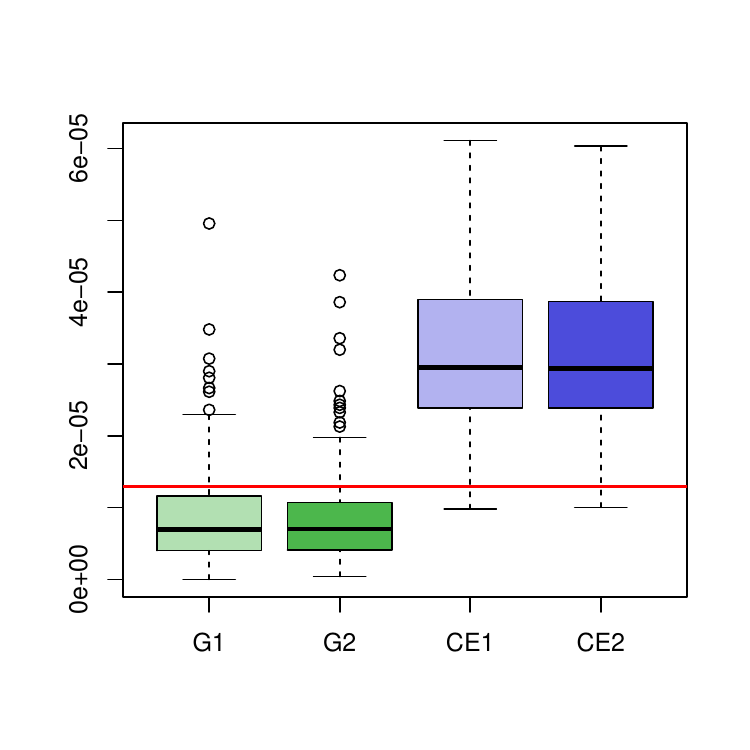}
	\includegraphics[width=0.3\textwidth]{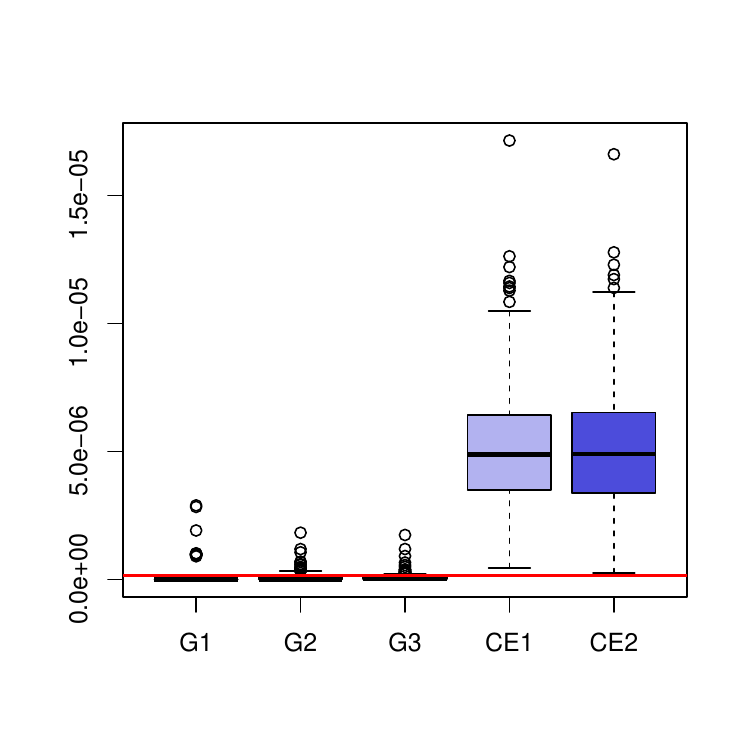}\\
	\includegraphics[width=0.3\textwidth]{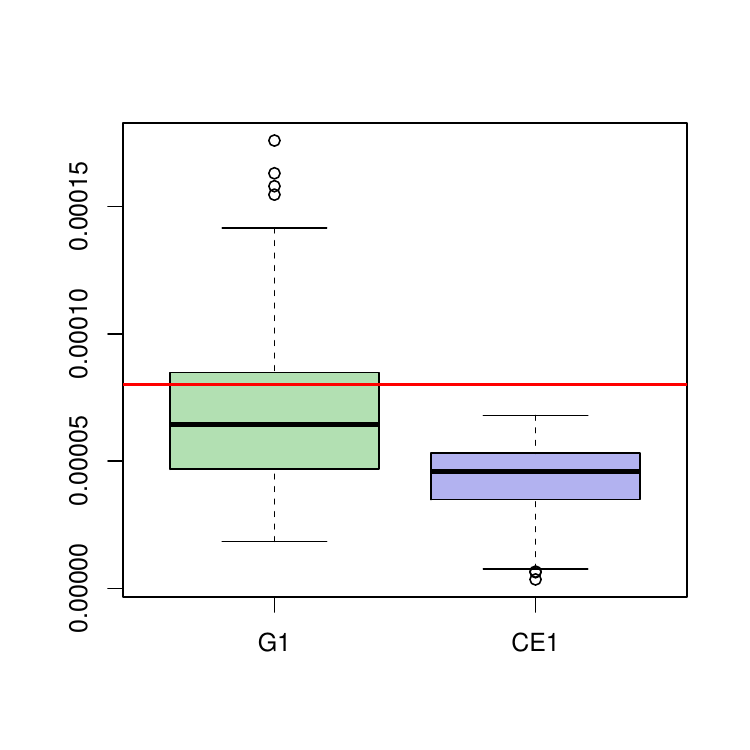}
	\includegraphics[width=0.3\textwidth]{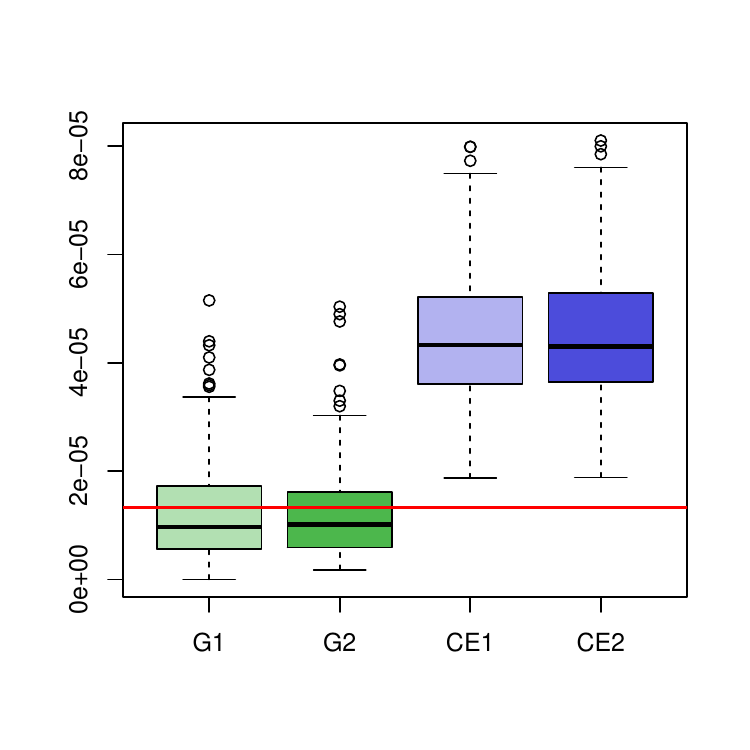}
	\includegraphics[width=0.3\textwidth]{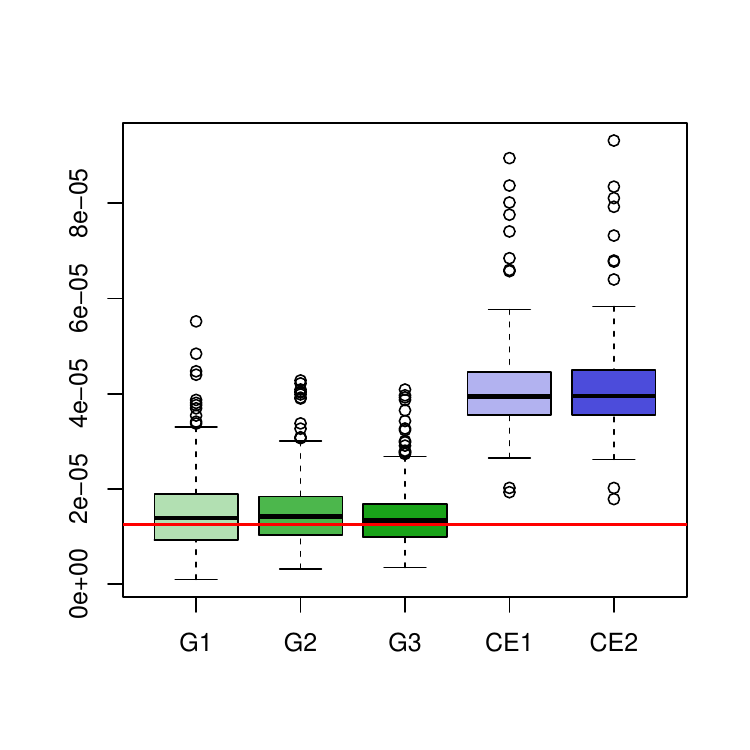}\\
	\includegraphics[width=0.3\textwidth]{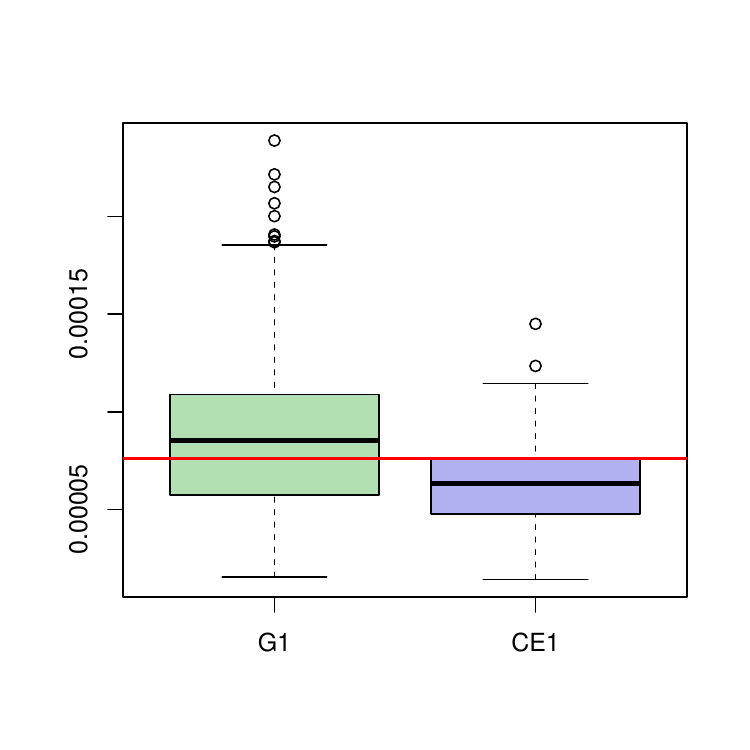}
	\includegraphics[width=0.3\textwidth]{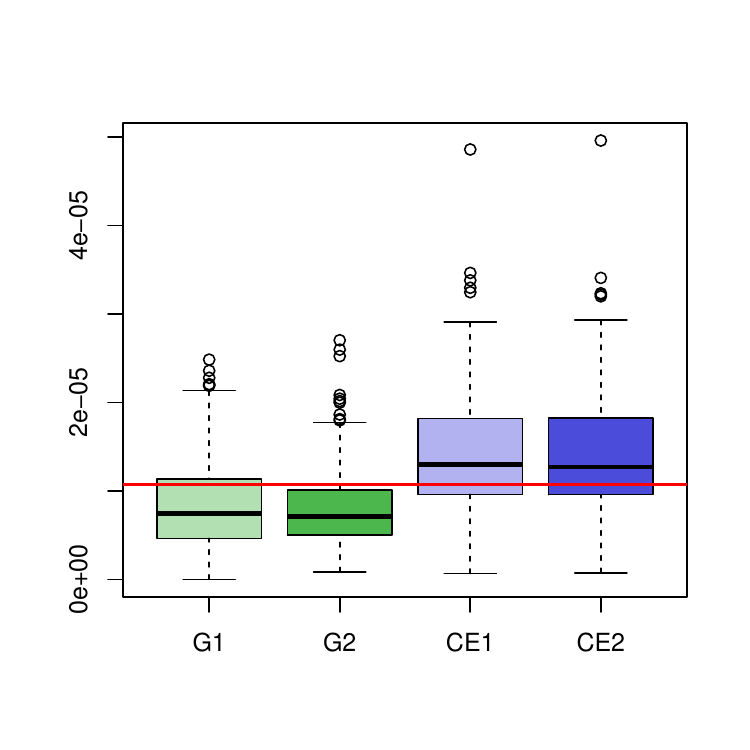}
	\includegraphics[width=0.3\textwidth]{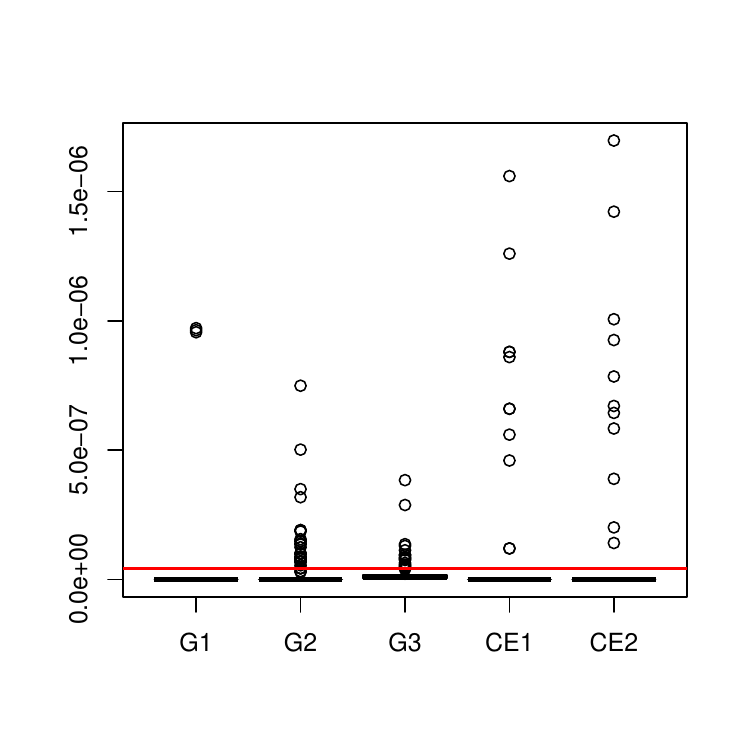}
	\caption{Boxplots of the estimated probabilities for $d=3$. From left to right columns represent $\Pr(\bm{X} \in B_1),\Pr(\bm{X} \in B_2),\Pr(\bm{X} \in B_3)$, respectively. From top to bottom, datasets are (I), (II), (III), respectively. Green boxplots, labelled G1, G2, G3 give results from the geometric approach: G1 is calculated from $\bm{X}|R'>1$; G2 and G3 are calculated from $\bm{X}|R'>k_j$, $j=1,2$, where $k_j$ is determined as a large value such that the sets $B_2$ or $B_3$ lie in the region $\{\bm{x}:x_1+x_2>k_j r_0(x_1/(x_1+x_2))\}$. Dark blue boxplots, labelled CE1, CE2 give results from the conditional extremes model: CE1 is calculated from $\bm{X}|X_1>6.9$; CE2 is calculated from $\bm{X}|X_1>8$. True values of the probabilities are indicated by horizontal red lines.}
	\label{fig:3dboxplots}
\end{figure}

Figure~\ref{fig:3dboxplots} displays boxplots of the estimated probabilities using the two methods. In most cases, the geometric approach exhibits relatively low bias, particularly in comparison to CE, which is typically biased down for $\Pr(\bm{X} \in B_1)$ and up for $\Pr(\bm{X} \in B_2), \Pr(\bm{X} \in B_3)$. In conditional extreme value modelling, dependence structures are defined pairwise, so while any pair of variables $(X_i,X_j)$ can theoretically have mass on $\mathbb{B}_{\{i,j\}}$ or $\mathbb{B}_{\{i\}}$ and $\mathbb{B}_{\{j\}}$, the methodology cannot usually capture more complex higher-order structures well. The structure of distribution (III) is the simplest, with only variables $X_1,X_2$ exhibiting simultaneous extremes, and CE is correspondingly more successful in this case. For $\Pr(\bm{X} \in B_3)$ and distributions (I) and (III), additional boxplots are provided in Section~\ref{sec:simstudyextra} of the supplement. These demonstrate that the geometric approach labelled G3 provides the best estimate in both cases, but \editm{underestimates the probability}. In contrast we can see from Figure~\ref{fig:3dboxplots} that estimates of this probability for distribution (I) are biased strongly upwards for CE, while for distribution (III) only 5.5\% of estimates for CE are positive \edit{at either threshold}.

\section{Data analyses}
\label{sec:data}
We use our new modelling approach to analyze two multivariate environmental datasets. The first is wave data from Newlyn, UK, included because of its extensive previous analysis in the literature. The second is a set of river flow data from \citet{Simpsonetal20}.

\subsection{Newlyn wave data}
\label{sec:wave}
This dataset of 2894 measurements of wave height (metres), surge (metres) and period (seconds), denoted here as $(X_H,X_S,X_P)$, was originally analyzed in \citet{ColesTawn94} using a model that assumed \edit{multivariate regular variation} with all mass of the spectral measure on $\mathbb{B}_{\{H,S,P\}}$. The full trivariate dataset has subsequently been analyzed in \citet{Bortotetal01}, who assumed a censored multivariate Gaussian model, and \citet{ColesPauli02}, whose model was able to accommodate the situation where the spectral measure places mass on some faces of the simplex, but was otherwise quite restrictive.

The first step is to transform each marginal to exponential, which is done using a semi-parametric estimate of the distribution function for each variable $X_{j}$:
\begin{align}
\widehat{F}_{j}(x) &=\begin{cases}
\tilde{F}_{j}(x), & x\leq u_{j},\\
1-\phi_{u,j}\left[1+\xi_{j}(x-u_{j})/\sigma_{j}\right]^{-1/\xi_{j}}_+, & x >u,
\end{cases}
\label{eq:marginaltf}
\end{align}
where $\tilde{F}_{j}$ is the empirical df, $u_{j}$ is a high threshold, $\phi_{u,j} = \Pr(X_{j}>u_{j})$, and the form above $u_{j}$ is the \edit{generalized Pareto} distribution with scale $\sigma_{j}>0$ and shape $\xi_{j}$. We take the thresholds $u_{H}, u_{S}$ and $u_{P}$ to be the 95\% quantiles of the respective distributions.

To get an initial idea of the extremal dependence structure, we use the \citet{Simpsonetal20} \editm{methodology and calculate} $\tau_C(\delta)$ for a range of values of $\delta$. These estimates suggest that the spectral measure places mass on the faces $\mathbb{B}_{\{H\}},\mathbb{B}_{\{S\}},\mathbb{B}_{\{P\}}, \mathbb{B}_{\{H,S\}}$, which fits with the assessment in \citet{Bortotetal01} and \citet{ColesPauli02}.

To calculate the threshold $r_0(\bm{w})$, we use the rolling-windows procedure described in Section~\ref{sec:threshold}, with $\tau=0.95$. We then fit model~\eqref{eq:rwgamma} with three forms for $g$: (i) the asymmetric logistic gauge function with the structure given by $\tau_C(\delta)$, (ii) gauge corresponding to the Gaussian distribution, and (iii) an additive mixture of the Gaussian and asymmetric logistic gauges, as described in Section~\ref{sec:additivemixing}. The respective AIC values are 374.9, 365.5 and 369.5.

\begin{figure}[htbp]
	\centering
	\includegraphics[width=0.3\textwidth]{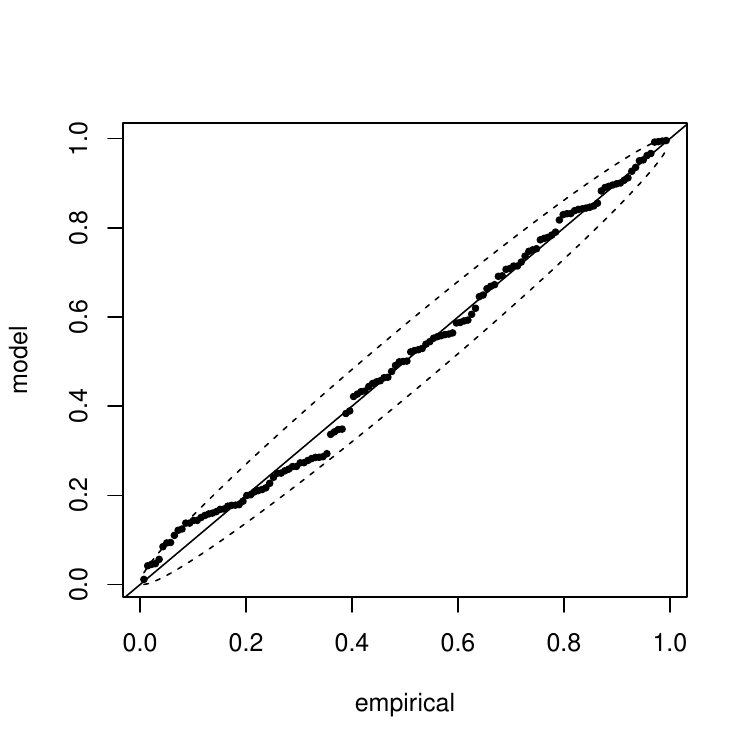}
\includegraphics[width=0.3\textwidth]{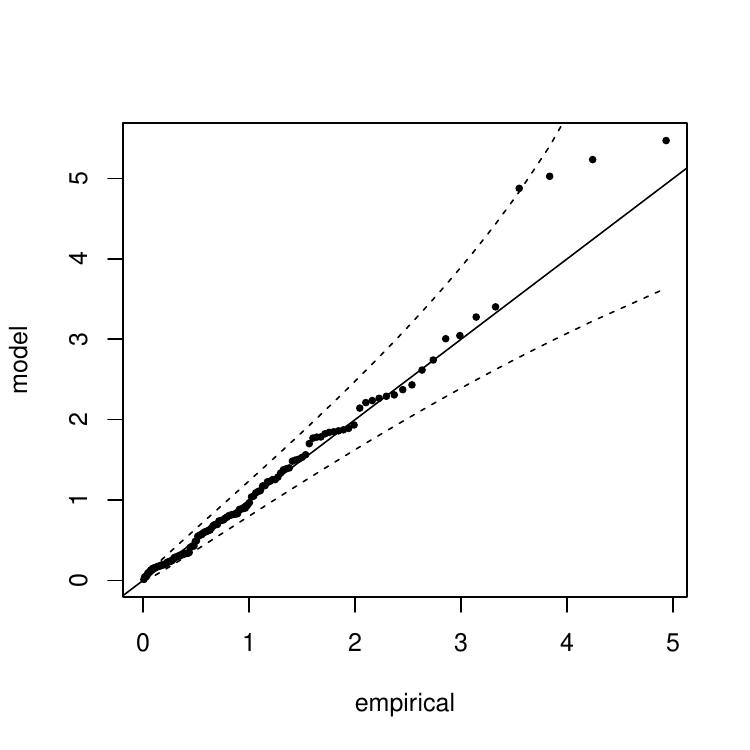}
\includegraphics[width=0.3\textwidth]{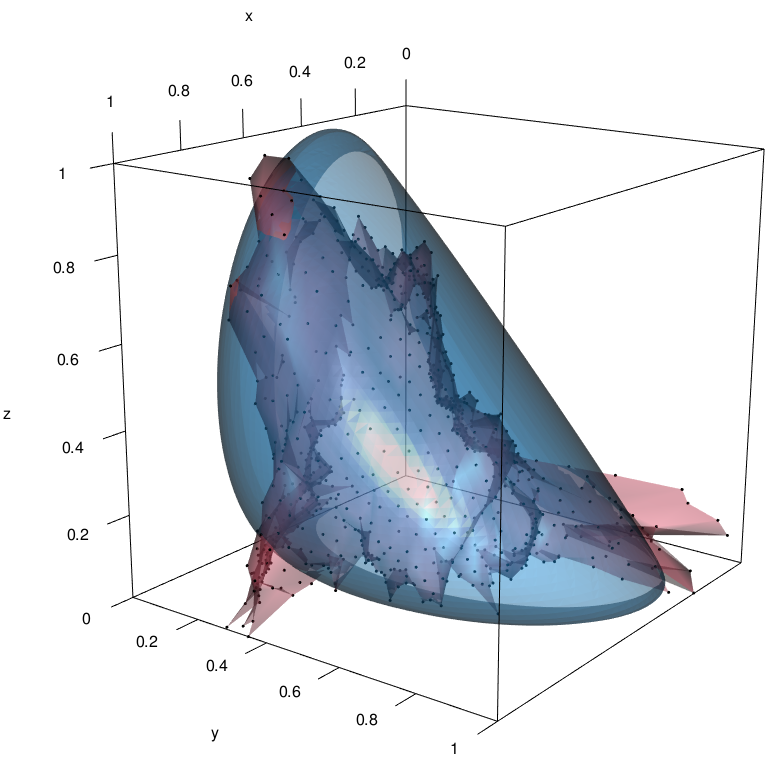}
	\caption{Left and centre: PP and exponential QQ plots for the fitted truncated gamma model with the Gaussian gauge function. Right: unit level set of the empirical gauge function (red) and fitted Gaussian gauge function (blue) for the Newlyn wave data.}
	\label{fig:wave-pp-qq-gauge}
\end{figure}

In spite of the structure suggested by the estimated $\tau_C(\delta)$ values, the AIC indicates a preference for the Gaussian gauge function. The \editm{maximum likelihood estimates} are $(\widehat{\alpha},\widehat{\theta}_{HP},\widehat{\theta}_{HS},\widehat{\theta}_{PS}) = (0.79, 0.70, 0.65, 0.30)$, \editm{where $\theta_{jk}$ are the Gaussian correlation parameters in the gauge function.} The data have been filtered to give approximate temporal independence, so we estimate Hessian-based standard errors as $(0.74,0.12, 0.12, 0.18)$. Figure~\ref{fig:wave-pp-qq-gauge} displays the PP plot for this fit as described in Section~\ref{sec:modelchecking}, as well as the same plot transformed onto the exponential scale to emphasize the upper tail, indicating no lack of fit. We also compare the empirical gauge $\hat{g}(\bm{w})$ and the fitted Gaussian gauge function  $g(\bm{w};\widehat{\bm{\theta}})$ in the right panel of Figure~\ref{fig:wave-pp-qq-gauge}. The empirical gauge is relatively ``jagged'' and variable due to the manner of its calculation, but there is \edit{broad} correspondence between its overall shape and that of the fitted gauge. \edit{Interestingly, the fit of the asymmetric logistic gauge returns a parameter estimate of $\widehat{\gamma}_{HS} = 1$, which is on the boundary of the parameter space. This could be indicative of wave height and surge not displaying exceptionally strong dependence, but also because there are restrictions on the shape of the limit set arising from the asymmetric logistic distribution, and this parameter estimate provides the best overall fit to all data simultaneously.}

\begin{figure}[htpb]
	\centering
	\includegraphics[width=0.3\textwidth]{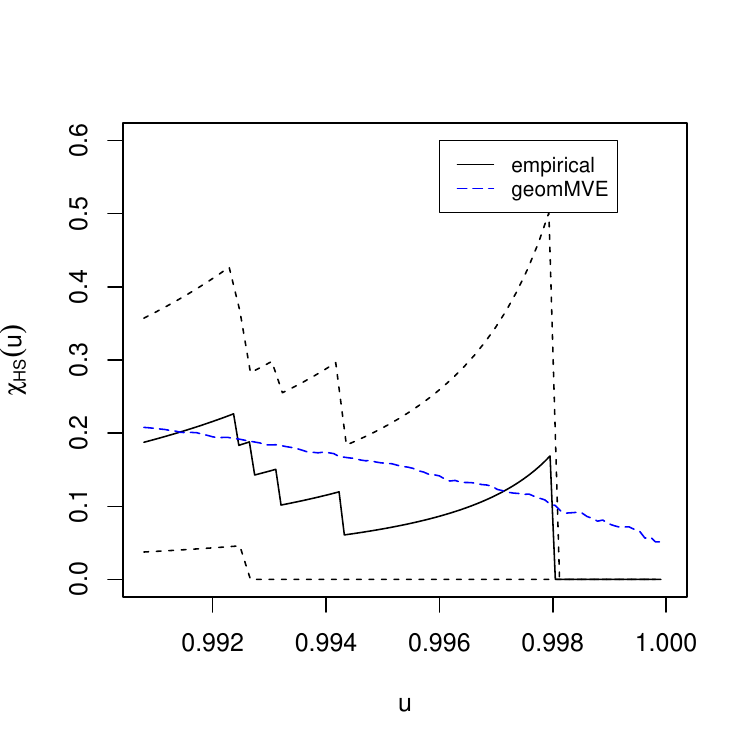}
	\includegraphics[width=0.3\textwidth]{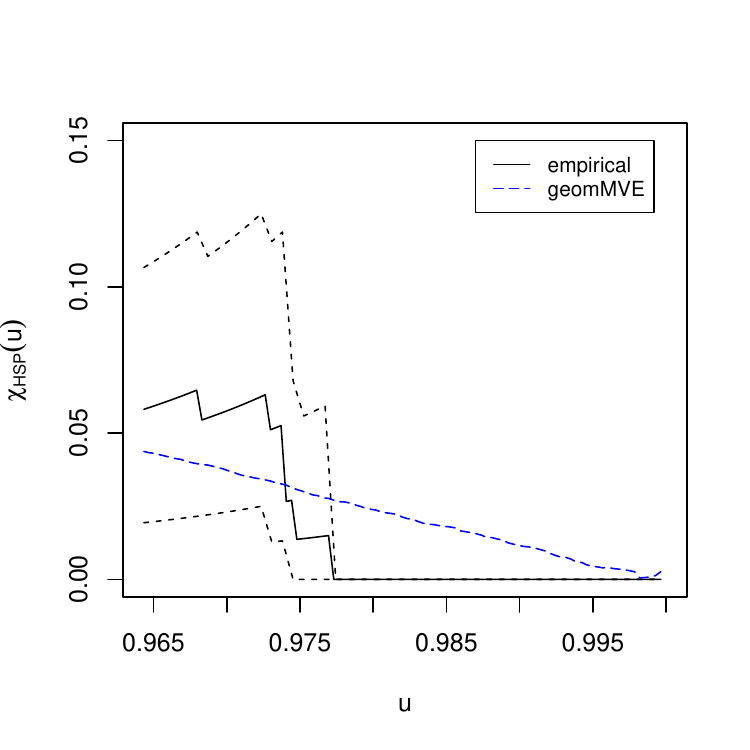}
		\includegraphics[width=0.3\textwidth]{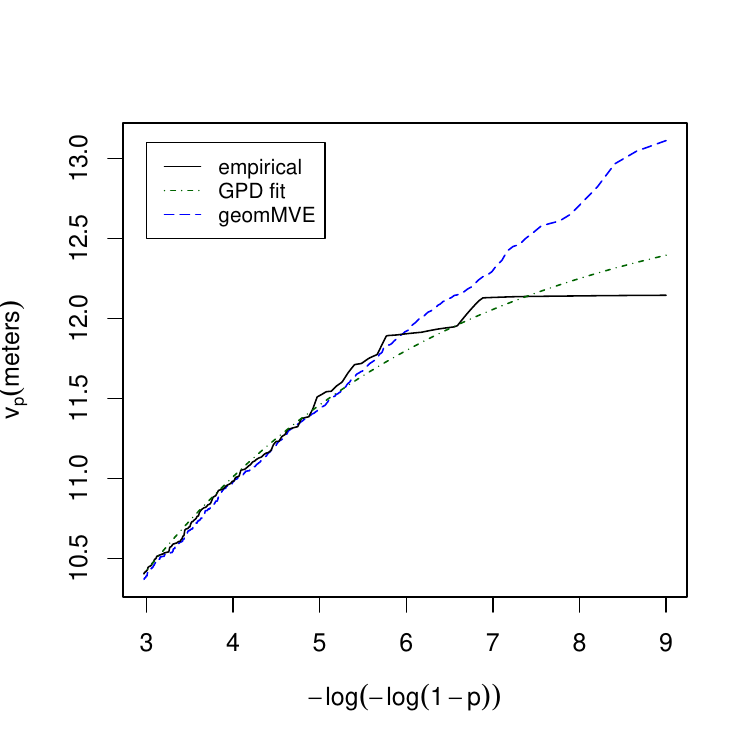}
	\caption{Left: Estimates of $\chi_{HS}(u)$, $u \in (0.99,1)$; centre: estimates of $\chi_{HSP}(u)$, $u \in (0.965,1)$. Black solid lines represent empirical estimates, dotted lines 95\% confidence intervals, and blue dashed lines the model-based estimate for the Newlyn wave dataset. Right: plot of quantiles $v_p$ of the \editm{structure} variable $V$, \edit{representing sea wall height}.}
	\label{fig:wave-chi-plot}
\end{figure}

As a further diagnostic, we compare empirical and model-based estimates of the sub-asymptotic joint \emph{tail dependence coefficient}. For $X_{j} \sim F_{j}$, this is defined by
\begin{align}
\chi_{C}(u) = \frac{1}{1-u}\Pr\left(F_j(X_{j})>u, \forall j \in C\right), \qquad u\in(0,1),~~ C \subseteq\{H,S,P\}. \label{eq:chiu}
\end{align}
The empirical estimator of $\chi_{C}(u)$ is obtained by replacing each distribution function and joint probability with its empirical counterpart, while the model-based estimate is calculated using simulation from the fitted model as described in Section~\ref{sec:Prediction}, and suitable sets $B$. In Figure~\ref{fig:wave-chi-plot} we consider $\chi_{HSP}(u)$ and $\chi_{HS}(u)$, meaning $B=(-\log(1-u),\infty)^3$ and $B=(-\log(1-u),\infty)^2 \times (0,\infty)$, respectively. The range over which the model-based tail dependence coefficients can be calculated depends on the values of $\bm{X}$ constituting the extreme region $\{R'>1\}$. There is good agreement with the empirical estimates, with the model-based estimates \editm{allowing extrapolation} beyond the range of the data.

Finally we consider analysis of the structure variable outlined in \citet{ColesTawn94}. They introduce the overtopping discharge rate $Q(v;\bm{X}_{HSP})$ for a sea wall of height $v$ as
\begin{align*}
Q(v;\bm{X}_{HSP}) &= a_1 X_S X_P \exp\left\{-\frac{a_2\left(v-X_S-l\right)}{X_P {X_H^\ast}^{{1}/{2}}}\right\}, & X_H^\ast &= X_H\left(1-\exp\left\{-\frac{\left(l+X_S\right)^2}{2X_H^2}\right\}\right)^{{1}/{2}}.
\end{align*}
The value $X^\ast_H$ is introduced to approximate the actual off-shore wave height, since measurements are taken on-shore. \edit{The goal is to estimate the sea wall height $v_p$ (in metres) for which the overtopping discharge rate is expected to exceed $0.002 m^3 s^{-1}$ \edit{per metre of sea wall} with probability $p$}. That is, setting $V=Q^{-1}(0.002; \bm{X}_{HSP})$, we solve $\Pr(V>v_p) = p$ for $v_p$ using realizations of $V$ generated through simulation and reverse marginal transformation. \edit{Specifically, we generate new realizations of $\bm{X}_{HSP}$, and hence $V$, in the tail region of our model by simulating on exponential margins and inverting equation~\eqref{eq:marginaltf}. Outside of the tail region, we use the empirical distribution of $V$.} As in \citet{Bortotetal01}, we fix $a_1=0.25$, $a_2=26$, and $l=4.3$. \edit{The right panel of} Figure~\ref{fig:wave-chi-plot} displays the obtained values $v_p$, with empirical quantiles and those calculated from fitting the \edit{generalized Pareto} distribution directly to the tail of $V$ (the so-called ``structure variable approach'') for comparison. For very small $p$, the return levels obtained from the geometric model are larger than those from the \edit{generalized Pareto} fit. They are comparable to those obtained in \citet{Bortotetal01}, but much lower than those in \citet{ColesTawn94}, whose model incorrectly assumes that the spectral measure places mass on $\mathbb{B}_{\{H,S,P\}}$.

\subsection{River flow data}
\label{sec:river}
We now apply our modelling approach on 12,327 measurements of daily mean river flow (m$^3$/s) from four gauging stations in the north west of England. The data were previously explored in \citet{Simpsonetal20}, where focus lay on determining the support of the spectral measure, but not subsequent modelling of the variables, due to lack of suitable models that could account for complex structures. We opt to consider four out of the five locations initially used in order to keep the number of parameters reasonable; further discussion on dimensionality can be found in Section~\ref{sec:discussion}. The four stations, labelled 1, 2, 3,  4, correspond to those labelled A, B, C, D in \citet{Simpsonetal20}.

Margins are standardized using equation~\eqref{eq:marginaltf}. We then use the \citet{Simpsonetal20} methodology, which suggests that the spectral measure may place mass on the faces $\mathbb{B}_{\{2\}},\mathbb{B}_{\{4\}},\mathbb{B}_{\{1,4\}},\mathbb{B}_{\{1,3,4\}},$ and $\mathbb{B}_{\{1,2,3,4\}}$ of the simplex $\mathcal{S}_{3}$. \editm{We fit the model with the corresponding asymmetric logistic gauge function, a Gaussian gauge function, and an additive mixture of the two. The AIC values are 2666, 2601 and 2609, respectively.} Once again, the model with the Gaussian gauge is preferred, in apparent conflict with the estimated structure of the spectral measure, though we note this is also subject to uncertainty. \edit{Parameter estimates and approximate standard errors are given in Table}~\ref{tab:riverparam}. To account for temporal dependence of river flows, \edit{standard errors} are found via \edit{use of a block bootstrap on the original data series}, with block length 20.
 The asymmetric logistic gauge, while able to capture the structure of different groups of variables being co-extreme, appears too inflexible to capture other aspects of the dependence. The additively mixed model is an attempt to alleviate this problem, but leads to a large number of parameters without a sufficient improvement in fit to compensate for them.

\begin{table}[h]
	\centering
	\caption{\edit{Parameter estimates and approximate block bootstrap-based standard errors for the river flow data. Parameter $\theta_{jk}$ represents the Gaussian gauge correlation parameter between sites $j,k$.}}
\begin{tabular}{l|lllllll}
	Parameter & $\alpha$ & $\theta_{12}$ & $\theta_{13}$ & $\theta_{14}$ & $\theta_{23}$ & $\theta_{24}$ & $\theta_{34}$  \\\hline
	Estimate & 2.46& 0.83& 0.90& 0.80& 0.90& 0.57& 0.62\\
	Standard error&0.62& 0.11& 0.14& 0.14& 0.14& 0.16& 0.14
\end{tabular}
\label{tab:riverparam}
\end{table}

%Parameter estimates are $(\widehat{\alpha},\widehat{\bm{\theta}})=(2.46, 0.83, 0.90, 0.80, 0.90, 0.57, 0.62)$, with approximate standard errors $(0.62, 0.11, 0.14, 0.14, 0.14, 0.16, 0.14)$. 

\begin{figure}[htpb]
	\centering
	\includegraphics[width=0.32\textwidth]{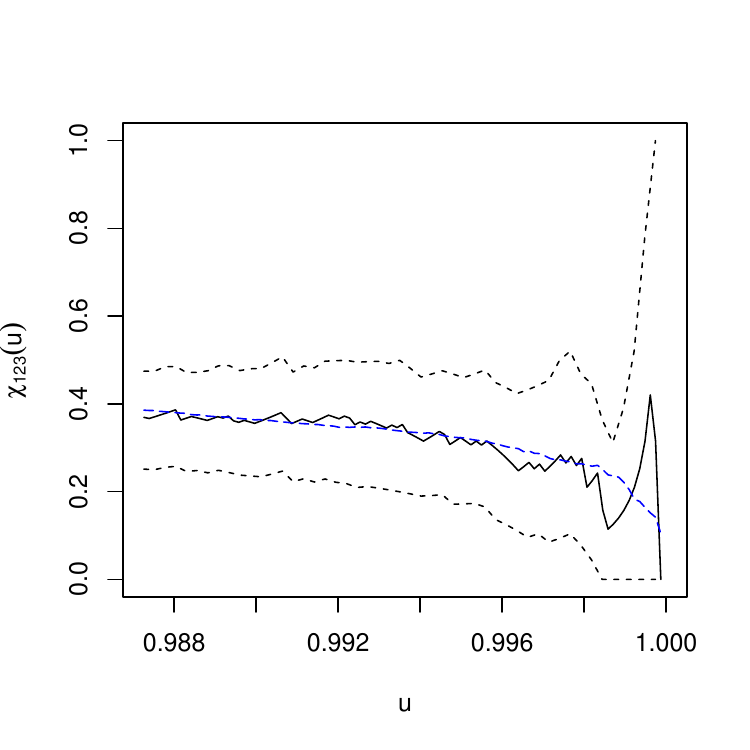}
	\includegraphics[width=0.32\textwidth]{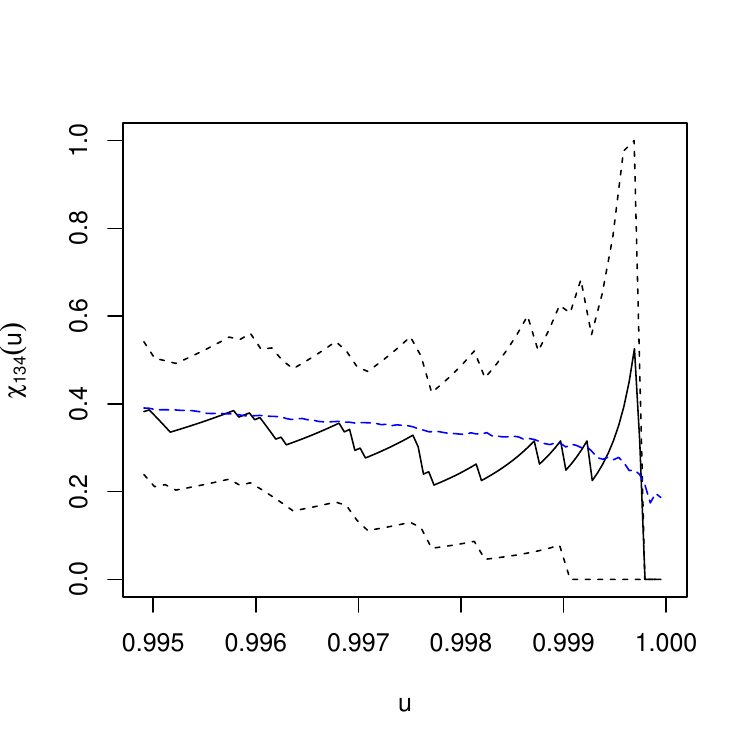}
	\includegraphics[width=0.32\textwidth]{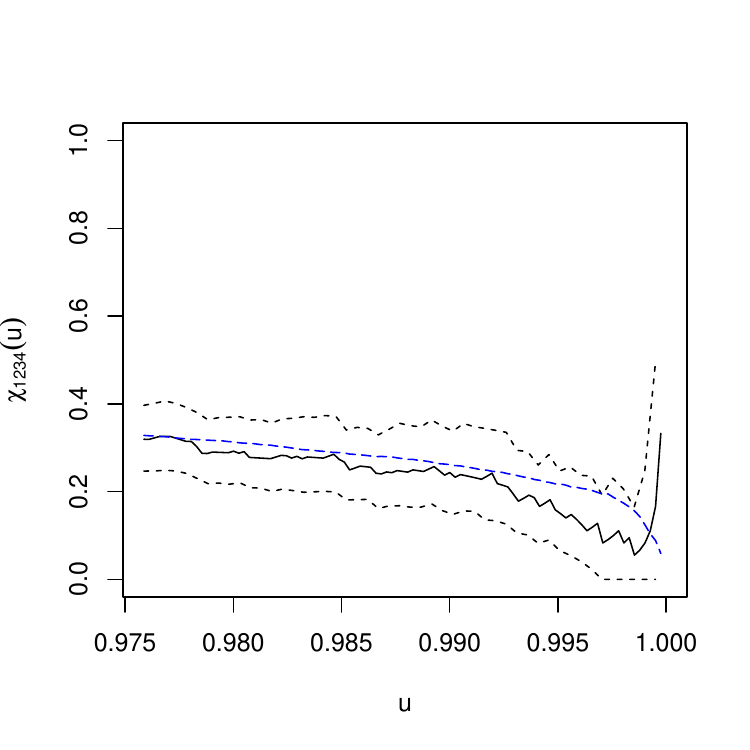}
	\caption{Empirical (black) estimates of $\chi_C(u)$ with 95\% confidence interval (dashed), and model-based estimate (blue) for $C=\{1,2,3\}$, $\{1,3,4\}$ and $\{1,2,3,4\}$ (left to right).}
	\label{fig:river-chi}
\end{figure}

Figure~\ref{fig:river-chi} displays coefficients $\chi_{123}(u)$, $\chi_{134}(u)$ and $\chi_{1234}(u)$, defined analogously to~\eqref{eq:chiu}. If $H$ places mass on $\mathbb{B}_{\{1,2,3,4\}}$, then each of these coefficients has a positive limit as $u \to 1$, but at observable levels, the model-based estimates from the Gaussian gauge all represent a good fit to the data.  Plots of $\chi_C(u)$ for the remaining groups of variables are given in Section~\ref{sec:dataextra} of the supplement, along with the PP plot, showing no lack of fit. \edit{In the limit as $u\to 1$, estimates of $\chi_C(u)$ from the geometric model with Gaussian gauge will all be zero. However, the inference that $H$ places mass on $\mathbb{B}_{\{1,2,3,4\}}$, and other faces, is subject to uncertainty. From the plots in Figure~\ref{fig:river-chi}, it is difficult to determine whether the limits of  $\chi_{123}(u)$, $\chi_{134}(u)$ and $\chi_{1234}(u)$ as $u\to1$ are indeed positive or zero, and as a consequence whether a gauge function that reflects $H(\mathbb{B}_{\{1,2,3,4\}})>0$ is truly preferable. Nonetheless, this framework offers the chance to test these models and assumptions in a way that was not previously possible.}

\section{Discussion}
\label{sec:discussion}

We have presented a new approach to multivariate extreme value modelling, based on estimation of the shape of the limit set of a sample cloud of data points in light-tailed margins. The methodology allows for modelling datasets with complicated extremal dependence structures, whereby different groups of variables may be co-extreme, as well as extrapolation into parts of the multivariate tail where only some variables are large. 

By offering models for complex dependence structures with non-simultaneous extremes, this approach paves the way for more useful higher dimensional extreme value modelling. Recent literature on multivariate extremes that is targeted at higher dimensions typically involves making strong simplifying assumptions on the dependence structure. For example, the extremal graphical models outlined in \citet{EngelkeHitz20} require an assumption that the spectral measure $H$ places all mass on $\mathbb{B}_{\{1,\ldots,d\}}$. 

In this work, we demonstrated the methods up to dimension $d=4$. The main challenges for routine application of the methods for $d$ larger than 3 or 4 lie in calculation of the threshold function $r_0(\bm{w})$, and specification of flexible gauge functions. The former could potentially be addressed by adapting the additive quantile regression approach of \citet{Fasioloetal21} to incorporate basis functions whose support is the simplex $\mathcal{S}_{d-1}$. Addressing the latter challenge requires ways to build flexible and parsimonious gauge functions, which is a topic of current work. In particular, we note that models fitted in Section~\ref{sec:data} had the ability to capture the complex structures suggested by the \citet{Simpsonetal20} methodology, but the best fits were obtained through models that were more flexible in other aspects. This led to the conclusion that the model with the Gaussian gauge function was preferred for both datasets, which is likely a consequence of being able to capture a range of strengths of dependences across different groups of variables; in contrast, the asymmetric logistic gauge function treats groups of variables that do not exhibit simultaneous extremes as effectively independent. We note also that estimates of the faces $\mathbb{B}_C$ on which $H$ places mass are themselves subject to uncertainty, which is not easily quantifiable thanks to the requirement to select tuning parameters. Conflicts between the estimated structure and the selected gauge function may therefore not be too concerning, provided the diagnostics for the model are adequate.

\edit{A further challenge with our methodology for dimensions $d \geq 5$ is the use of the empirical distribution for the angles $\bm{W}$. We anticipate that considering (semi-)parametric forms for this distribution will be needed as part of adapting the methods to higher dimensions.}

We have focused here primarily on positive dependence as it is common in many datasets and simplifies the presentation. For datasets exhibiting any form of negative dependence, the limit set shapes are more descriptive in Laplace, rather than exponential, margins. For example, we mentioned for the multivariate Gaussian case that the continuous convergence to $g(\bm{x})$ fails when some component of $\bm{x}$ is zero; this is not an issue in Laplace margins, where the limit set lies in the region $[-1,1]^d$ rather than $[0,1]^d$, and similarly for the $t_{\nu}$ distribution. Moving from the positive quadrant to $\mathbb{R}^d$ requires defining the angles $\bm{W}$ differently, but otherwise a similar approach could be applied, and represents a natural next step in developing this methodology.

\subsection*{Acknowledgements}
\edit{We thank two reviewers and an associate editor for helpful comments that have improved the manuscript.} We are grateful for \edit{EPSRC grant EP/X010449/1 supporting Jennifer Wadsworth, and} EPSRC DTP funding EP/W523811/1 supporting Ryan Campbell.

\subsection*{Code and data}
Code and data for the analyses in Section~\ref{sec:data} are available as supplementary material. The river flow dataset originated from the UK Centre for Ecology \& Hydrology at \url{nrfa.ceh.ac.uk}. An R package \texttt{geometricMVE} for implementing the methodology \editm{presented} in the article is available as supplementary material and at \url{http://www.lancaster.ac.uk/~wadswojl/geometricMVE.html}. \edit{Interactive versions of 3d plots are available at the same URL.} 
%\bigskip
%\vfill
%\newpage

%\newpage
%\bibliographystyle{apalike}
%\bibliography{GeometricBib}

%%%%%%%%%%%%%%%%%%%%%%%%%%%%%%%%%%%%%%%%%%%%%%%%%%%%%%%%%%%%%%%%%%%%%%%%%%%%%%%%%%%%%%%%%%%
%%%%%%%%%%%%%%%%%%%%%%%%%%%%%%%%%%%%%%%%%%%%%%%%%%%%%%%%%%%%%%%%%%%%%%%%%%%%%%%%%%%%%%%%%%%

\newpage
\begin{center}
	{\Large\bf Supplementary Material}
\end{center}
\vspace{-1cm}

\appendix

\section{Example limit sets}
\label{sec:examplelimitsets}
\begin{figure}[htbp]
	\centering
	\includegraphics[width=0.3\textwidth]{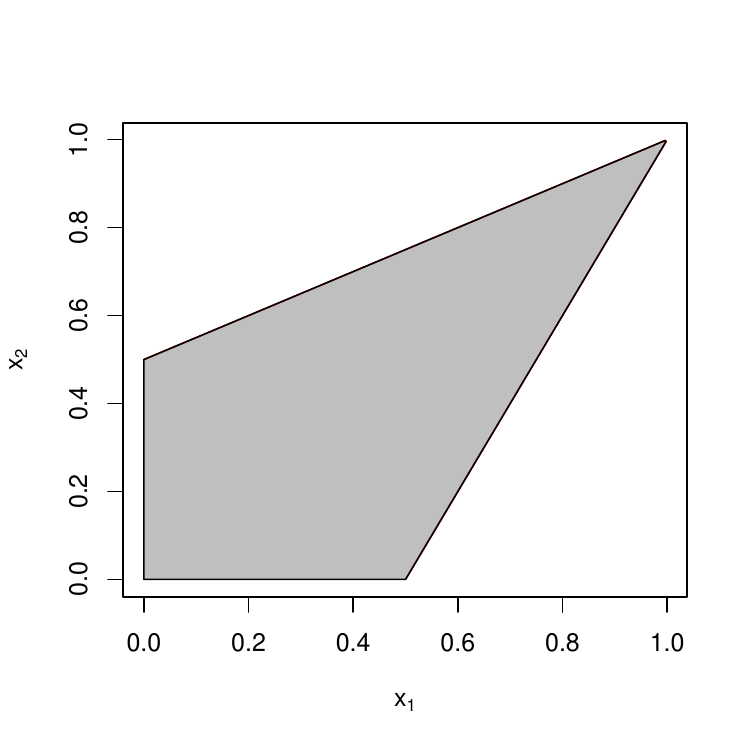}
	\includegraphics[width=0.3\textwidth]{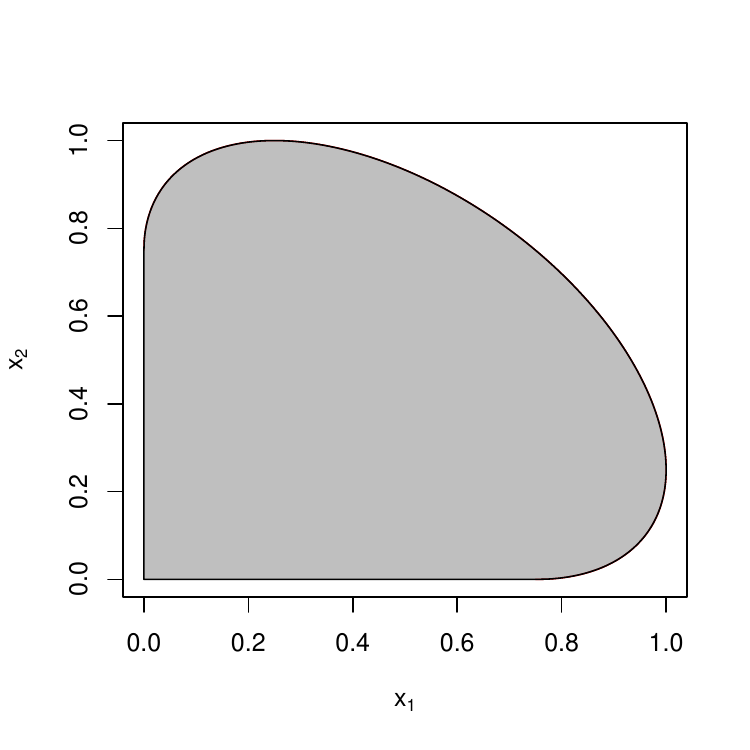}
	\includegraphics[width=0.3\textwidth]{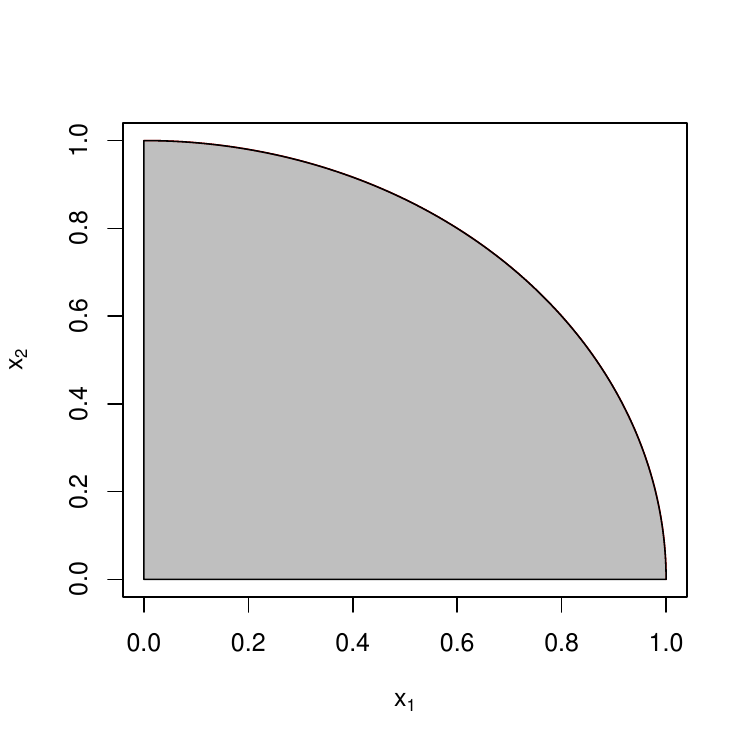}
	\caption{Illustration of limit sets \edit{(grey shaded region)} arising from the logistic, Gaussian, and inverted logistic distributions (L-R) in dimension $d=2$. Red lines represent unit level sets of the gauge function. Each dependence parameter is equal to 0.5.}
	\label{fig:2dgaugeexamples}
\end{figure}

\begin{figure}[h]
	\centering
	\includegraphics[width=0.32\textwidth]{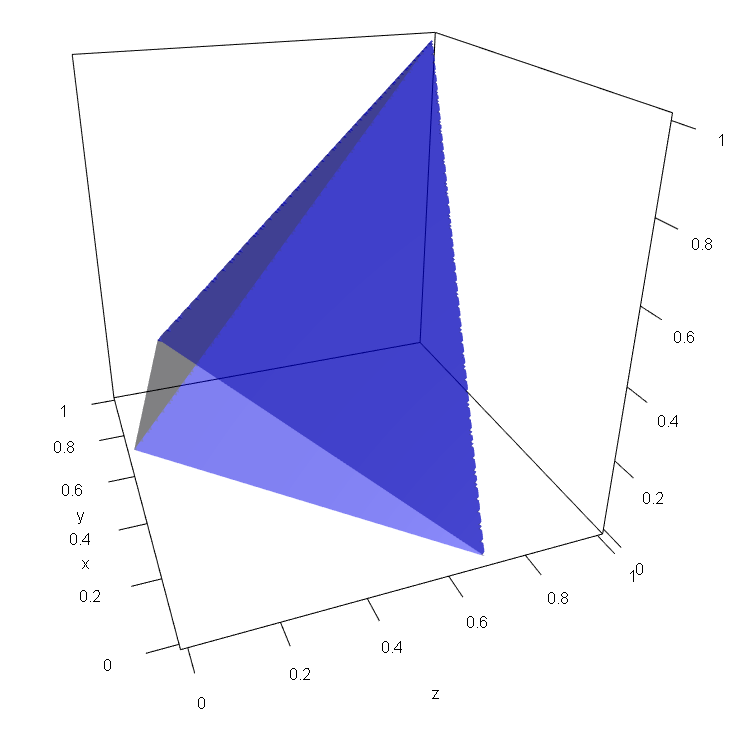}
	\includegraphics[width=0.32\textwidth]{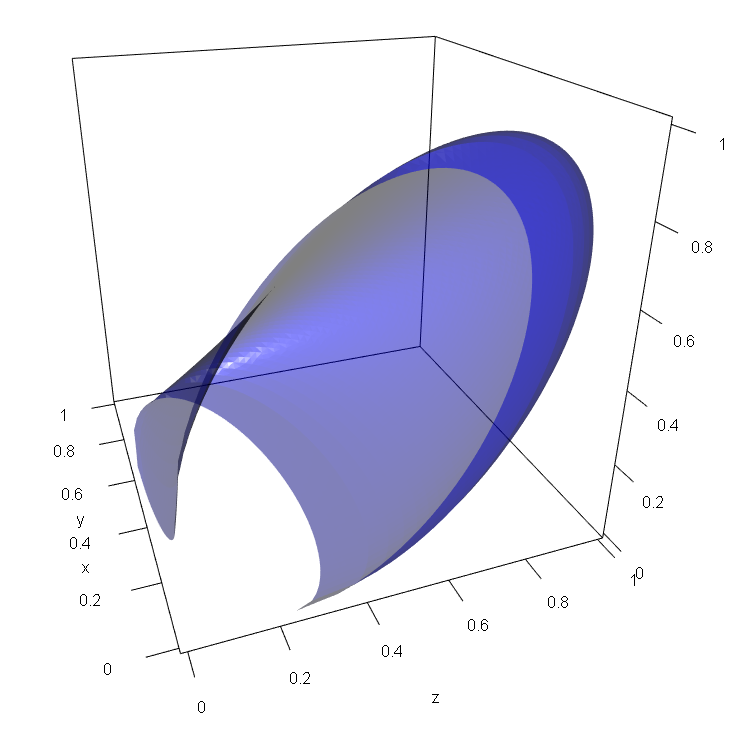}
	\includegraphics[width=0.32\textwidth]{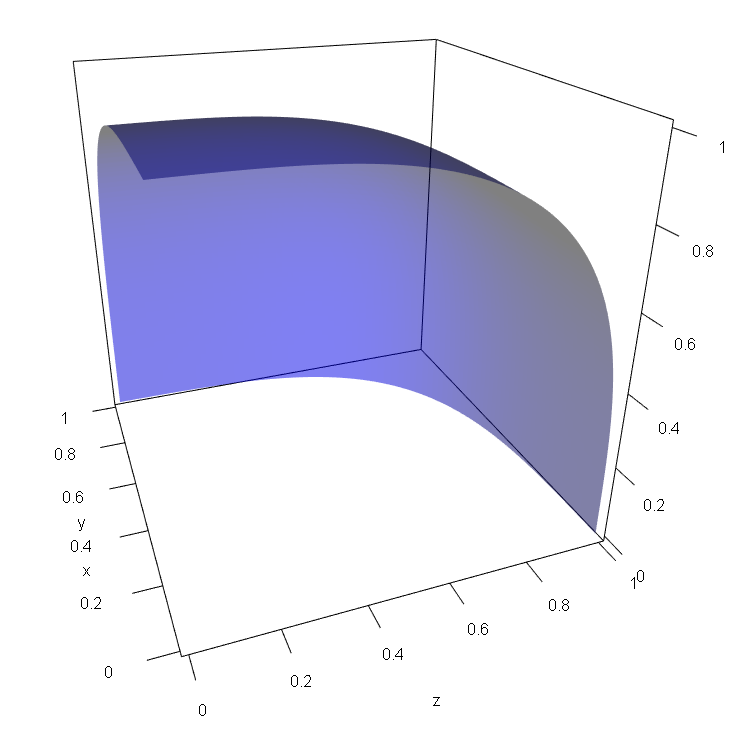}
	\caption{Illustration of limit sets (region between blue surface and planes $x_j=0$) arising from the logistic, Gaussian, and inverted logistic distributions (L-R) in dimension $d=3$. Blue surfaces represent unit level sets of the gauge function. Dependence parameters are set to: $\gamma=0.7$, $\bm{\rho}=(0.2,0.5,0.8)$ and $\gamma=0.3$, respectively.}
	\label{fig:3dgaugeexamples}
\end{figure}

Figures~\ref{fig:2dgaugeexamples} and~\ref{fig:3dgaugeexamples} display example illustrations of limit sets in exponential margins for three dependence structures in dimension $d=2$ and $d=3$ respectively. Equations for the gauge functions of these limit sets can be found in Section~\ref{sec:Examples} of the main paper, or Section~\ref{sec:rwderivations} of the supplement.

\newpage 

\section{Gauge functions and conditional distributions of \texorpdfstring{$R|\bm{W}=\bm{w}$}{rgivenw}}
\label{sec:rwderivations}
Here we provide detailed calculations of the gauge functions such that we can establish the asymptotic behavior of $f_{R|\bm{W}}(r|\bm{w})$ as $r \to \infty$. In each case, we begin with the relevant density in exponential margins and calculate the asymptotic behavior of $f_{\bm{X}}(t\bm{x})$ as $t \to \infty$; this is subsequently used to establish results for $f_{R|\bm{W}}(r|\bm{w})$ as $r \to \infty$. We recall that the notation for ordered values is $x_{(1)}\geq x_{(2)} \geq \cdots \geq x_{(d)}>0$, and similarly $w_{(1)}\geq w_{(2)} \geq \cdots \geq w_{(d)}>0$.

\paragraph{Logistic distribution}
The $d$-dimensional logistic distribution with unit Fr\'echet margins has density
\begin{align*}
f_{\bm{Z}_F}(\bm{z}) = \exp\{-V(\bm{z})\}\sum_{\pi \in \Pi}\prod_{s \in \pi} -V_s(\bm{z}),
\end{align*}
where $V: \mathbb{R}^d_+ \to \mathbb{R}_+$ is the homogeneous of order $-1$ \emph{exponent function}. For the logistic distribution, this is 
\begin{align*}
V(\bm{z}) = \left(\sum_{j=1}^d z_j^{-1/\gamma}\right)^\gamma, \qquad \gamma \in (0,1].
\end{align*}
The transformation to exponential margins is given by $z_j = [-\log(1-e^{-x_j})]^{-1}$. Expanding this to give the asymptotic behavior for large $x_j$ yields $z_j(x_j) = e^{x_j}+1/2+O(e^{-x_j})$. We can therefore express the density in exponential margins as
\begin{align*}
f_{\bm{X}}(t\bm{x}) = \exp\{-V(e^{t\bm{x}}+1/2+O(e^{-t{\bm{x}}}))\}\sum_{\pi \in \Pi}\prod_{s \in \pi} V_s(e^{t\bm{x}}+1/2+O(e^{-t\bm{x}})) \times e^{t \sum_{j=1}^d x_j}[1+O(e^{-2tx_{(d)}})].
\end{align*}

Firstly consider the contribution $\exp\{-V(e^{t\bm{x}}+1/2+O(e^{-t{\bm{x}}}))\}$. By homogeneity
\begin{align*}
V(e^{t\bm{x}}+1/2+O(e^{-t{\bm{x}}})) & = e^{-t x_{(d)}} V(e^{t(\bm{x}-x_{(d)})}+e^{-tx_{(d)}}/2+O(e^{-t{(\bm{x}+x_{(d)})}}))\\
& =  e^{-t x_{(d)}} [c+o(1)], \qquad t \to \infty,
\end{align*}
where $c$ is a constant that equals 1 if $x_{(d)}<x_{(d-1)}$. Consequently, 
\begin{align*}
\exp\{-V(e^{t\bm{x}}+1/2+O(e^{-t{\bm{x}}}))\} = 1+ O(e^{-t x_{(d)}}).
\end{align*}
Next consider the partial derivatives $V_{s}(\bm{z})$. We have
\begin{align*}
V_s(\bm{z}) \propto \left(\prod_{j \in s }z_j\right)^{-1/\gamma-1}\left(\sum_{j=1}^d z_j^{-1/\gamma}\right)^{\gamma-|s|},
\end{align*}
and therefore 
\begin{align*}
V_s(e^{t\bm{x}}+1/2+O(e^{-t\bm{x}})) &\propto \left(\prod_{j \in s}[e^{tx_j}+1/2+O(e^{-t x_j})]\right)^{-1/\gamma-1}\left(\sum_{j=1}^d [e^{tx_j}+1/2+O(e^{-t x_j})]^{-1/\gamma}\right)^{\gamma-|s|}\\
&= e^{-t(1/\gamma+1)\sum_{j\in s} x_j}[1+O(e^{-t \min_{j \in s} x_j})] \\&~~~~\times e^{-t x_{(d)}(1-|s|/\gamma)}[1+O(e^{-t(x_{(d-1)}-x_{(d)})/\gamma})+O(e^{-t x_{(d)}/\gamma})]\\
& = e^{-t(1/\gamma+1)\sum_{j\in s} x_j-tx_{(d)}(1-|s|/\gamma)}[1+O(e^{-t(x_{(d-1)}-x_{(d)})/\gamma})+O(e^{-t x_{(d)}/\gamma})].
\end{align*}
For each partition $\pi \in \Pi$,
\begin{align*}
\prod_{s \in \pi} V_s(e^{t\bm{x}}+1/2+O(e^{-t\bm{x}})) &\propto e^{-t(1/\gamma+1)\sum_{j=1}^d x_j-t \sum_{s \in \pi}x_{(d)}(1-|s|/\gamma)} [1+O(e^{-t(x_{(d-1)}-x_{(d)})/\gamma})+O(e^{-t x_{(d)}/\gamma})]\\
&=e^{-t[(1/\gamma+1)\sum_{j=1}^d x_j+x_{(d)}(|\pi|-d/\gamma)]}[1+O(e^{-t(x_{(d-1)}-x_{(d)})/\gamma})+O(e^{-t x_{(d)}/\gamma})].
\end{align*}
Combining all of these results yields
\begin{align*}
f_{\bm{X}}(t\bm{x}) \propto \sum_{\pi \in \Pi}e^{-t[(1/\gamma)\sum_{j=1}^d x_j+ x_{(d)}(|\pi|-d/\gamma)]}[1+O(e^{-t(x_{(d-1)}-x_{(d)})/\gamma})+O(e^{-t x_{(d)}})].
\end{align*}
The gauge function comes from taking $\min_{\pi \in \Pi} [(1/\gamma)\sum_{j=1}^d x_j+ x_{(d)}(|\pi|-d/\gamma)]$, which clearly occurs for $\pi$ with $|\pi|=1$, i.e. $\pi = \{\{1,\ldots,d\}\}$. Hence $g(\bm{x}) = (1/\gamma)\sum_{j=1}^d x_j+ x_{(d)}(1-d/\gamma)$.

Turning to the conditional distribution of $R|\bm{W}=\bm{w}$, we have
\begin{align*}
f_{R|\bm{W}}(r|\bm{w}) \propto r^{d-1} e^{-rg(\bm{w})}[1+O(e^{-r(w_{(d-1)}-w_{(d)})/\gamma})+O(e^{-rw_{(d)}})], \qquad r \to \infty.
\end{align*}

\paragraph{Negative logistic MGPD}
MGPDs have marginal scale and shape parameters, and when these are all set to 1 and 0 respectively, the marginal distributions are exponential conditionally upon being positive. That is, if $\bm{Z}$ follows a MGPD with unit-scale and zero-shape parameters, the marginal distribution functions $F_{j}(z)$ are
\begin{align*}
F_j(z) & = \begin{cases}
\Pr(Z_j \leq z), & z <0,\\
c_j + (1-c_j)(1-e^{-z}), & z \geq 0,
\end{cases}
\end{align*}
with $c_j = \Pr(Z_j \leq 0)$. To translate to exponential margins we solve $x_j=-\log(1-F_j(z_j))$, which leads to $z_j(x_j)= x_j +\log (1-c_j) +$, $x_j >-\log(1-c_j)$.

Following calculations in \citet{Kiriliouketal18}, the unit-scale zero-shape MGPD density associated to the negative logistic max-stable distribution is
\begin{align*}
f(\bm{z}) &\propto e^{\gamma\sum_{j=1}^d z_j} \left(\sum_{j=1}^d  e^{\gamma  z_j}\right)^{-(d+1/\gamma)}, \qquad \gamma>1,
\end{align*}
and so on the region $\{\bm{x}: x_j>-\log(1-c_j)/t, j=1,\ldots, d\}$
\begin{align*}
f_{\bm{X}}(t\bm{x}) &\propto e^{\gamma\sum_{j=1}^d tx_j + \gamma\sum_{j=1}^d \log(1-c_j)} \left(\sum_{j=1}^d  e^{\gamma t x_j +\gamma\log(1-c_j)}\right)^{-(d+1/\gamma)},\\
& \propto e^{t\gamma\sum_{j=1}^d x_j-t(1+d\gamma)x_{(1)}} \left[1+O(e^{t(x_{(2)}-x_{(1)})\gamma})\right],
\end{align*}
so that $g(\bm{x}) = (1+d\gamma)x_{(1)} - \gamma\sum_{j=1}^d x_j$, and
\begin{align*}
f_{R|\bm{W}}(r|\bm{w}) \propto r^{d-1} e^{-rg(\bm{w})}\left[1+O(e^{r(w_{(2)}-w_{(1)})\gamma})\right],
\end{align*}
on the region $\{r> \min_{1 \leq j \leq d} -\log(1-c_j)/w_j\}$. Outside of this region we require knowledge of the distribution of $Z_j|Z_j<0$, $j=1,\ldots,d$, which is harder to summarize in general.

\paragraph{Dirichlet MGPD}
In this case
\begin{align*}
f(\bm{z}) &\propto e^{\sum_{j=1}^d \theta_j z_j} \left(\sum_{j=1}^d  e^{ z_j}\right)^{-\left(\sum_{j=1}^d \theta_j+1\right)},\qquad \theta_1,\ldots,\theta_d>0,
\end{align*}
and so on the region $\{\bm{x}: x_j>-\log(1-c_j)/t, j=1,\ldots, d\}$
\begin{align*}
f_{\bm{X}}(t\bm{x}) &\propto e^{\sum_{j=1}^d t\theta_j x_j+\sum_{j=1}^d \theta_j\log(1-c_j)} \left(\sum_{j=1}^d  e^{t x_j + \log(1-c_j)}\right)^{-\left(\sum_{j=1}^d \theta_j+1\right)},\\
& \propto e^{t\sum_{j=1}^d \theta_j x_j-t(1+\sum_{j=1}^d \theta_j)x_{(1)}} \left[1+O(e^{t(x_{(2)}-x_{(1)})})\right], 
\end{align*}
so that $g(\bm{x}) =(1+\sum_{j=1}^d \theta_j)x_{(1)} - \sum_{j=1}^d \theta_j x_j$, and
\begin{align*}
f_{R|\bm{W}}(r|\bm{w}) \propto r^{d-1} e^{-rg(\bm{w})}\left[1+O(e^{r(w_{(2)}-w_{(1)})})\right],
\end{align*}
on the region $\{r> \min_{1 \leq j \leq d} -\log(1-c_j)/w_j\}$.

\paragraph{Inverted max-stable distributions}
Recall the form of the density is
\begin{align*}
f_{\bm{X}}(\bm{x}) = \exp\{-l(\bm{x})\}\sum_{\pi \in \Pi}\prod_{s \in \pi} l_s(\bm{x}).
\end{align*}
The derivatives $l_s(\bm{x})$ are homogeneous of order $1-|s|$, and so 
\begin{align*}
f_{\bm{X}}(t\bm{x}) = \exp\{-t l(\bm{x})\}\sum_{\pi \in \Pi}\prod_{s \in \pi} t^{1-|s|}l_s(\bm{x}).
\end{align*}
The leading-order term in the summation therefore comes from the partition $\pi=\{\{1\},\{2\},\ldots,\{d\}\}$, with $|s|=1$ for all $s \in \pi$. Second-order behavior comes from the $d$ partitions containing one set with $|s|=2$ and all others with $|s|=1$. We therefore have
\begin{align*}
f_{\bm{X}}(t\bm{x}) = \exp\{-t l(\bm{x})\}l_{\{1\}}(\bm{x})\cdots l_{\{d\}}(\bm{x})[1+O(t^{-1})],
\end{align*}
so that $g(\bm{x}) = l(\bm{x})$, and
\begin{align*}
f_{R|\bm{W}}(r|\bm{w}) \propto r^{d-1}\exp\{-r g(\bm{w})\}[1+O(r^{-1})].
\end{align*}
\paragraph{Multivariate Gaussian distribution}
The multivariate Gaussian density with Gaussian margins is
\begin{align*}
f_{\bm{Z}_G}(\bm{z}) \propto \exp\left\{-\frac{1}{2}\bm{z}^\top \Sigma^{-1} \bm{z}\right\} = \exp\left\{-\frac{1}{2}\sum_{j=1}^d \sum_{k=1}^d z_{j}z_{k} \omega_{jk}\right\},
\end{align*}
where $\Omega = (\omega_{jk})_{j,k} = \Sigma^{-1}$ is the precision matrix. Transforming to exponential margins, we obtain
\begin{align}
f_{\bm{X}}(\bm{x}) \propto \exp\left\{-\frac{1}{2}\sum_{j=1}^d \sum_{k=1}^d z_{j}(x_j)z_{k}(x_k) \omega_{jk}\right\} \prod_{j=1}^d \frac{1-\Phi(z_j(x_j))}{\phi(z_{j}(x_j))}, \label{eq:GaussExp}
\end{align}
where $z_j(x_j)$ is found through solving $x_j = -\log(1-\Phi(z_j))$, with $\phi, \Phi$ the standard univariate normal density and df, respectively. Since we are interested in $tx_j$, $t \to \infty$, we exploit Mills' ratio for the solution. Dropping the component index, and writing $z_t=z(tx)$, this gives
\begin{align*}
tx = -\log\left\{\frac{\phi(z_t)}{z_t}[1+O(z_t^{-2})]\right\} = \frac{z_t^2}{2}+\frac{1}{2}\log2\pi + \log z_t +O(z_t^{-2}),
\end{align*}
which is rearranged to give
\begin{align*}
z_t = (2tx)^{1/2} - (2tx)^{-1/2}\frac{\log 4\pi t x}{2} + O\left(\frac{(\log t)^2}{t^{3/2}}\right).
\end{align*}
We firstly deal with the Jacobian expression in density~\eqref{eq:GaussExp}. Again via Mills' ratio,
\begin{align*}
\prod_{j=1}^d \frac{1-\Phi(z_j(x_j))}{\phi(z_{j}(x_j))} &= \prod_{j=1}^d z_j(t x_j)^{-1}[1+O(z_j(t x_j)^{-2})]\\
&= 2^{-d/2}t^{-d/2}\prod_{j=1}^d x_j^{-1/2} [1+O(\log t / t)].
\end{align*}
Next consider the terms in the quadratic form of the exponent:
\begin{align*}
\omega_{jk}z_{j}(tx_j)z_{k}(tx_k) &=\omega_{jk} \left\{2t(x_jx_k)^{1/2}-\left(\frac{x_j}{x_k}\right)^{1/2}\frac{\log 4\pi t x_k}{2}-\left(\frac{x_k}{x_j}\right)^{1/2}\frac{\log 4\pi t x_j}{2}\right\} + O\left(\frac{(\log t)^2}{t}\right) \\
&= \omega_{jk} \left\{2t(x_jx_k)^{1/2}-\left[\left(\frac{x_j}{x_k}\right)^{1/2}+\left(\frac{x_k}{x_j}\right)^{1/2}\right]\frac{\log t }{2}\right.\\&~~~~~~~~~~\left.-\left(\frac{x_j}{x_k}\right)^{1/2}\frac{\log 4\pi x_k}{2}-\left(\frac{x_k}{x_j}\right)^{1/2}\frac{\log 4\pi x_j}{2}\right\} + O\left(\frac{(\log t)^2}{t}\right). 
\end{align*}
Therefore
\begin{align*}
-\frac{1}{2}\sum_{j,k}\omega_{jk}z_{j}(tx_j)z_{k}(tx_k) = -t(\bm{x}^{1/2})^\top\Sigma^{-1}\bm{x}^{1/2} + (\bm{x}^{1/2})^\top\Sigma^{-1}\bm{x}^{-1/2}\frac{\log t}{2} + k(\bm{x}) + O\left(\frac{(\log t)^2}{t}\right),
\end{align*}
where $k(\bm{x})$ does not depend on $t$. Putting everything together, with $g(\bm{x}) = (\bm{x}^{1/2})^\top\Sigma^{-1}\bm{x}^{1/2}$,
\begin{align*}
f_{R|\bm{W}}(r|\bm{w}) &\propto r^{d/2-1+\frac{1}{2} (\bm{w}^{1/2})^\top\Sigma^{-1}\bm{w}^{-1/2}}[1+O(\log r/r)] \exp\left\{-rg(\bm{w})+O\left(\frac{(\log r)^2}{r}\right)\right\}\\
&= r^{d/2-1+\frac{1}{2} (\bm{w}^{1/2})^\top\Sigma^{-1}\bm{w}^{-1/2}} \exp\left\{-rg(\bm{w})\right\}\left[1+O\left(\frac{(\log r)^2}{r}\right)\right].
\end{align*}
We therefore observe that the conditional distribution of $R|\bm{W}=\bm{w}$ has the gamma form, but with shape parameter $\alpha(\bm{w}) = d/2 + (\bm{w}^{1/2})^\top\Sigma^{-1}\bm{w}^{-1/2} /2$, rather than $d$ as in all other examples calculated here. 

\paragraph{Multivariate $t_{\nu}$ distribution (positive dependence)}
The multivariate $t$ distribution with $\nu$ degrees of freedom exhibits both positive and negative dependence. After transformation to exponential marginals, the negative dependence is manifested in the limit set by inclusion of sections on the planes $\{x_j=0\}$, $j=1,\ldots,d$. We focus here on the shape of the limit sets for $\bm{x}>\bm{0}$ only, which captures the positive dependence in the tail.

The density with centred $t_{\nu}$ margins and dispersion matrix $\Sigma $ is
\begin{align*}
f_{\bm{Z}_{T}}(\bm{z}) \propto \left[1+\frac{\bm{z}^\top \Sigma^{-1} \bm{z}}{\nu}\right]^{-(\nu+d)/2} = \left[1+\frac{\sum_{j=1}^d\sum_{k=1}^d \omega_{jk} z_jz_k}{\nu}\right]^{-(\nu+d)/2},
\end{align*}
with $\Omega = (\omega_{jk})_{j,k}$ the inverse dispersion matrix. Transforming to exponential margins gives
\begin{align}
f_{\bm{X}}(\bm{x}) \propto \left[1+\frac{\sum_{j=1}^d\sum_{k=1}^d \omega_{jk} z_j(x_j)z_k(x_k)}{\nu}\right]^{-(\nu+d)/2} \prod_{j=1}^d \frac{1-F_{Z_T}(z_j(x_j))}{f_{Z_T}(z_j(x_j))}, \label{eq:tExp}
\end{align}
where $z_j(x_j)$ is the solution to $x_j=-\log(1-F_{Z_T}(z_j))$, and $f_{Z_T}, F_{Z_T}$ represent the marginal density and distribution function of the $t_\nu$ distribution. Again, we are interested in large values of $x_j$ and $z_j$: \citet{Soms76} gave an expansion for the ratio of the univariate survival function to density, from which we can deduce that
\begin{align*}
\frac{1-F_{Z_T}(z_j(t x_j))}{f_{Z_T}(z_j(t x_j))} = \frac{z_j(t x_j)}{\nu}+ O(z_j(t x_j)^{-1}).
\end{align*}
Dropping the component index and writing $z_t = z(tx)$, we have
\begin{align*}
tx = -\log\left\{f_{Z_T}(z_t)\left[\frac{z_t}{\nu}+O(z_t^{-1})\right]\right\} = c+\nu\log z_t +O(z_{t}^{-2}),
\end{align*}
where $c$ is a constant depending on $\nu$. For a new constant $c'$, this is rearranged to give
\begin{align*}
z_t = c' e^{tx/\nu}[1+O(e^{-2tx/\nu})].
\end{align*}
To find the asymptotic behavior of~\eqref{eq:tExp} we firstly consider the Jacobian term:
\begin{align*}
\prod_{j=1}^d \frac{1-F_{Z_T}(z_j(tx_j))}{f_{Z_T}(z_j(tx_j))} &= \prod_{j=1}^d \frac{z_j(t x_j)}{\nu}[1+ O(z_j(t x_j)^{-2})] \propto e^{t\sum_{j=1}^d x_j / \nu} [1+O(e^{-2 t x_{(d)}/\nu})].
\end{align*}
Considering the kernel, we have
\begin{align*}
\left[1+\sum_{j=1}^d\sum_{k=1}^d \frac{\omega_{jk}}{\nu} z_j(tx_j)z_k(tx_j)\right]^{-\frac{\nu+d}{2}} & = \left[1+\sum_{j=1}^d\sum_{k=1}^d \frac{\omega_{jk}c'^2}{\nu} e^{t(x_j+x_k)/\nu} + O(e^{t(x_{(1)}-x_{(d)})/\nu})\right]^{-\frac{\nu+d}{2}}\\& \propto e^{-\frac{2t x_{(1)}}{\nu}\frac{\nu+d}{2}}\left[1+O(e^{t(x_{(2)}-x_{(1)})/\nu})\right] \\&= e^{-t x_{(1)}(1+d/\nu)}\left[1+O(e^{t(x_{(2)}-x_{(1)})/\nu})\right].
\end{align*}
Combining both expressions,
\begin{align*}
f_{\bm{X}}(t\bm{x}) \propto  e^{-t [(1+d/\nu)x_{(1)}-\sum_{j=1}^d x_j/\nu]}\left[1+O(e^{t(x_{(2)}-x_{(1)})/\nu}) +O(e^{-2 t x_{(d)}/\nu})\right].
\end{align*}
Therefore $g(\bm{x}) = (1+d/\nu)x_{(1)}-\sum_{j=1}^d x_j/\nu$ and 
\begin{align*}
f_{R|\bm{W}}(r|\bm{w}) \propto r^{d-1} e^{-rg(\bm{w})}\left[1+O(e^{r(w_{(2)}-w_{(1)})/\nu}) +O(e^{-2 r w_{(d)}/\nu})\right].
\end{align*}

\paragraph{Clayton and inverted Clayton copulas}
The Clayton copula with parameter $\gamma>0$ has distribution function in uniform margins
\begin{align*}
F_{\bm{U}}(\bm{u}) = \left(\sum_{j=1}^d u_j^{-\gamma}-d+1\right)^{-1/\gamma}.
\end{align*}
The corresponding density is
\begin{align*}
f_{\bm{U}}(\bm{u}) \propto \left(\prod_{j=1}^d u_j\right)^{-\gamma-1}\left(\sum_{j=1}^d u_j^{-\gamma} -d+1\right)^{-1/\gamma-d}.
\end{align*}
The density in exponential margins is $f_{\bm{X}}(\bm{x}) = e^{-\sum_{j=1}^d x_j} f_{\bm{U}}(1-e^{-\bm{x}})$, and so
\begin{align*}
f_{\bm{X}}(t\bm{x}) &\propto e^{-\sum_{j=1}^d t x_j} \left(\prod_{j=1}^d [1-e^{-tx_j}]\right)^{-(\gamma+1)} \left(\sum_{j=1}^d (1-e^{-x_j})^{-\gamma} -d+1\right)^{-1/\gamma-d}\\
&\propto e^{-\sum_{j=1}^d t x_j} \left[1+O(e^{-t x_{(d)}})\right].
\end{align*}
The gauge function is therefore $g(\bm{x}) = \sum_{j=1}^d x_j$, and
\begin{align*}
f_{R|\bm{W}}(r|\bm{w}) \propto r^{d-1} e^{-rg(\bm{w})}[1+O(e^{-r w_{(d)}})].
\end{align*}
If the random vector $\bm{U}$ follows a Clayton copula with uniform margins, then the random vector $1-\bm{U}$ follows an inverted Clayton copula with uniform margins. Its density is $f_{\bm{U}}(1-\bm{u})$. In exponential margins
\begin{align*}
f_{\bm{X}}(t\bm{x}) &\propto e^{t (\gamma+1)\sum_{j=1}^d x_j - t \sum_{j=1}^d x_j} \left(\sum_{j=1}^d e^{tx_j/\gamma} - d + 1\right)^{-1/\gamma-d}\\
&=  e^{t \gamma\sum_{j=1}^d x_j -t(1+d\gamma) x_{(1)}} \left(\sum_{j=1}^d e^{t(x_j-x_{(1)})/\gamma} + (1-d)e^{-t x_{(1)}/\gamma}\right)^{-1/\gamma -d}\\
&= e^{t \gamma\sum_{j=1}^d x_j -t(1+d\gamma) x_{(1)}} \left(1+O(e^{t(x_{(2)}-x_{(1)})/\gamma})\right).
\end{align*}
The gauge function is therefore $g(\bm{x}) = (1+d\gamma)x_{(1)}-\sum_{j=1}^d x_j \gamma$, and
\begin{align*}
f_{R|\bm{W}}(r|\bm{w}) \propto r^{d-1} e^{-rg(\bm{w})}[1+O(e^{-r(w_{(2)}-w_{(1)})})].
\end{align*}

\paragraph{Vine copula from \citet{NoldeWadsworth21}}
\citet{NoldeWadsworth21} give an example of a gauge function derived from a particular vine copula construction. Vine copulas are specified by pairs of bivariate copulas: in this case we take the two base pairs to be independence (between $(X_1,X_2)$) and inverted Clayton with parameter $\beta>0$ (between $(X_2,X_3)$), and use the inverted Clayton with parameter $\gamma>0$ to model the dependence between $(X_3|X_2,X_1|X_2)$.

Let $c_{1,2}, c_{2,3}$ and $c_{1|2,1|3}$ denote the densities of the respective copulas in standard uniform margins. The joint density in exponential margins is
\begin{align*}
f_{\bm{X}}(\bm{x}) &= e^{-(x_1+x_2+x_3)}c_{1,2}\left(1-e^{-x_1},1-e^{-x_2}\right)c_{2,3}\left(1-e^{-x_2},1-e^{-x_3}\right)c_{1|2,3|2}\left(F_{1|2}(x_1|x_2),F_{3|2}(x_3|x_2)\right),
\end{align*}
where $F_{1|2}(x_1|x_2) = \Pr(X_1 \leq x_1 |X_2 = x_2) = 1-e^{-x_1}$, and 
\begin{align*}
F_{X_3|X_2}(x_3|x_2) &= \Pr\left(X_3\leq x_3 | X_2=x_2\right) = 1- e^{(\beta+1) x_2}\left[e^{\beta x_2} + e^{\beta x_3} - 1\right]^{-\left(\frac{1}{\beta}+1\right)},
\end{align*}
so that
\begin{align*}
F_{X_3|X_2}(t x_3|t x_2) &= 1-e^{-t(\beta+1)[\max(x_2,x_3)-x_2]}[1+O(e^{t\beta[\min(x_2,x_3)-\max(x_2,x_3)]})].
\end{align*}
For the copula densities, $c_{1,2}=1$, while
\begin{align*}
c_{2,3}(u_2,u_3) = \left(1+\beta\right)(1-u_2)^{-(\beta+1)}(1-u_3)^{-(\beta+1)}\left[(1-u_2)^{-\beta}+(1-u_3)^{-\beta}-1\right]^{-\left(\frac{1}{\beta}+2\right)},
\end{align*}
and $c_{1|2,3|2}$ is of the same form but with parameter $\gamma$. We have
\begin{align*}
c_{2,3}\left(1-e^{-tx_2},1-e^{-tx_3}\right) &\propto e^{(\beta+1)tx_2}e^{(\beta+1)tx_3}\left[e^{\beta tx_2}+e^{\beta tx_3}-1\right]^{-\left(\frac{1}{\beta}+2\right)}\\& = e^{-\beta t \max(x_2,x_3)+(\beta+1)t\min(x_2,x_3)}[1+O(e^{\beta t (\min(x_2,x_3)-\max(x_2,x_3))})],
\end{align*}
and
\begin{align*}
c_{1|2,3|2}&(1-e^{-tx_1}, 1-e^{-t(\beta+1)[\max(x_2,x_3)-x_2]}[1+O(e^{t\beta[\min(x_2,x_3)-\max(x_2,x_3)]})]) \\
&\propto e^{(\gamma+1)tx_1+(\gamma+1)(\beta+1)t[\max(x_2,x_3)-x_2]}[1+O(e^{t\beta[\min(x_2,x_3)-\max(x_2,x_3)]})]\\
&~~~\times\left\{e^{\gamma t x_1} + e^{\gamma(\beta+1)t[\max(x_2,x_3)-x_2]}[1+O(e^{t\beta[\min(x_2,x_3)-\max(x_2,x_3)]})] -1\right\}^{-\left(\frac{1}{\gamma}+2\right)}\\
& = e^{(\gamma+1)tx_1+(\gamma+1)(\beta+1)t[\max(x_2,x_3)-x_2]-(2\gamma+1)t\max(x_1,(\beta+1)[\max(x_2,x_3)-x_2])}\\
&~~~\times [1+O(e^{t\beta[\min(x_2,x_3)-\max(x_2,x_3)]}) + O(e^{t\gamma\{\min(x_1,(\beta+1)[\max(x_2,x_3)-x_2])-\max(x_1,(\beta+1)[\max(x_2,x_3)-x_2])\}})]
\end{align*}

Combining all components,
\begin{align*}
f_{\bm{X}}(t\bm{x}) &\propto e^{-t[x_1+x_2+x_3 +\beta \max(x_2,x_3)- (\beta+1)\min(x_2,x_3)-(\gamma+1)\{x_1+(\beta+1)[\max(x_2,x_3)-x_2]\}+(2\gamma+1)\max(x_1,(\beta+1)[\max(x_2,x_3)-x_2])]}\\
&~~~\times [1+O(e^{t\beta[\min(x_2,x_3)-\max(x_2,x_3)]}) + O(e^{t\gamma\{\min(x_1,(\beta+1)[\max(x_2,x_3)-x_2])-\max(x_1,(\beta+1)[\max(x_2,x_3)-x_2])\}})].
\end{align*}
Simplifying the gauge function we get
\begin{align*}
g(\bm{x}) =& (1+\beta)\max(x_2,x_3) -\beta \min(x_2,x_3) - \gamma x_1 -(\gamma+1)(\beta+1)\left(\max(x_2,x_3)-x_2\right)\\
&+(2\gamma+1)\max(x_1,(\beta+1)(\max(x_2,x_3)-x_2)),
\end{align*}
and
\begin{align*}
f_{R|\bm{W}}(r|\bm{w}) &\propto r^{d-1}e^{-rg(\bm{w})}\\ &~~~\times [1+O(e^{r\beta[\min(w_2,w_3)-\max(w_2,w_3)]})+\\ &~~~~~~~~O(e^{r\gamma\{\min(w_1,(\beta+1)[\max(w_2,w_3)-w_2])-\max(w_1,(\beta+1)[\max(w_2,w_3)-w_2])\}})].
\end{align*}

\newpage
\section{Multivariate Gaussian case}
\label{sec:gauss}
\begin{figure}%[h!]
	\centering
	\includegraphics[width=0.25\textwidth]{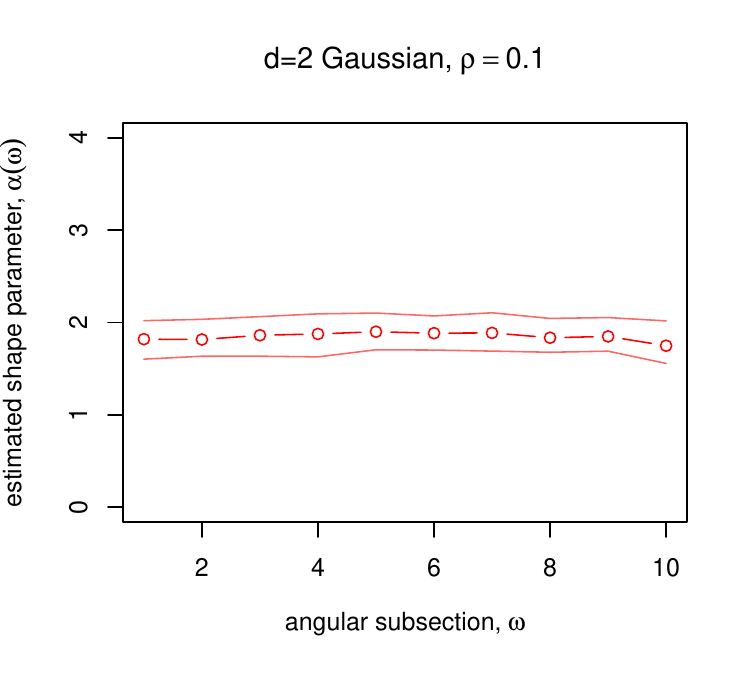}
	\includegraphics[width=0.25\textwidth]{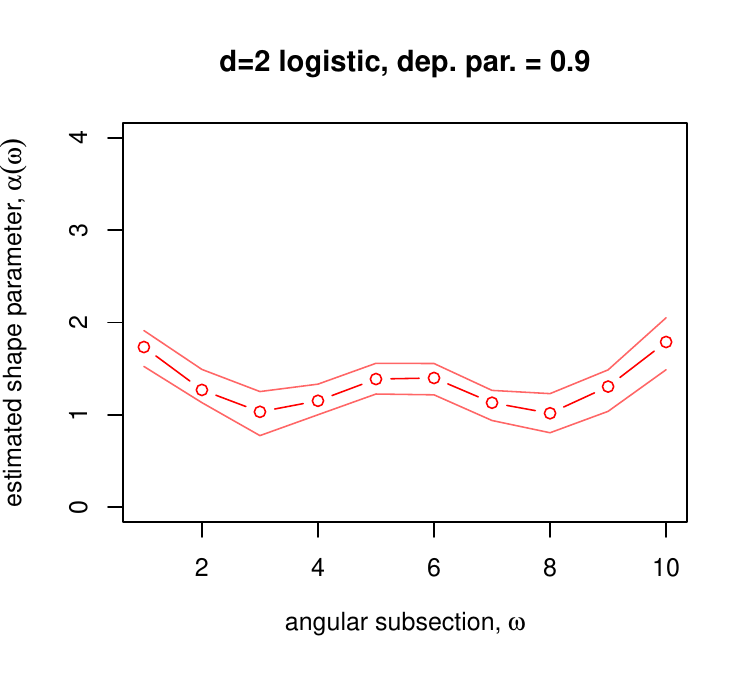}
	\includegraphics[width=0.25\textwidth]{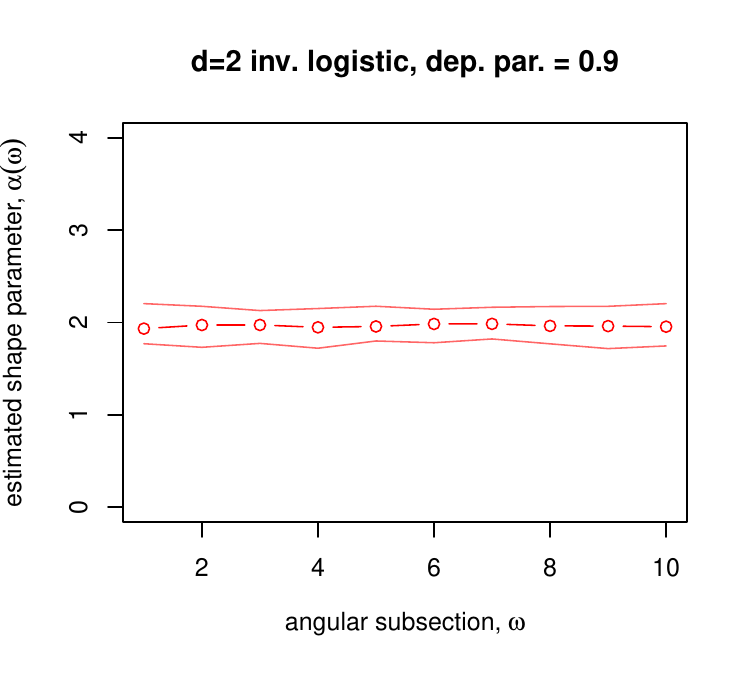}
	\includegraphics[width=0.25\textwidth]{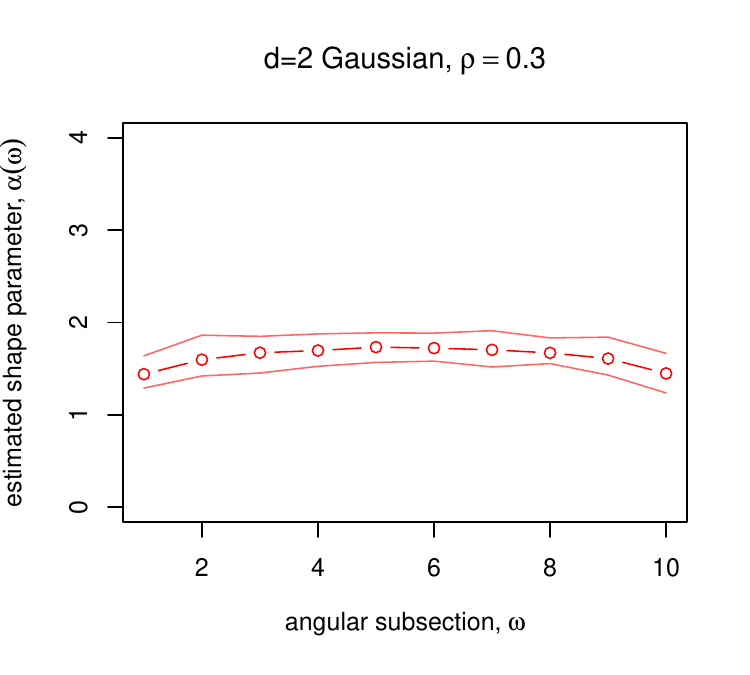}
	\includegraphics[width=0.25\textwidth]{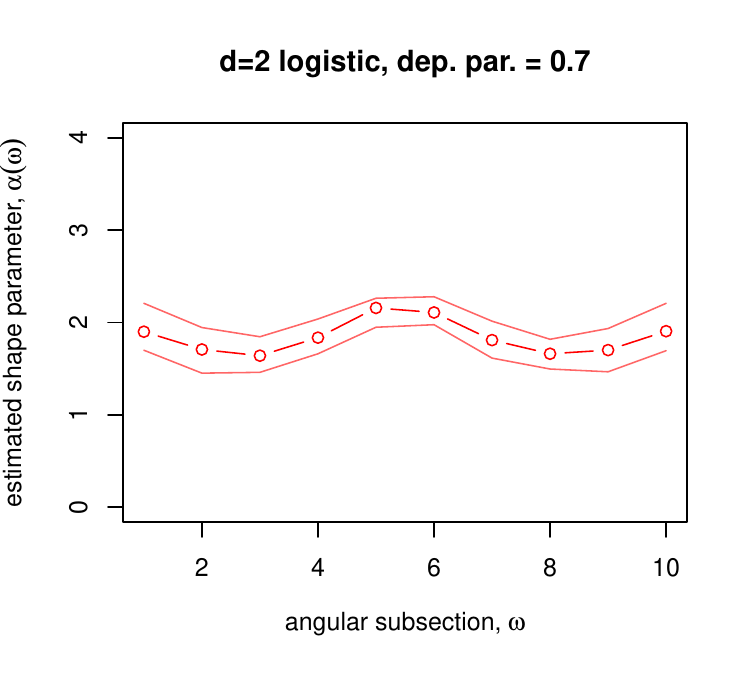}
	\includegraphics[width=0.25\textwidth]{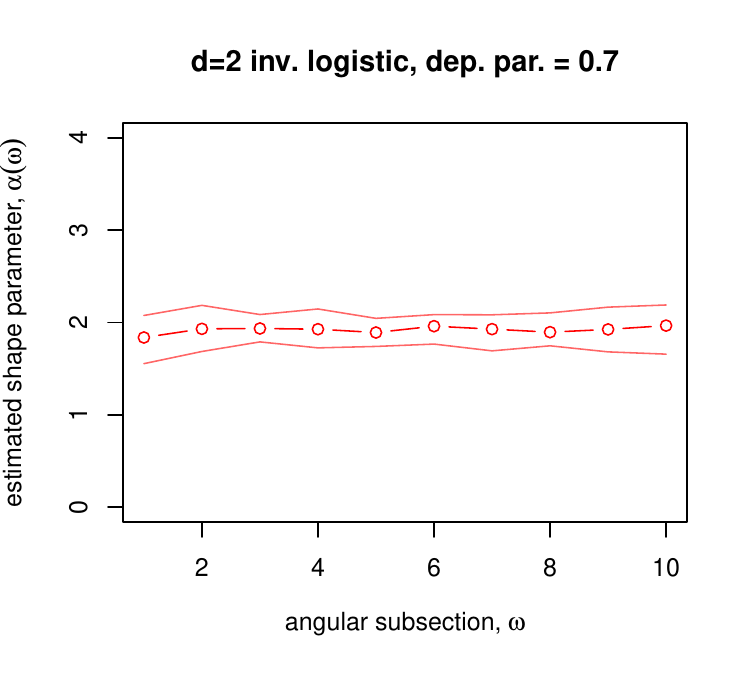}
	\includegraphics[width=0.25\textwidth]{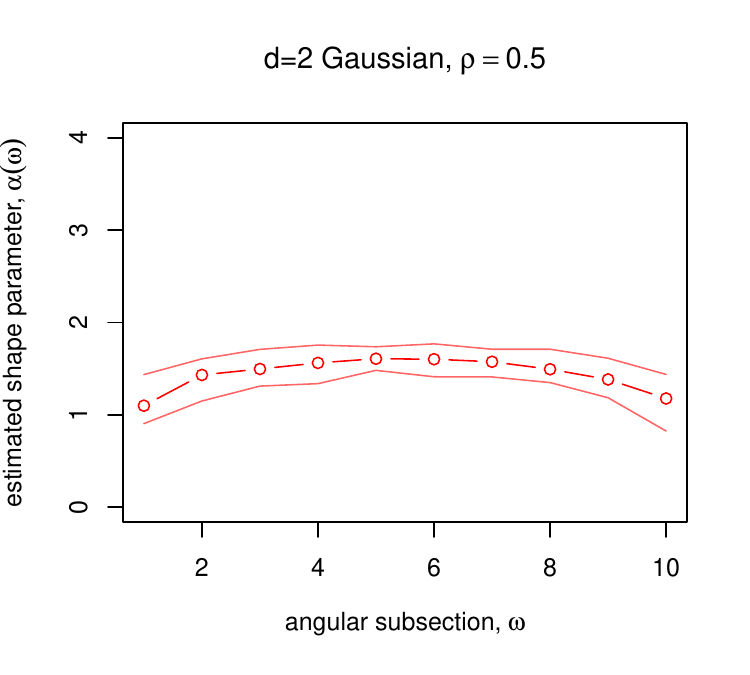}
	\includegraphics[width=0.25\textwidth]{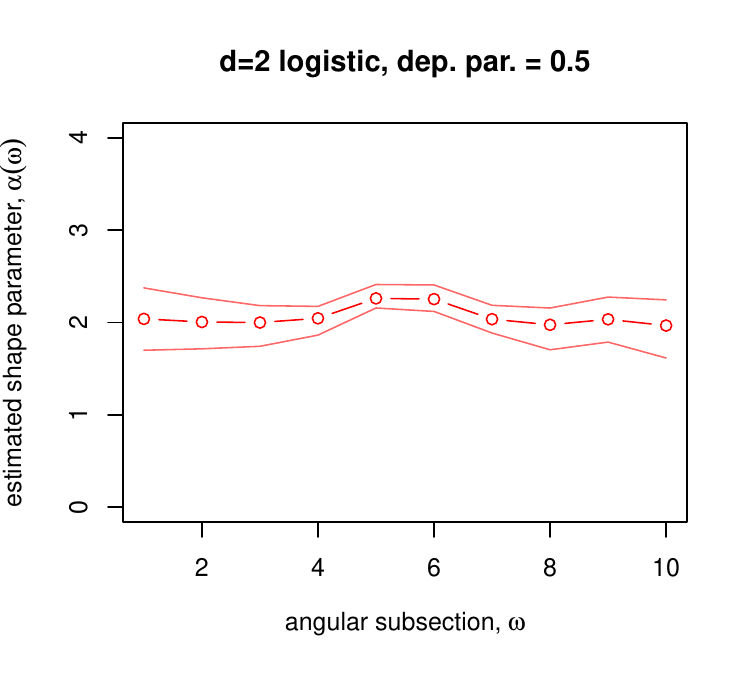}
	\includegraphics[width=0.25\textwidth]{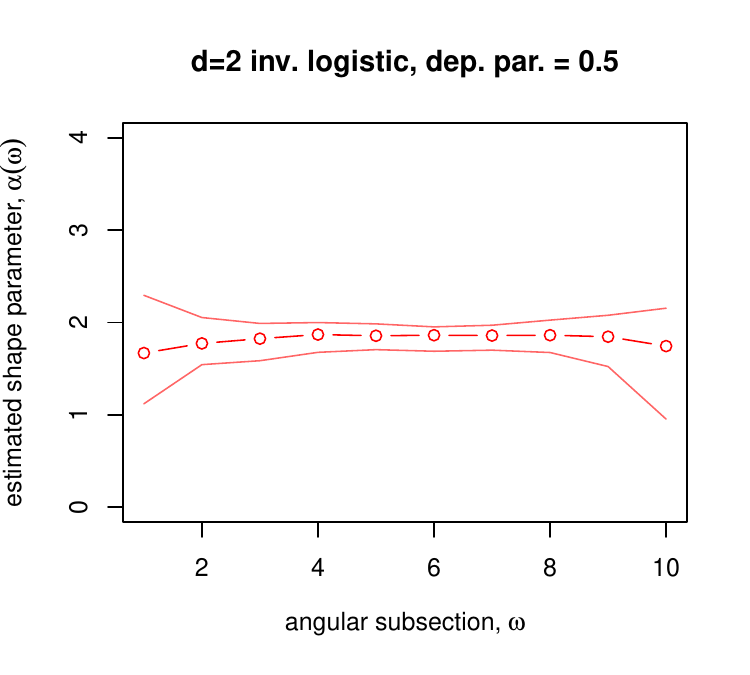}
	\includegraphics[width=0.25\textwidth]{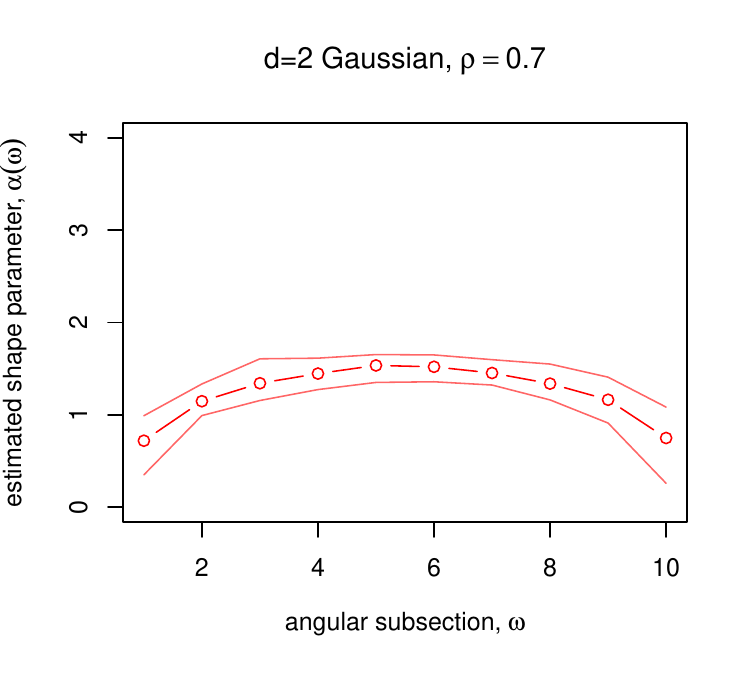}
	\includegraphics[width=0.25\textwidth]{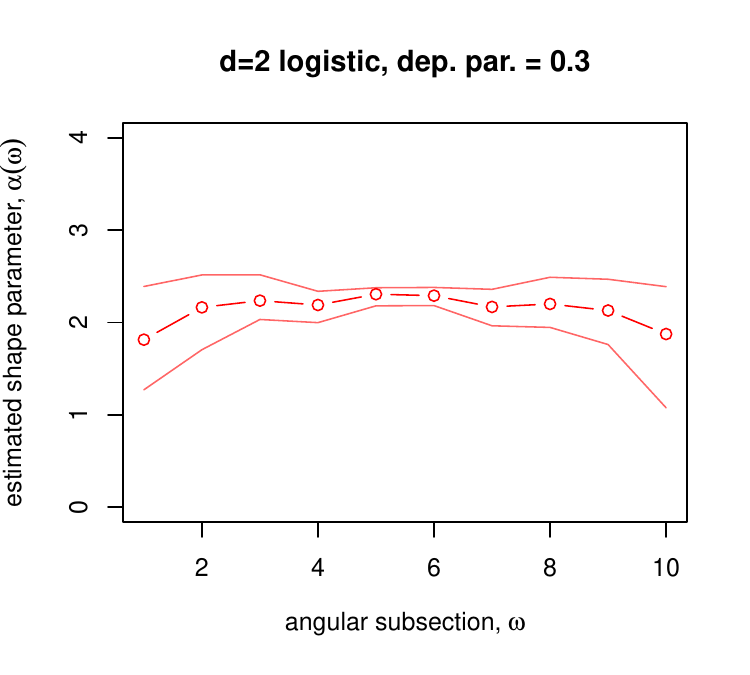}
	\includegraphics[width=0.25\textwidth]{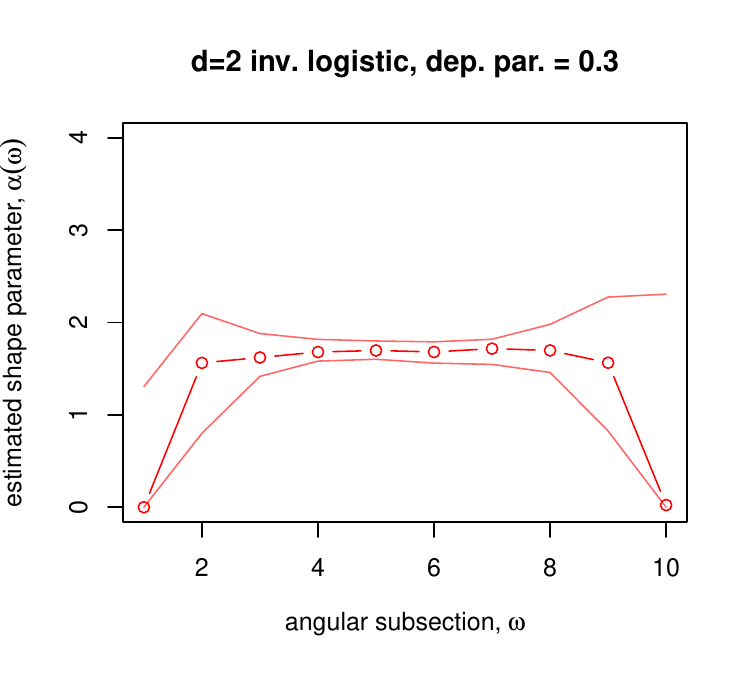}
	\includegraphics[width=0.25\textwidth]{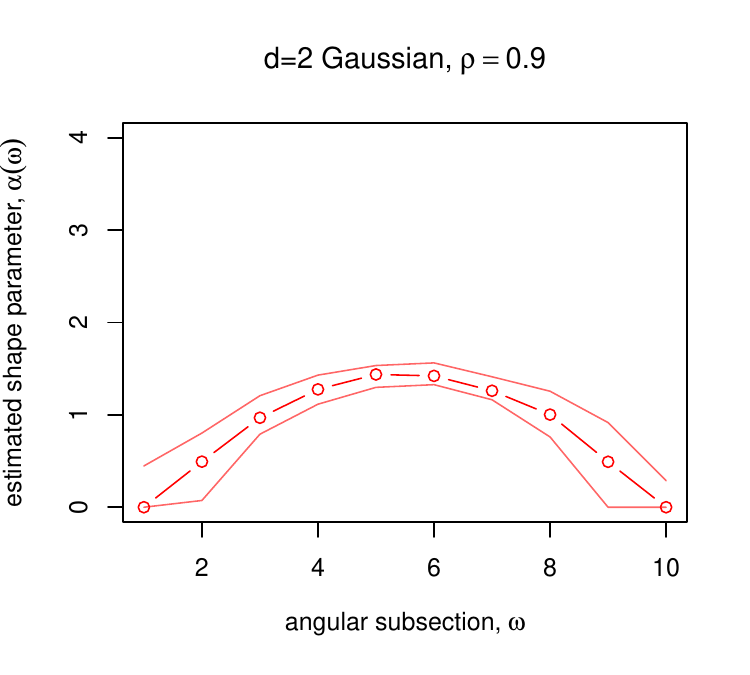}
	\includegraphics[width=0.25\textwidth]{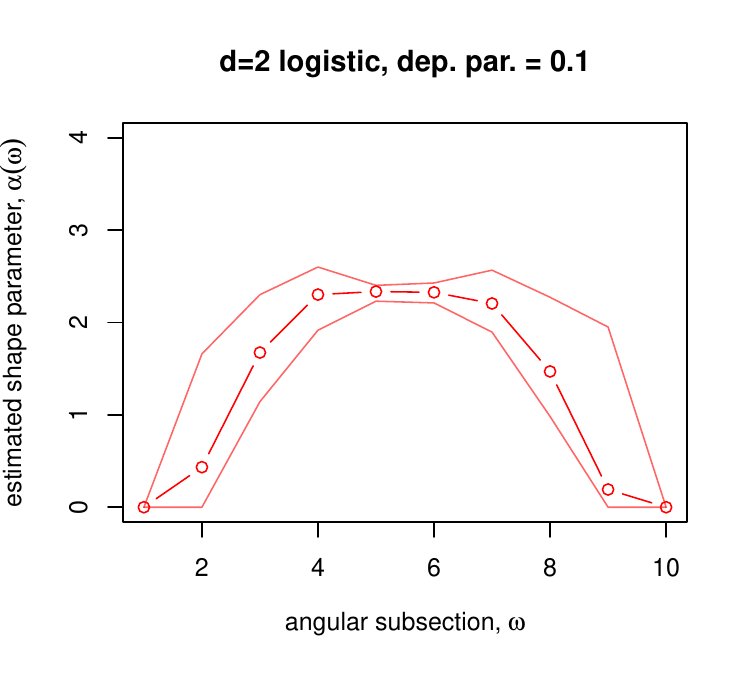}
	\includegraphics[width=0.25\textwidth]{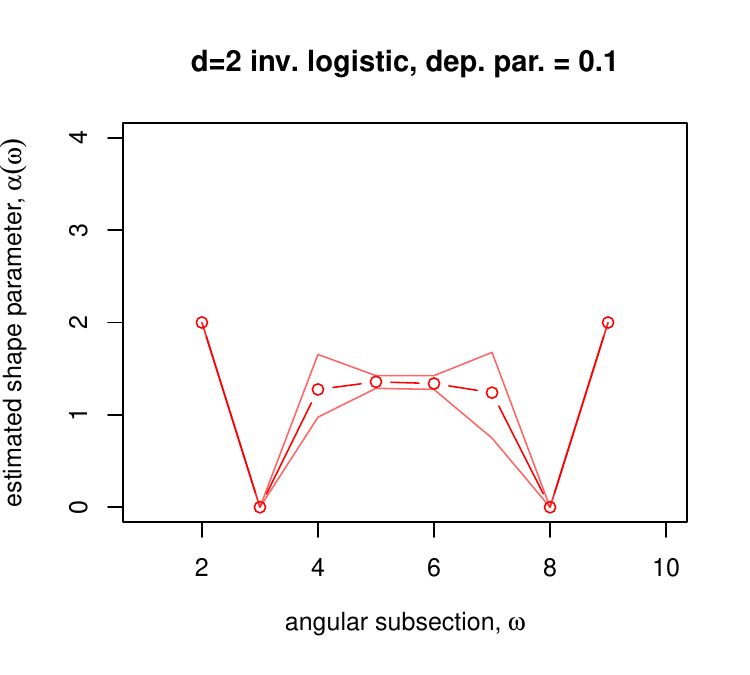}
	%	\raggedright
	\caption{
		\small{
			Shape parameter estimates across non-overlapping sections of the simplex $\mathcal{S}_{1}$ in the $d=2$ case. Gaussian data and gauge function are used for column 1, logistic in column 2, inverted logistic in column 3. From top to bottom, dependence parameters are such that joint dependence is increasing. Dots represent \edit{median} point estimates, and outer solid lines approximate 95\% pointwise confidence intervals.
		}
	}
	\label{fig:2d-shape-param-ests}
\end{figure}

\begin{figure}%[t!]
	\centering
	\includegraphics[width=0.25\textwidth]{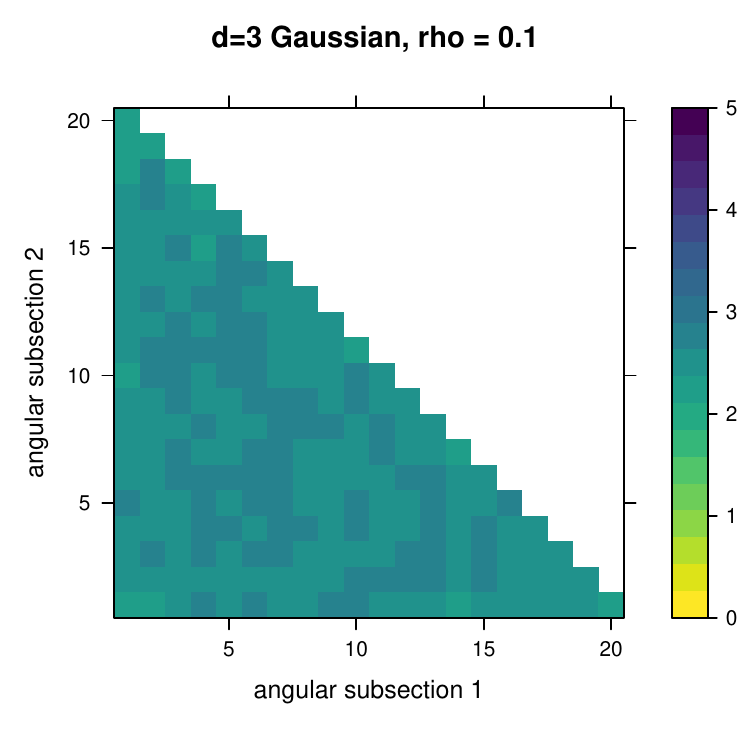}
	\includegraphics[width=0.25\textwidth]{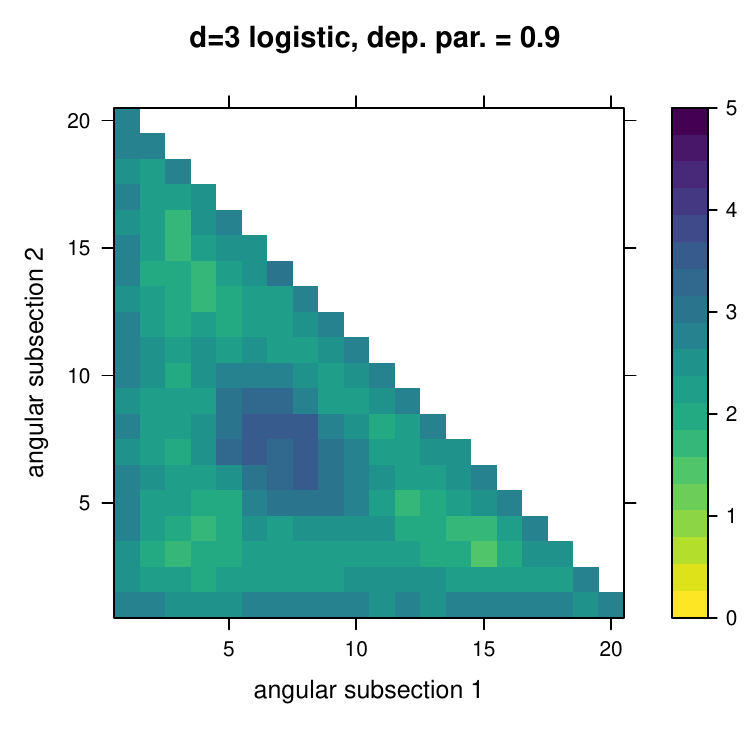}
	\includegraphics[width=0.25\textwidth]{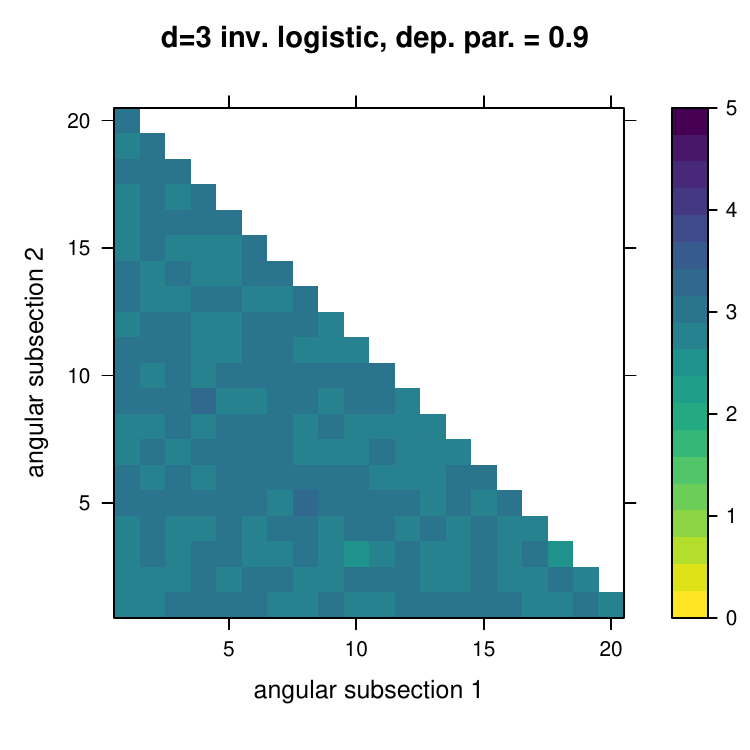}
	\includegraphics[width=0.25\textwidth]{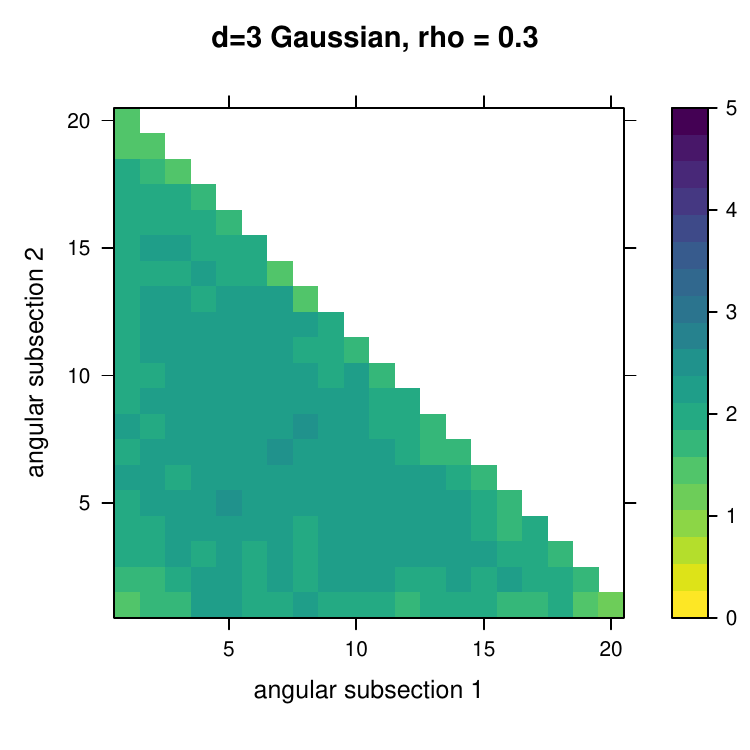}
	\includegraphics[width=0.25\textwidth]{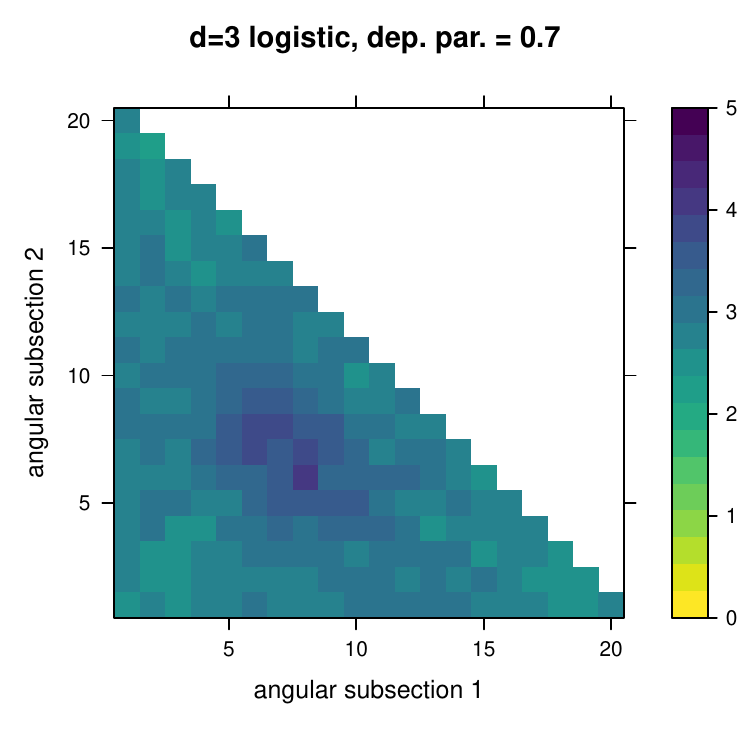}
	\includegraphics[width=0.25\textwidth]{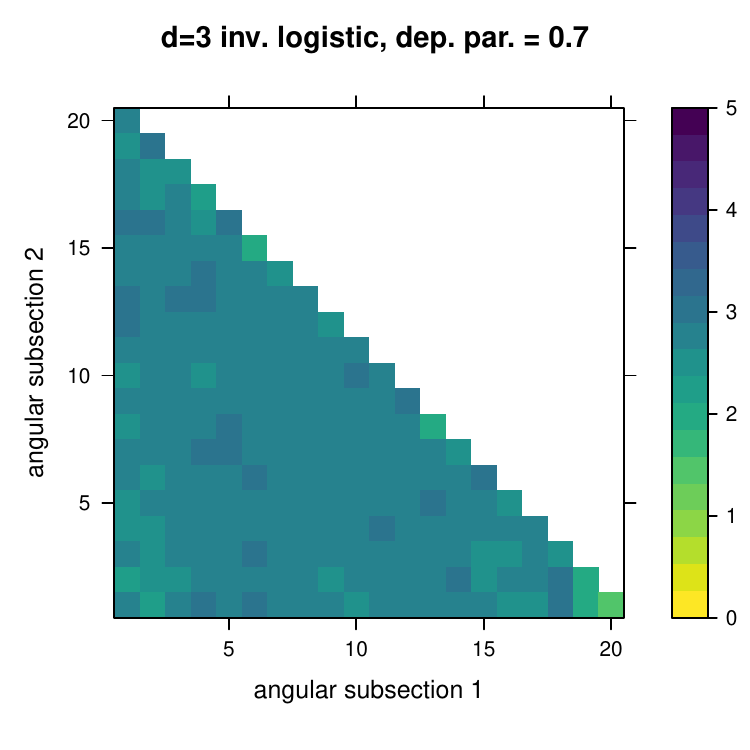}
	\includegraphics[width=0.25\textwidth]{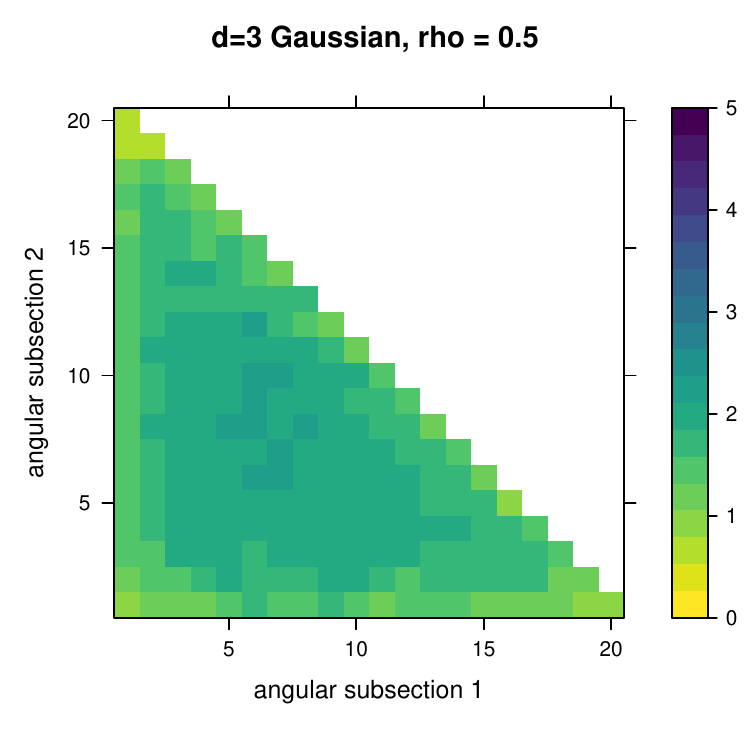}
	\includegraphics[width=0.25\textwidth]{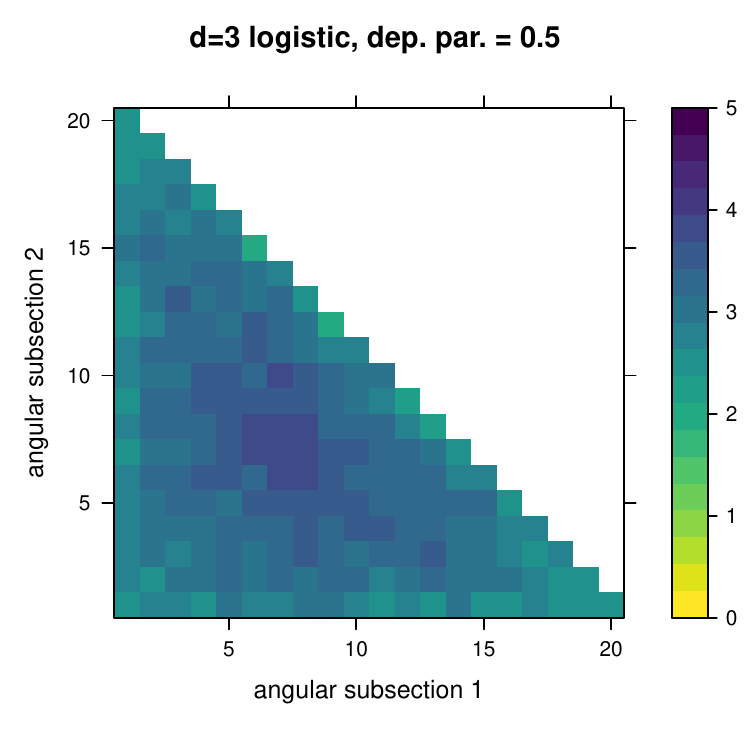}
	\includegraphics[width=0.25\textwidth]{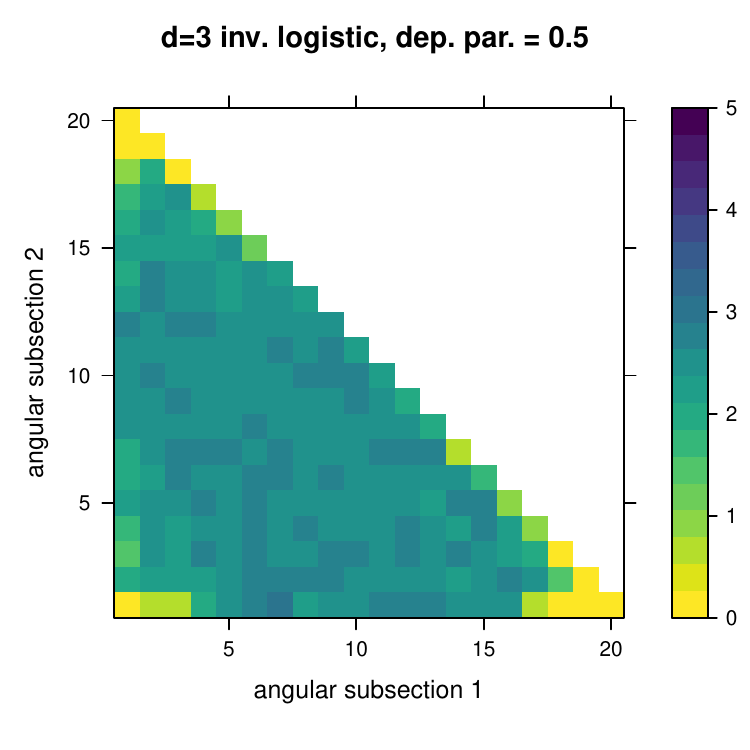}
	\includegraphics[width=0.25\textwidth]{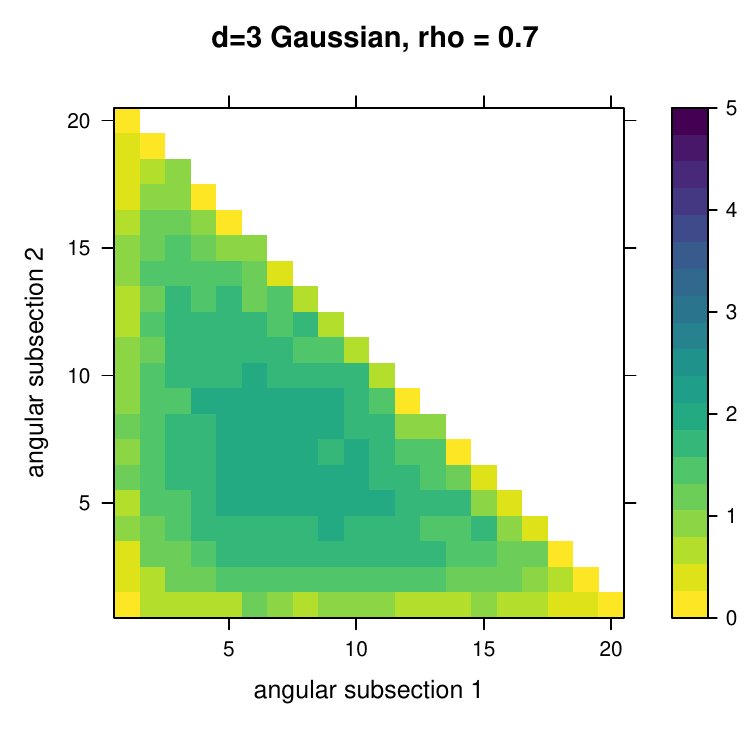}
	\includegraphics[width=0.25\textwidth]{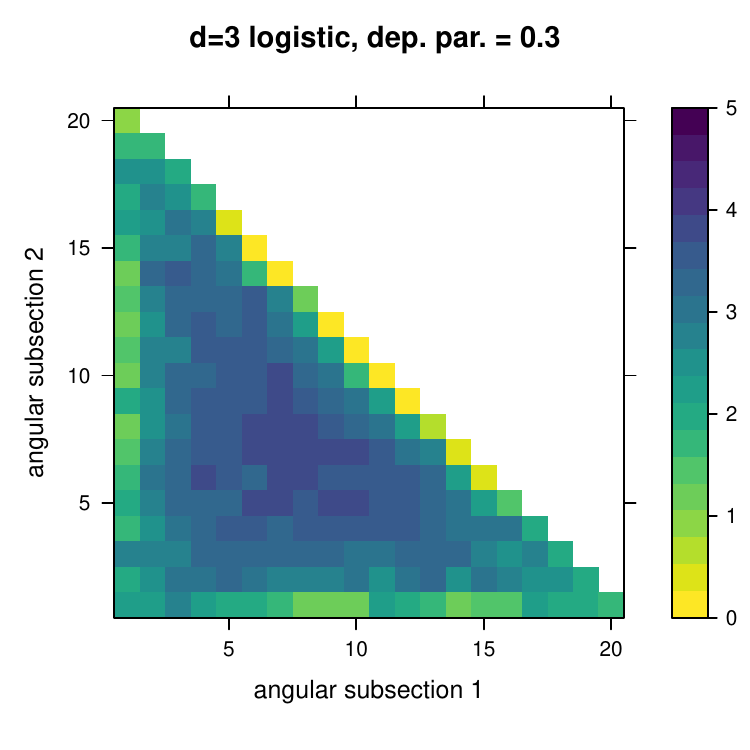}
	\includegraphics[width=0.25\textwidth]{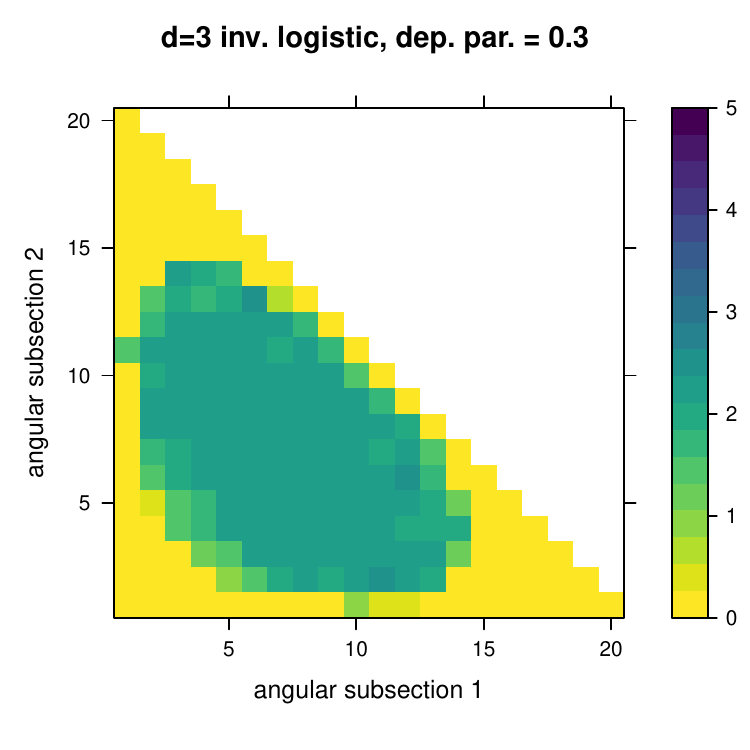}
	\includegraphics[width=0.25\textwidth]{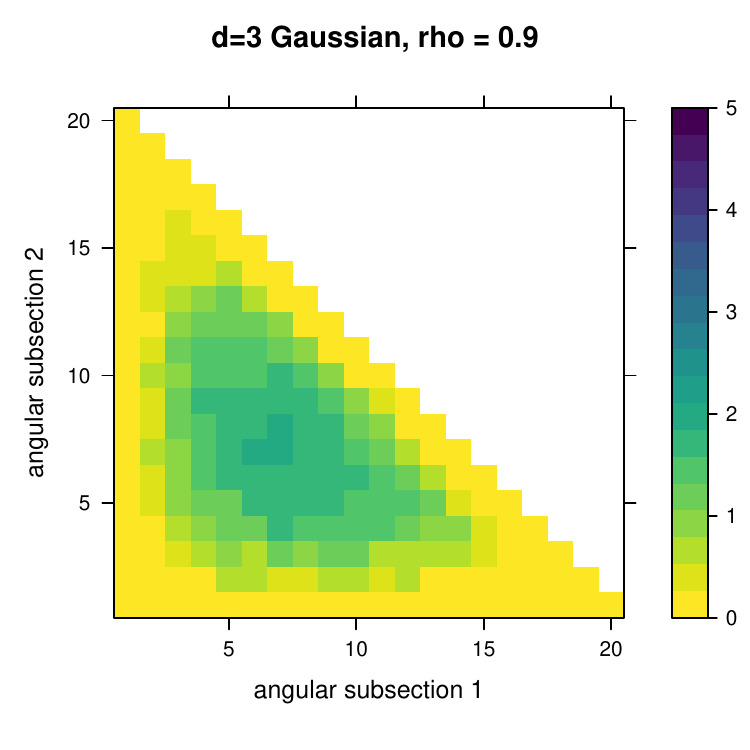}
	\includegraphics[width=0.25\textwidth]{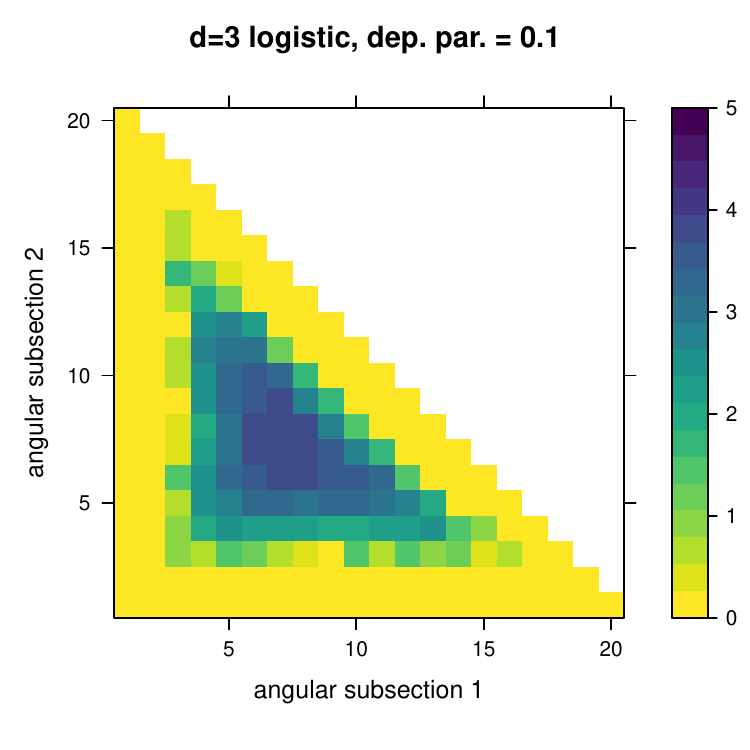}
	\includegraphics[width=0.25\textwidth]{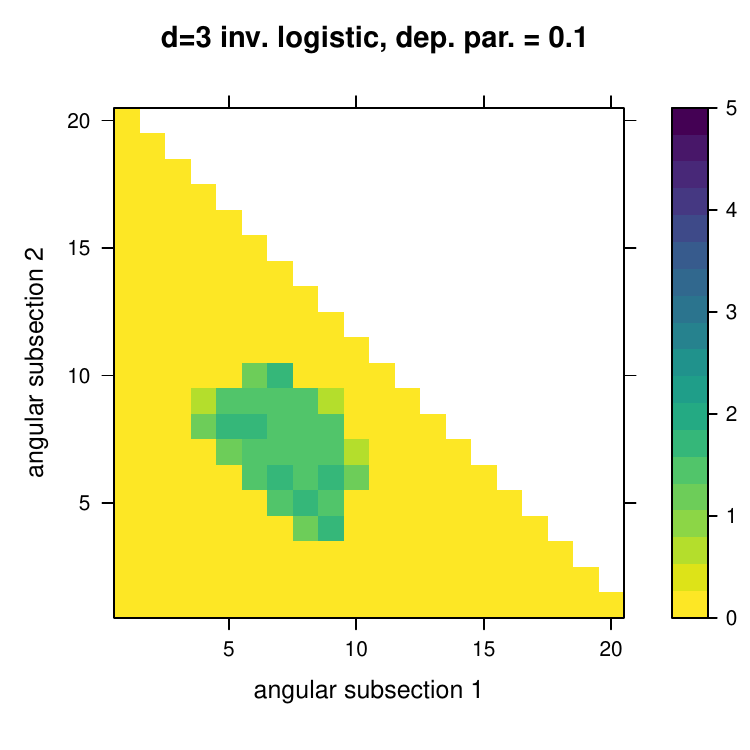}
	%	\raggedright
	\caption{
		%\small{
		\edit{Median} shape parameter estimates across non-overlapping sections of the simplex $\mathcal{S}_{2}$ in the $d=3$ case. Gaussian data and gauge function are used for column 1, logistic in column 2, inverted logistic in column 3. From top to bottom, the dependence parameters are such that joint dependence is increasing. \edit{Pixels in yellow represent estimates $\widehat{\alpha}=0$.}
		%}	
	}
	\label{fig:3d-shape-param-ests}
\end{figure}

In this section we consider the conditional distribution $R|\bm{W}=\bm{w}$ for the multivariate Gaussian dependence structure in more detail. In Section~\ref{main-sec:Examples} of the main paper, and Section~\ref{sec:rwderivations} of the supplement, the asymptotic form of this distribution is shown to have a gamma form with shape parameter given by a function of the angle $\bm{w}$,
\begin{equation*}\label{eq:gaussian-shape}
\alpha(\bm{w}) = \frac{d}{2} + \frac{1}{2}\left(\bm{w}^{{1}/{2}}\right)^\top\Sigma^{-1}\bm{w}^{-{1}/{2}}.
\end{equation*}
There are two concerns with this shape parameter: (i) whether its variation with $\bm{w}$ indicates the need for a more complex model than that in equation~(5), where the shape is assumed constant, and (ii) the fact that $\alpha(\bm{w}) \leq 0$ for some values of $\bm{w}$. We investigate these issues in turn, using a single correlation across all pairwise variables to define our covariance matrix,
\begin{equation*}\label{eq:cov-mat}
\Sigma_{ij} = \begin{cases}1&;\;\;i=j\\
\rho&;\;\;i\neq j\end{cases}, \qquad i,j=1,\dots,d.
\end{equation*}
Figures~\ref{fig:2d-shape-param-ests} and~\ref{fig:3d-shape-param-ests} display local estimates of the shape parameter $\alpha$ under the truncated gamma model~(5) for $R|\{\bm{W}=\bm{w},R>r_0(\bm{w})\}$ for $d=2,3$, respectively. In each case the angles $\bm{w}$ are restricted to a small section of the simplex, and the rate parameter is fixed at the ``true'' value $g(\bm{w})$. For comparison, we also perform the same procedure for the logistic and inverted logistic dependence structures, over a variety of dependence strengths. \edit{For each distribution the total sample size, over all angular subsections, is 500,000. In Figure~\ref{fig:2d-shape-param-ests} we plot the median estimates and pointwise 95\% CIs based on 100 repetitions. In Figure~\ref{fig:3d-shape-param-ests} we plot the median estimates based on 100 repetitions.}

For the Gaussian case, we observe that estimates $\widehat{\alpha}$ remain relatively constant on the simplex $\mathcal{S}_{d-1}$ in practice for $d=2,3$. In particular, the shape parameter estimates \edit{generally} vary no more with the angle $\bm{w}$ when compared to the logistic and inverted logistic setting. The evidence suggests that for many distributions the shape parameter will vary with the angle $\bm{w}$ in practice to some degree. This is likely due to the fact that the rate of convergence of $R|\bm{W}=\bm{w}$ to the gamma form can depend on $\bm{w}$ either explicitly (as in the logistic case), or practically through a constant term (as in the inverted logistic case). When the dependence is strong for the inverted logistic distribution there are large parts of the simplex where there is insufficient data to estimate the local model.

%	\newpage
%	\newpage
%	\clearpage

\begin{figure}%[t!]
	\centering
	\includegraphics[width=0.4\textwidth]{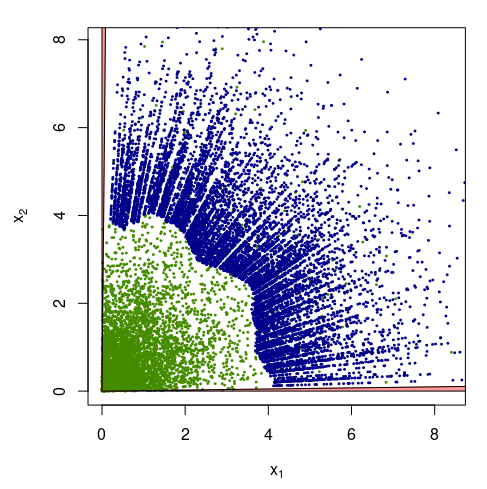}
	\includegraphics[width=0.4\textwidth]{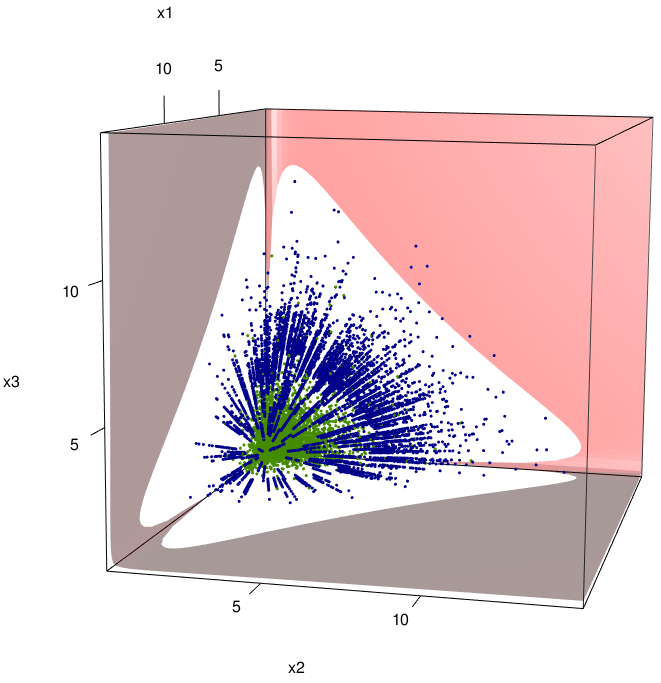}
	\includegraphics[width=0.4\textwidth]{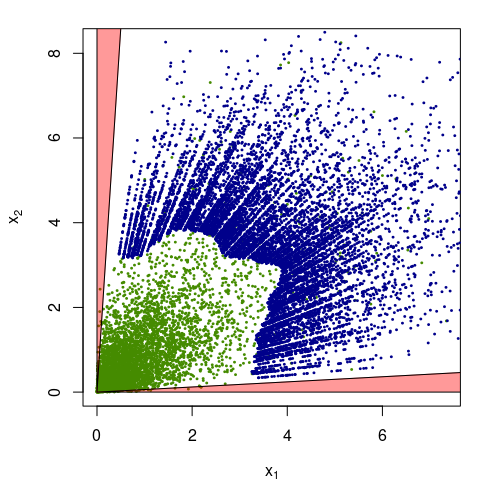}
	\includegraphics[width=0.4\textwidth]{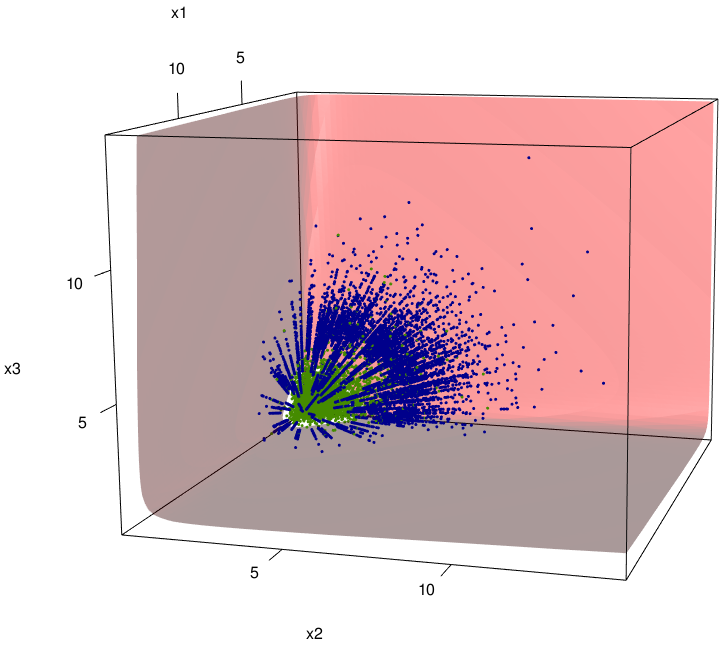}
	\includegraphics[width=0.4\textwidth]{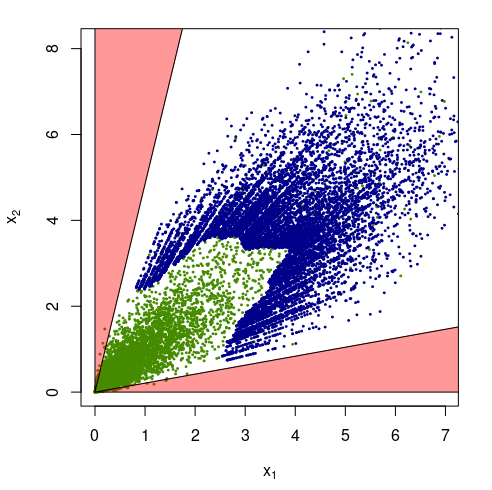}
	\includegraphics[width=0.4\textwidth]{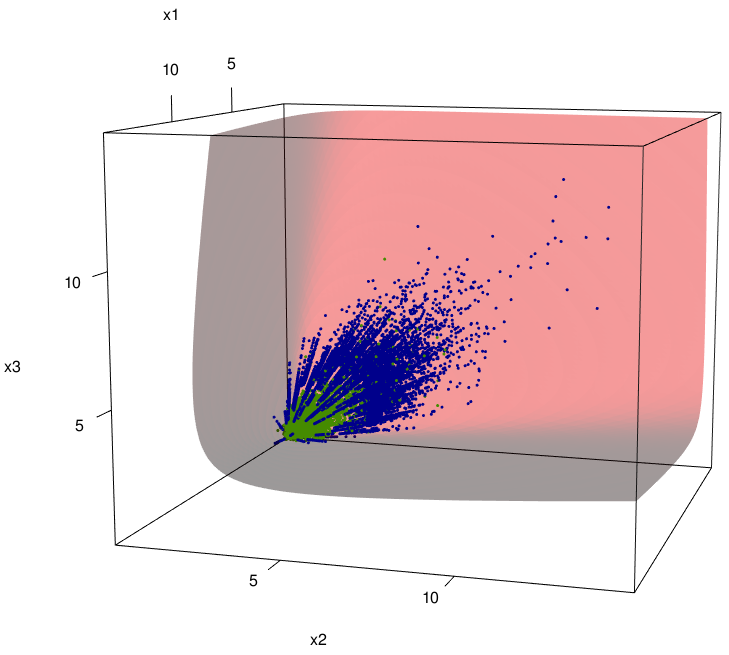}
	\caption{The region $\left\{(r,\bm{w})\in (0,\infty)\times\mathcal{S}_{d-1} : \alpha(\bm{w})<0\right\}$ for correlation $\rho=0.4$, 0.7, and 0.9 with $d=2$ on the left (red region) and $d=3$ on the right (volume between coordinate planes and red surface). Interactive versions of the plots for $d=3$ can be found at \url{www.lancaster.ac.uk/~wadswojl/geometricMVE.html}.}
	\label{fig:2d-3d-problem-region}
\end{figure}

\begin{figure}%[b!]
	\centering
	\includegraphics[width=0.45\textwidth]{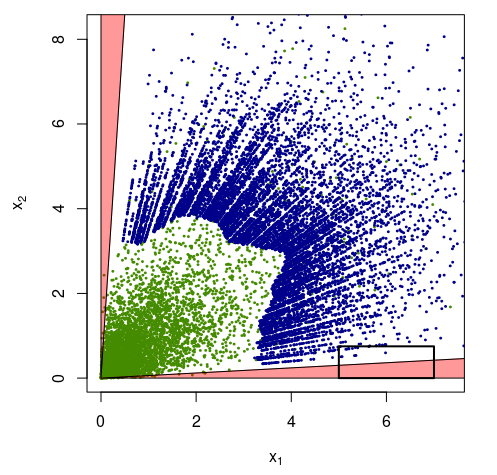}
	\includegraphics[width=0.45\textwidth]{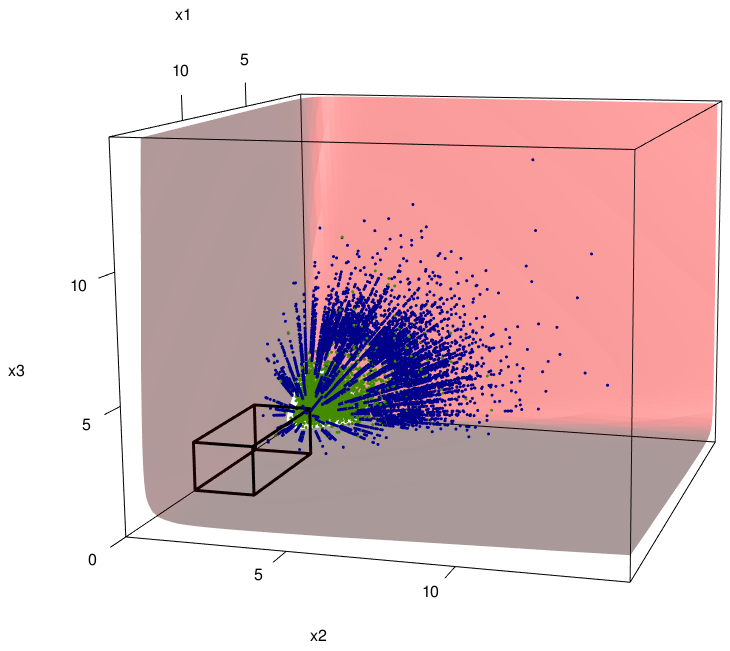}
	\caption{Rectangular regions with black borders are the extremal sets that we estimate the probability of lying in. Left: $d=2$, right: $d=3$.}
	\label{fig:regions-prob-est}
\end{figure}

\begin{figure}%[h!]
	\centering
	\includegraphics[width=0.45\textwidth]{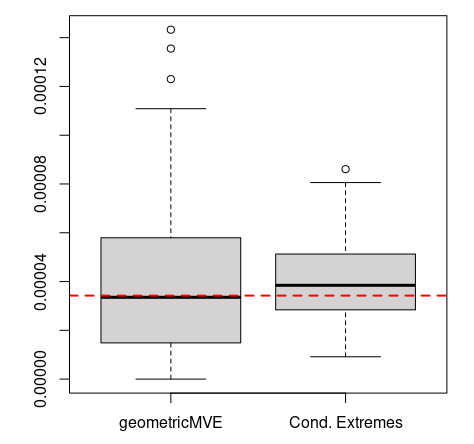}
	\includegraphics[width=0.45\textwidth]{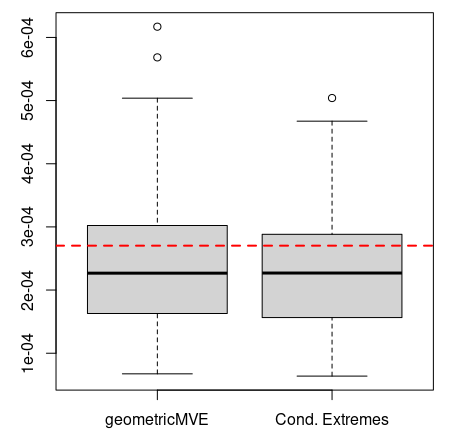}
	\caption{Probability estimates of lying in a predefined extremal region. Left: $d=2$, right: $d=3$. True probability values are displayed by the red dashed line.}
	\label{fig:prob-est-boxplots}
\end{figure}

The second issue with the shape parameter is that there will be values of $\bm{w}$ on the simplex $\mathcal{S}_{d-1}$ such that $\alpha(\bm{w})\leq 0$. Figure~\ref{fig:2d-3d-problem-region} illustrates these regions, demonstrating that they appear to be small in volume and thus not important, especially as the dimension $d$ increases.  In both the $d=2$ and $d=3$ cases, we see that the region is almost negligible for $\rho<0.7$. Furthermore, simulations from our models often do not produce points in these regions because when the dependence is high, there are very few points $\bm{W}$ near these boundaries that are accompanied by large values of $R$.
\newpage
%	\clearpage
Finally, we offer evidence that extrapolation and probability estimation in these potentially problematic regions is not an issue when compared to other contemporary multivariate extremes methods. In both the $d=2$ and $d=3$ setting, we fix $\rho=0.7$ and consider rectangular sets which overlap with the regions where $\alpha(\bm{w}) \leq 0$. For $d=2$ we estimate the probability of lying in the set $(5,7)\times(0,0.75)$, and for $d=3$ we estimate the probability of lying in the set $(5,10)\times(0,2)\times(0,2)$ (see Figure~\ref{fig:regions-prob-est}).
The resulting probability estimates are presented in the boxplots provided in Figure~\ref{fig:prob-est-boxplots}. We compare our geometric approach to the conditional extremes model of \cite{HeffernanTawn04}. For $d=3$ the results are comparable in terms of bias and variance, while for $d=2$, our method is unbiased but with slightly higher variability than conditional extremes.

\newpage

\section{Gauge function corresponding to the asymmetric logistic distribution}
\label{sec:asylog}
The asymmetric logistic distribution \citep{Tawn90} is a max-stable distribution with exponent function
\begin{align}
V(\bm{z}) = \sum_{C \in 2^D\setminus \emptyset} \left[\sum_{j \in C} \left(\frac{\theta_{j,C}}{z_j}\right)^{1/\gamma_C}\right]^{\gamma_C}, \label{eq:VAL}
\end{align}
where $2^D\setminus \emptyset$ is the power set of $D=\{1,\ldots,d\}$. The parameters satisfy $\gamma_C \in (0,1]$ and for each $j$ there is a marginal condition that $\sum_{C \in 2^D\setminus \emptyset} \theta_{j,C} = 1$, with $\theta_{j,C} =0$ if $j \not\in C$. Notice that for singleton sets the values $\gamma_{\{1\}},\ldots, \gamma_{\{d\}}$ are irrelevant; for convenience we assume that each of these is equal to 1.

The parameters $\theta_{j,C}$ play no role in determining the structure of the gauge function, except for where they lead to certain sets of variables being discounted as not taking extreme values simultaneously. We therefore consider a modification of $V$ that allows us to derive the associated gauge function in a simpler manner. Define
\begin{align}
V^*(\bm{z}) = \sum_{C \in 2^D\setminus \emptyset} \theta_C \left[\sum_{j \in C} \left(\frac{1}{z_j}\right)^{1/\gamma_C}\right]^{\gamma_C}, \label{eq:VstarAL}
\end{align}
with $\gamma_C$ as before, and
\begin{align*}
\theta_C = \begin{cases}
1 & \mbox{Variables in $C$ can be simultaneously extreme}\\
0 & \mbox{Variables in $C$ cannot be simultaneously extreme}.
\end{cases}
\end{align*}
In other words, $\theta_C=1$ when the corresponding spectral measure $H$ places mass on $\mathbb{B}_C$. For each variable $j=1,\ldots,d$ there must be at least one $\theta_C=1$ for $j \in C$. The distribution function $\exp\{-V^*(e^{\bm{x}})\}$ is a multivariate max-stable distribution with asymmetric logistic type dependence and Gumbel margins with non-zero location and unit scale. Distributions with unit-scale Gumbel margins have the same limit sets as distributions with exactly unit exponential margins \citep{NoldeWadsworth21}. There is one further difference between the models defined by exponent functions~\eqref{eq:VAL} and~\eqref{eq:VstarAL}: when $\theta_C=0$ in~\eqref{eq:VstarAL}, this ``switches off'' the group of variables corresponding to $C$ with no effect on other groups. With~\eqref{eq:VAL}, if $\theta_{j,C}=0$ for a single $j \in C$, then the set of variables that can be simultaneously extreme corresponds to $\theta_{C\setminus\{j\}}$, meaning that there could be two (or more) different $\gamma$ parameters corresponding to the same set of variables. The function $V^*$ in~\eqref{eq:VstarAL} is restricted to a single $\gamma$ parameter per group of variables.

We define the collection of sets $\mathcal{C}_J$ to be all of those containing the index set $J \subset\{1,\ldots,d\}$. For example, with $d=3$, $\mathcal{C}_{\{1\}} = \{\{1\},\{1,2\},\{1,3\},\{1,2,3\}\}$ and $\mathcal{C}_{\{1,2\}} = \{\{1,2\},\{1,2,3\}\}$. Here we outline the important steps in derivation of the gauge function for $d=3$.

The partial derivatives of $V^*(\bm{z})$ are, for $k,l \in \{1,2,3\}$:
\begin{align*}
V_{\{k\}}^*(\bm{z}) &= \sum_{C_k \in \mathcal{C}_{\{k\}}} \theta_{C_k}\kappa_{C_k} z_k^{-1/\gamma_{C_k}-1}\left[\sum_{j \in C_k}z_{j}^{-1/\gamma_{C_k}}\right]^{\gamma_{C_k}-1}\\
V_{\{k,l\}}^*(\bm{z}) &= \sum_{C_{kl} \in \mathcal{C}_{\{k,l\}}} \theta_{C_{k,l}}\kappa_{C_{k,l}} z_k^{-1/\gamma_{C_{kl}}-1}z_l^{-1/\gamma_{C_{kl}}-1}\left[\sum_{j \in C_{kl}}z_{j}^{-1/\gamma_{C_{kl}}}\right]^{\gamma_{C_{kl}}-2}\\
V_{\{1,2,3\}}^*(\bm{z}) &= \sum_{C_{123} \in \mathcal{C}_{\{1,2,3\}}}\theta_{C_{123}} \kappa_{C_{123}} z_1^{-1/\gamma_{C_{123}}-1}z_2^{-1/\gamma_{C_{123}}-1}z_3^{-1/\gamma_{C_{123}}-1}\left[\sum_{j \in C_{123}} z_{j}^{-1/\gamma_{C_{123}}}\right]^{\gamma_{C_{123}}-3}\\
&= \theta_{\{1,2,3\}}\kappa_{\{1,2,3\}} z_1^{-1/\gamma_{\{1,2,3\}}-1}z_2^{-1/\gamma_{{\{1,2,3\}}}-1}z_3^{-1/\gamma_{{\{1,2,3\}}}-1}\left[\sum_{j =1}^3 z_{j}^{-1/\gamma_{\{1,2,3\}}}\right]^{\gamma_{\{1,2,3\}}-3},
\end{align*}
with $\kappa_C \neq 0$ representing constant terms. Consequently,
\begin{align*}
V^*_{\{1\}}(e^{t\bm{x}})V^*_{\{2\}}(e^{t\bm{x}})V^*_{\{3\}}(e^{t\bm{x}}) & = \sum_{C_1 \in \mathcal{C}_{\{1\}}}\sum_{C_2 \in \mathcal{C}_{\{2\}}}\sum_{C_3 \in \mathcal{C}_{\{3\}}}\left\{ \theta_{C_1}\theta_{C_2}\theta_{C_3}\kappa_{C_1}\kappa_{C_2}\kappa_{C_3}\right.\\ &~~~\left.e^{-tx_1\left(\frac{1}{\gamma_{C_1}}+1\right)-tx_2\left(\frac{1}{\gamma_{C_2}}+1\right)-tx_3\left(\frac{1}{\gamma_{C_3}}+1\right)}\right.\\
&~~~\left.e^{-t\min_{j \in C_1}x_j\left(1-\frac{1}{\gamma_{C_1}}\right)-t\min_{j \in C_2}x_j\left(1-\frac{1}{\gamma_{C_2}}\right)-t\min_{j \in C_3}x_j\left(1-\frac{1}{\gamma_{C_3}}\right)}\right\}[1+o(1)],
\end{align*}

\begin{align*}
V^*_{\{j\}}(e^{t\bm{x}})V^*_{\{k,l\}}(e^{t\bm{x}})&= \sum_{C_j \in \mathcal{C}_{\{j\}}}\sum_{C_{kl} \in \mathcal{C}_{\{k,l\}}} \left\{\theta_{C_j}\theta_{C_{kl}}\kappa_{C_j}\kappa_{C_{kl}}e^{-tx_j\left(\frac{1}{\gamma_{C_j}}+1\right)-tx_k\left(\frac{1}{\gamma_{C_{kl}}}+1\right)-tx_l\left(\frac{1}{\gamma_{C_{kl}}}+1\right)}\right.\\
&~~~ \left. e^{-t\min_{i \in C_j}x_i\left(1-\frac{1}{\gamma_{C_j}}\right)-t\min_{i \in C_{kl}}x_i\left(1-\frac{2}{\gamma_{C_{kl}}}\right)}\right\}[1+o(1)],
\end{align*}

\begin{align*}
V_{\{1,2,3\}}^*(e^{t\bm{x}}) &=\sum_{C_{123} \in \mathcal{C}_{\{1,2,3\}}}\theta_{C_{123}}\kappa_{C_{123}}e^{-tx_1\left(\frac{1}{\gamma_{C_{123}}}+1\right)-tx_2\left(\frac{1}{\gamma_{C_{123}}}+1\right)-tx_3\left(\frac{1}{\gamma_{C_{123}}}+1\right) + t\min_{j\in C_{123}} x_j \left(1-\frac{3}{\gamma_{C_{123}}}\right)}\\
&= \theta_{\{1,2,3\}}\kappa_{\{1,2,3\}}e^{-tx_1\left(\frac{1}{\gamma_{\{1,2,3\}}}+1\right)-tx_2\left(\frac{1}{\gamma_{\{1,2,3\}}}+1\right)-tx_3\left(\frac{1}{\gamma_{\{1,2,3\}}}+1\right) + t\min_{j\in\{1,2,3\}} x_j \left(1-\frac{3}{\gamma_{\{1,2,3\}}}\right)}.
\end{align*}

The density of the distribution in non-centred Gumbel margins is
\begin{align*}
\exp\{-V^*(e^{t\bm{x}})\}e^{t\sum_{j=1}^d x_j}\sum_{\pi \in \Pi}\prod_{s \in \pi} -V_s(e^{t\bm{x}}),
\end{align*}
and the gauge function is derived through the minimum of all non-zero terms coming from the partial derivatives. To this end, we define $\mathcal{C}_{J}^+$ to be the collection of all sets that contain the index set $J$ and for which $\theta_J = 1$. For example, when $d=3$ and $\theta_{\{1\}}=\theta_{\{1,2,3\}}=0$ and all other $\theta_J=1$ then $\mathcal{C}_{\{1\}}^+=\{\{1,2\},\{1,3\}\}$. For $d=3$ this yields the following expression for $g(\bm{x})$:
\begin{align*}
\min&\left\{\min_{\substack{C_1 \in \mathcal{C}_{\{1\}}^+,C_2 \in \mathcal{C}_{\{2\}}^+,\\C_3 \in \mathcal{C}_{\{3\}}^+}}\left[\frac{x_1}{\gamma_{C_1}}+\frac{x_2}{\gamma_{C_2}}+\frac{x_3}{\gamma_{C_3}}+ \min_{j \in C_1}x_j\left(1-\frac{1}{\gamma_{C_1}}\right) + \min_{j \in C_2}x_j\left(1-\frac{1}{\gamma_{C_2}}\right) +\min_{j \in C_3}x_j\left(1-\frac{1}{\gamma_{C_3}}\right)\right], \right.\\
&~~\left.\min_{C_1 \in \mathcal{C}_{\{1\}}^+, C_{23} \in \mathcal{C}_{\{2,3\}}^+}\left[\frac{x_1}{\gamma_{C_1}}+\frac{x_2}{\gamma_{C_{23}}}+\frac{x_3}{\gamma_{C_{23}}}+ \min_{j \in C_1}x_j\left(1-\frac{1}{\gamma_{C_1}}\right) + \min_{j \in C_{23}}x_j\left(1-\frac{2}{\gamma_{C_{23}}}\right)\right],\right.\\
&~~\left.\min_{C_2 \in \mathcal{C}_{\{2\}}^+, C_{13} \in \mathcal{C}_{\{1,3\}}^+}\left[\frac{x_1}{\gamma_{C_{13}}}+\frac{x_2}{\gamma_{C_{2}}}+\frac{x_3}{\gamma_{C_{13}}}+ \min_{j \in C_{2}}x_j\left(1-\frac{1}{\gamma_{C_2}}\right) + \min_{j \in C_{13}}x_j\left(1-\frac{2}{\gamma_{C_{13}}}\right)\right],\right.\\
&~~\left.\min_{C_3 \in \mathcal{C}_{\{3\}}^+, C_{12} \in \mathcal{C}_{\{1,2\}}^+}\left[\frac{x_1}{\gamma_{C_{12}}}+\frac{x_2}{\gamma_{C_{12}}}+\frac{x_3}{\gamma_{C_3}}+ \min_{j \in C_{3}}x_j\left(1-\frac{1}{\gamma_{C_3}}\right) + \min_{j \in C_{12}}x_j\left(1-\frac{2}{\gamma_{C_{12}}}\right)\right],\right.\\
&~~\left. \min_{C_{123} \in \mathcal{C}_{\{1,2,3\}}^+} \left[\frac{x_1}{\gamma_{C_{123}}}+\frac{x_2}{\gamma_{C_{123}}}+\frac{x_3}{\gamma_{C_{123}}}+\min_{j \in C_{123}} x_j \left(1-\frac{3}{\gamma_{C_{123}}}\right)\right]
\right\}.
\end{align*}

From the derivation and form of this gauge function, we observe that the general form for any dimension $d$ is 
\begin{align*}
g(\bm{x}) = \min_{\pi \in \Pi} \min_{C_s \in \mathcal{C}_s^+: s \in \pi} \left[\sum_{s \in \pi}\left(\sum_{j \in s}\frac{x_j}{\gamma_{C_s}} + \min_{j \in C_s} x_j \left(1-\frac{|s|}{\gamma_{C_s}}\right)\right)\right].
\end{align*}

\newpage 
\section{Example limit sets obtained by mixing gauge functions}
\label{sec:mixinggauge}
\subsection{Mixing via minimization}
\begin{figure}[htbp]
	\centering
	\includegraphics[width=0.3\textwidth]{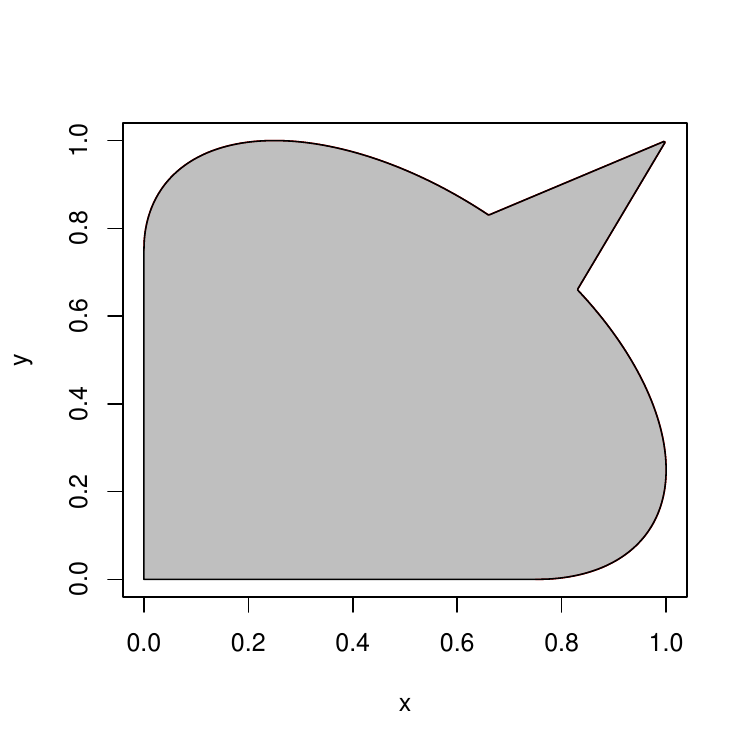}
	\includegraphics[width=0.3\textwidth]{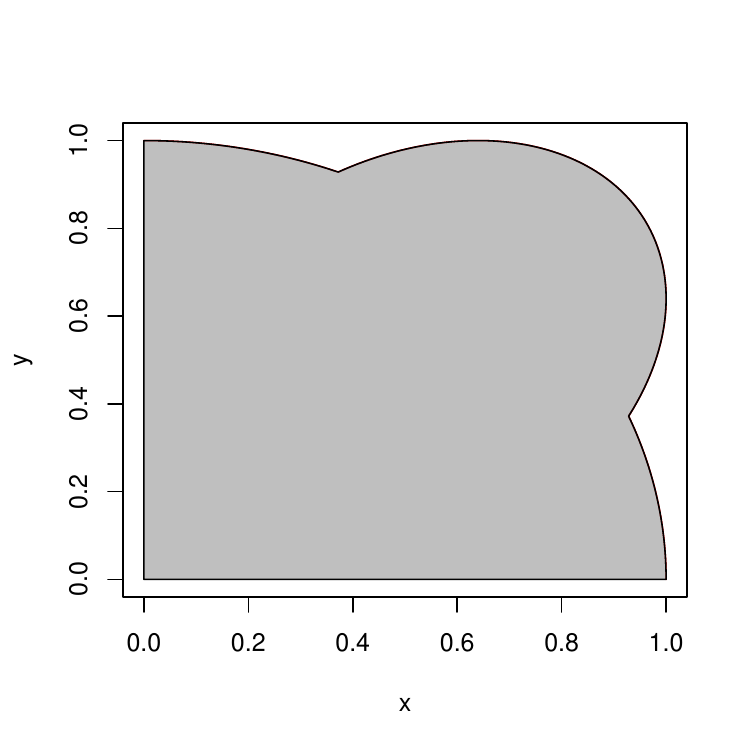}
	\includegraphics[width=0.3\textwidth]{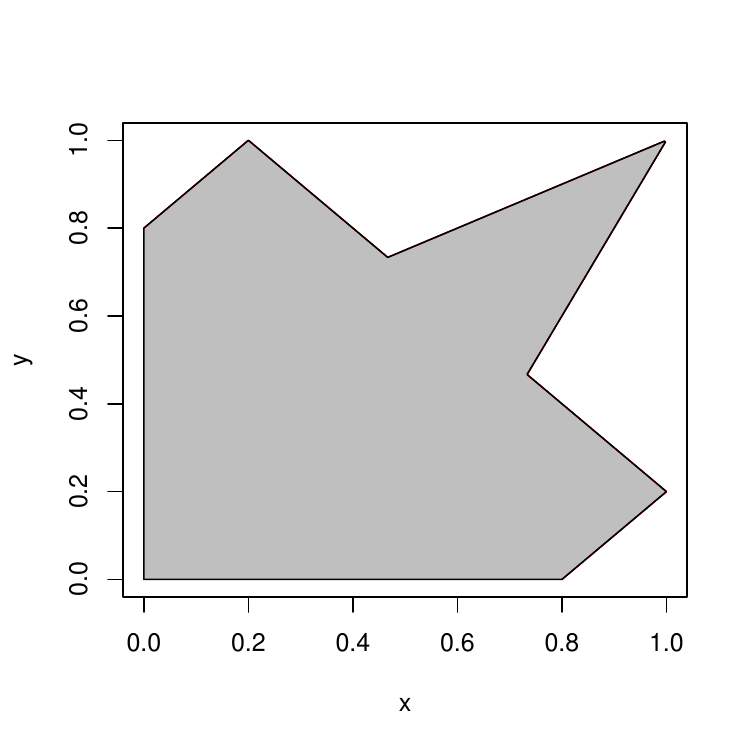}
	\caption{Illustration of limit sets \edit{(grey shaded region)} arising from taking the minimum of two component gauge functions. Left: $g^{[1]}(x_1,x_2) = (x_1+x_2)/\gamma+(1-2/\gamma)\min(x_1,x_2)$ with $\gamma=0.5$, $g^{[2]}(x_1,x_2) = (x_1+x_2-2\rho(x_1 x_2)^{1/2})/(1-\rho^2)$ with $\rho=0.5$; centre: $g^{[1]}(x_1,x_2) = (x_1^{1/\gamma}+x_2^{1/\gamma})^\gamma$ with $\gamma=0.5$, $g^{[2]}(x_1,x_2) = (x_1+x_2-2\rho(x_1 x_2)^{1/2})/(1-\rho^2)$  with $\rho=0.8$; right: $g^[1](x_1,x_2) = (x_1+x_2)/\gamma+(1-2/\gamma)\min(x_1,x_2)$ with $\gamma=0.5$, $g^{[2]}(x_1,x_2) = \max((x_1-x_2)/\theta,(x_2-x_1)/\theta,(x_1+x_2)/(2-\theta))$ with $\theta=0.8$.}
	\label{fig:2dgaugeminmix}
\end{figure}

Figure~\ref{fig:2dgaugeminmix} displays illustrations of limit sets that arise from mixing two gauge functions by minimization: $g(x_1,x_2) = \min\{g^{[1]}(x_1,x_2),g^{[2]}(x_1,x_2)\}$.

\subsection{Additive mixing}
\begin{figure}[htbp]
	\centering
	\includegraphics[width=0.3\textwidth]{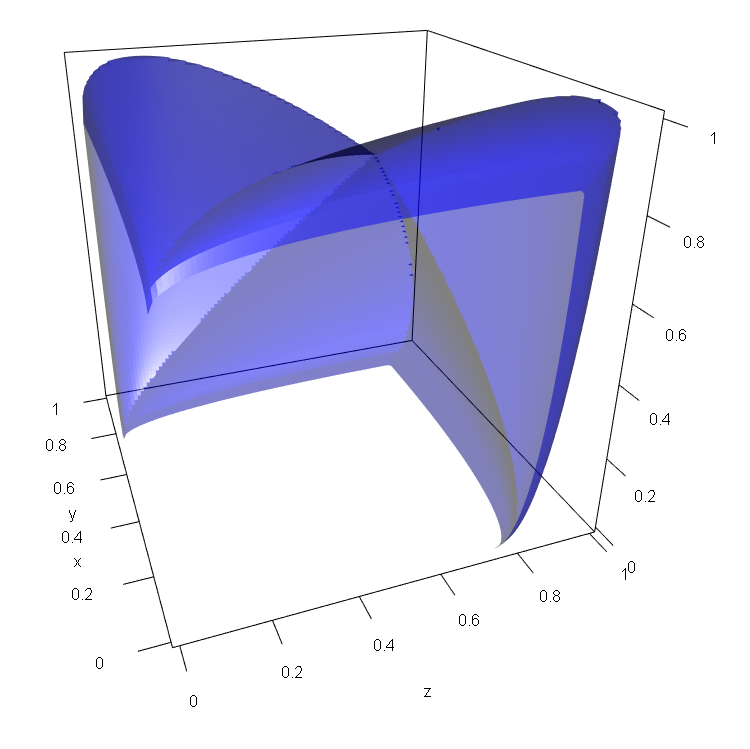}
	\includegraphics[width=0.3\textwidth]{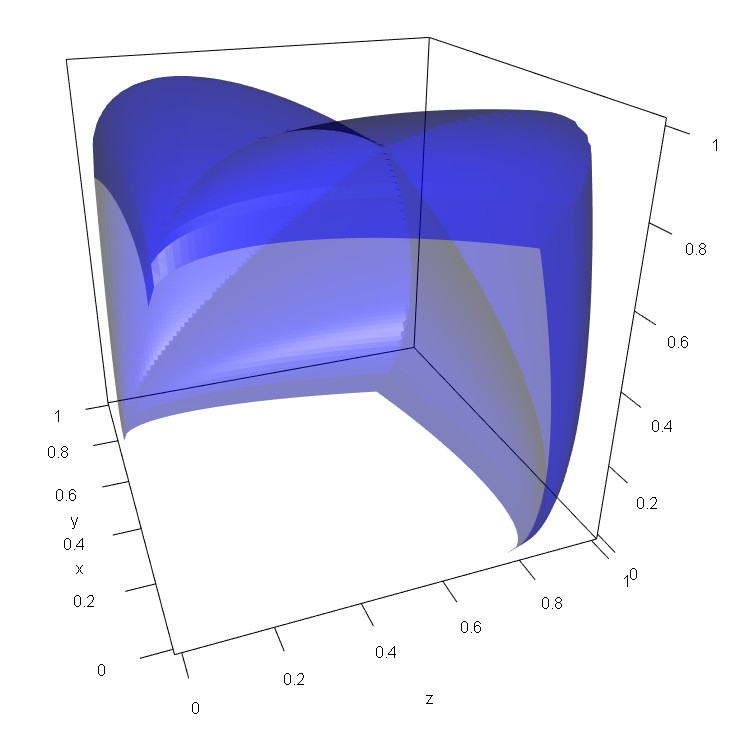}
	\includegraphics[width=0.3\textwidth]{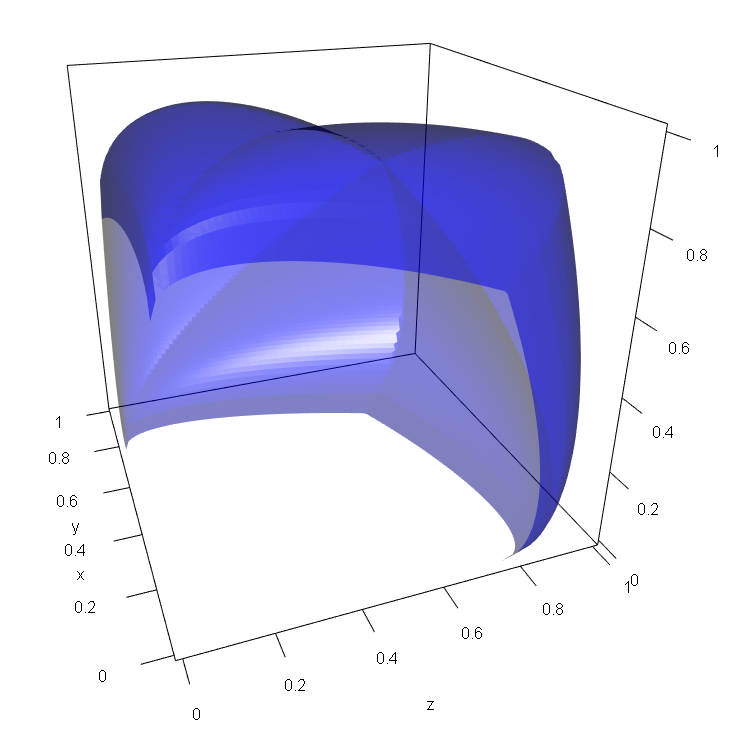}
	\caption{Illustration of limit sets  (region between blue surface and planes $x_j=0$) arising from additively mixing gauge functions corresponding to the Gaussian distribution with $\bm{\rho}=(0.5,0.5,0.5)$, and the asymmetric logistic distribution with spectral measure placing mass on $\mathbb{B}_{\{1,2\}},\mathbb{B}_{\{1,3\}},\mathbb{B}_{\{2,3\}}$, and parameters $\gamma_{\{1,2\}}=\gamma_{\{1,3\}}=\gamma_{\{2,3\}}=0.5$. From left to right the weights are $a_1=1,2,3$, with $g^{[1]}$ the Gaussian gauge.}
	\label{fig:3dgaugemix}
\end{figure}

Some examples of shapes obtainable by additively mixing gauge functions with $d=2$ are presented in Figure~\ref{fig:additivemixing} of the main manuscript. Figure~\ref{fig:3dgaugemix} displays some examples with $d=3$.

\newpage
\section{Additional simulation study figures}
\label{sec:simstudyextra}

%--------------

%--------------

\subsection{$d=2$}
\begin{figure}[htpb]
	\centering
	\includegraphics[width=0.4\textwidth]{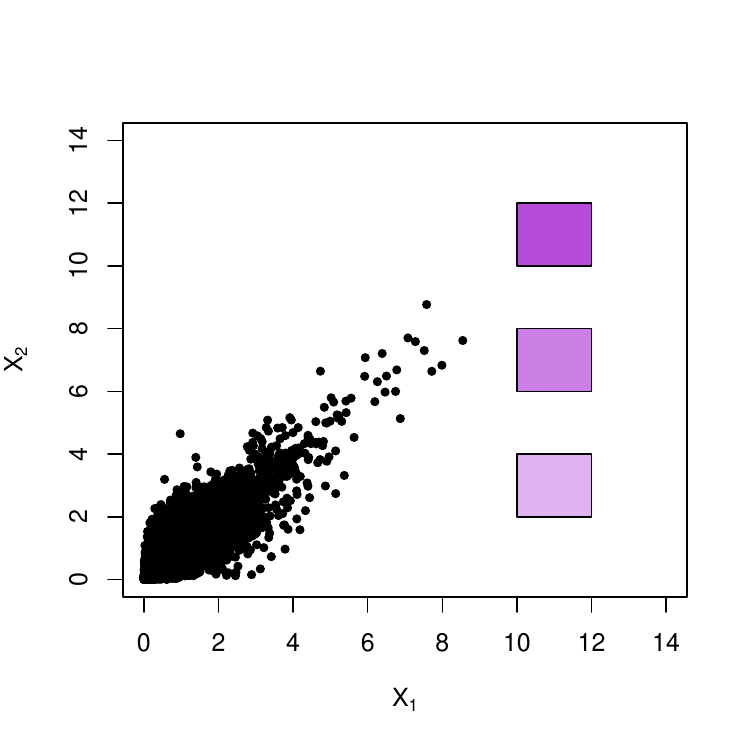}
	\includegraphics[width=0.4\textwidth]{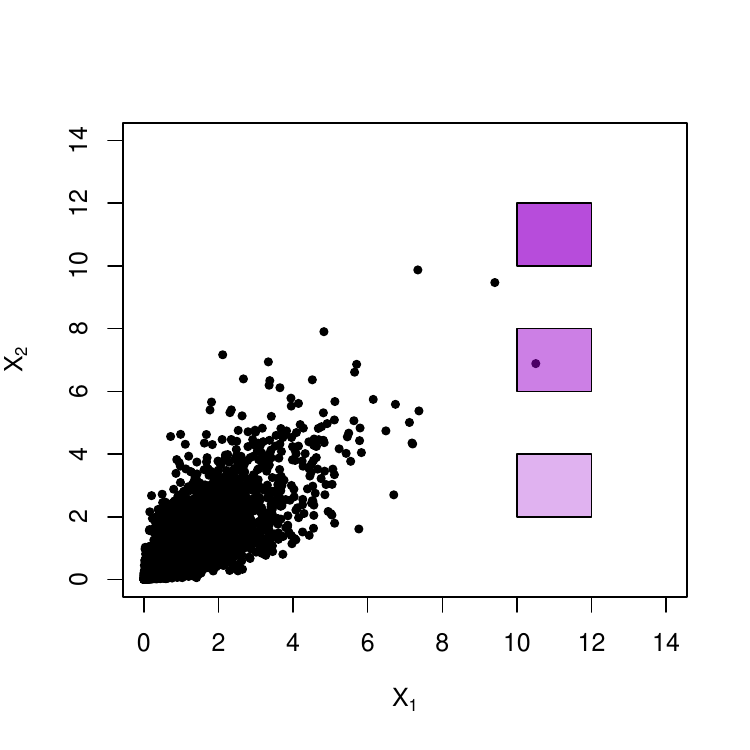}\\
	\includegraphics[width=0.4\textwidth]{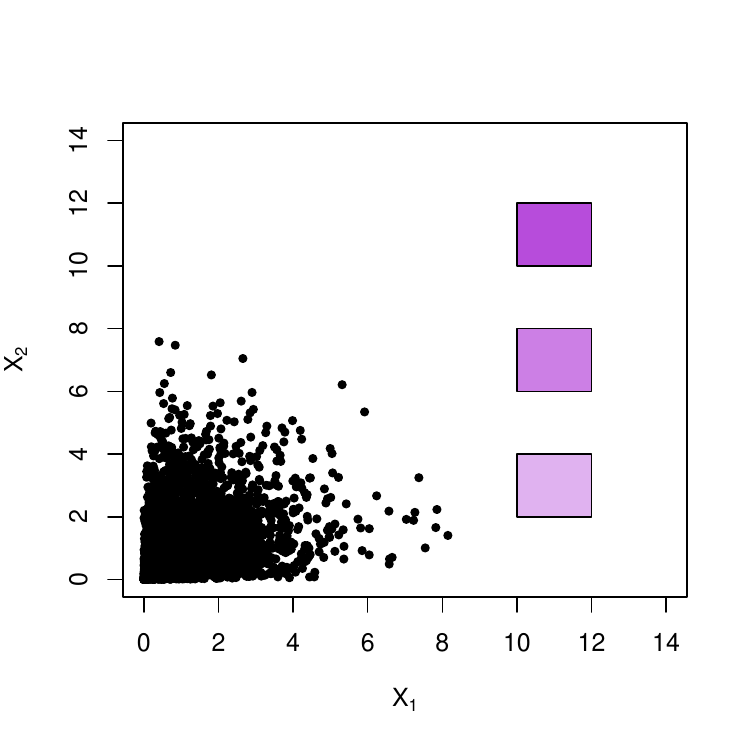}
	\includegraphics[width=0.4\textwidth]{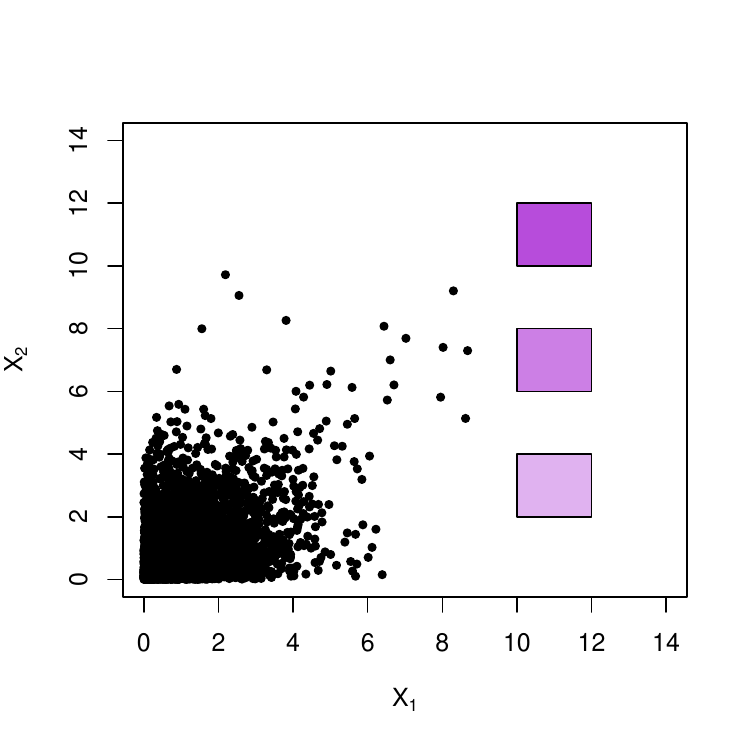}
	\caption{\edit{$d=2$}: Example data from distributions \edit{(I) (top left), (II) (top right), (III) (bottom left) and (IV) (bottom right)}, and illustration of sets $B_1,B_2,B_3$ (purple shading).}
	\label{fig:data_sets}
\end{figure}

Figure~\ref{fig:data_sets} shows examples of the \edit{four} datasets for the $d=2$ simulation study, and three sets of interest $B_1 = (10,12) \times (10,12)$, $B_2=(10,12) \times(6,8)$, and $B_3 = (10,12)\times(2,4)$. 

\begin{figure}[htbp]
	\centering
	\includegraphics[width=0.3\textwidth]{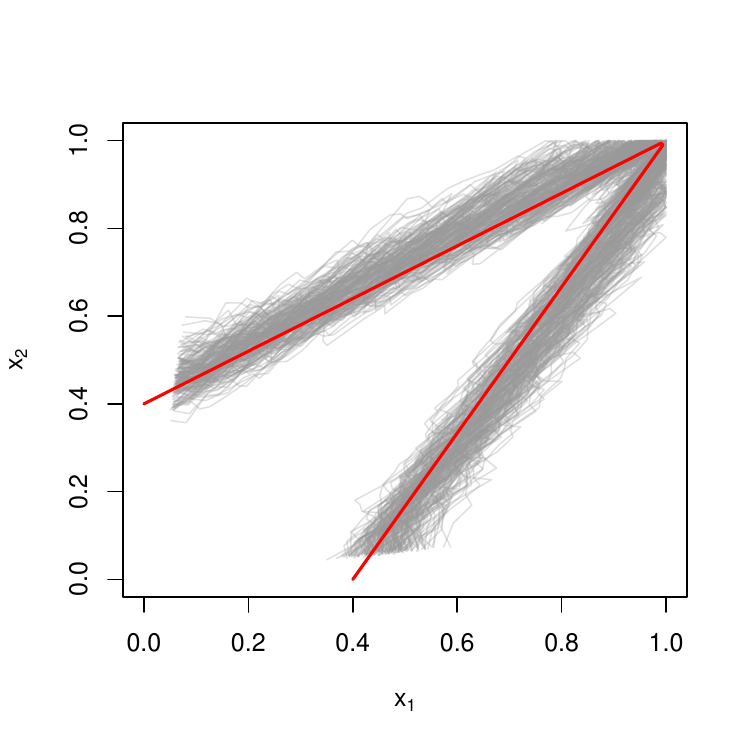}
	\includegraphics[width=0.3\textwidth]{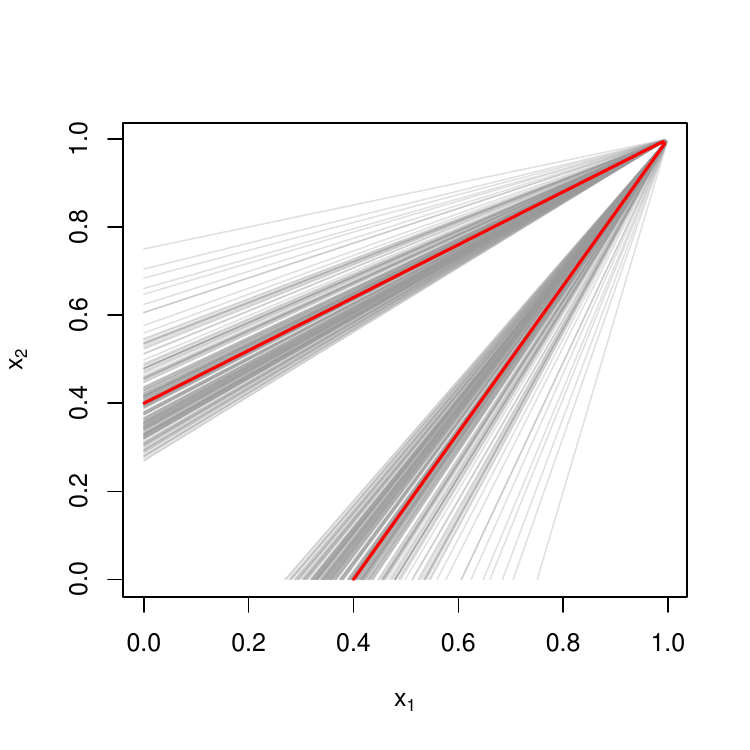}
	\includegraphics[width=0.3\textwidth]{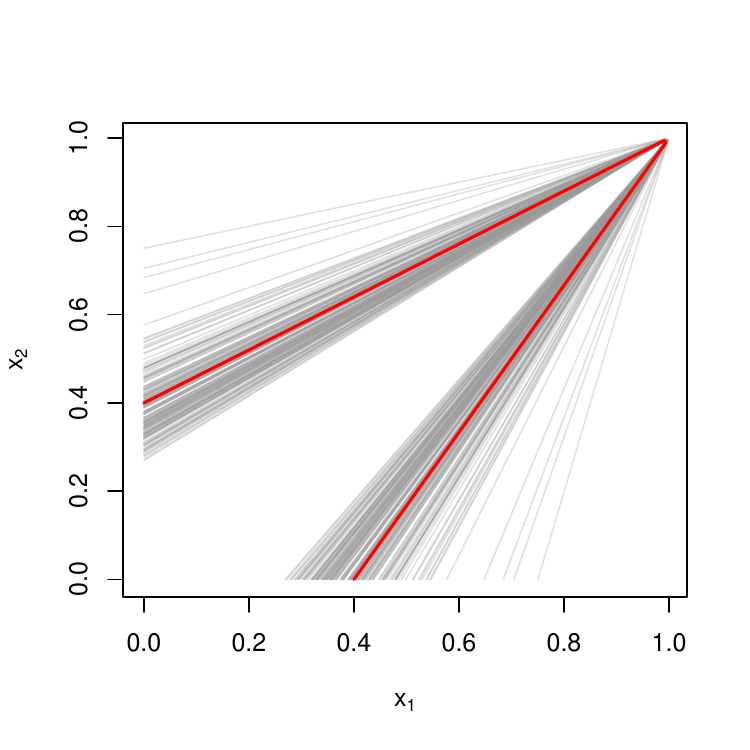}\\
	\includegraphics[width=0.3\textwidth]{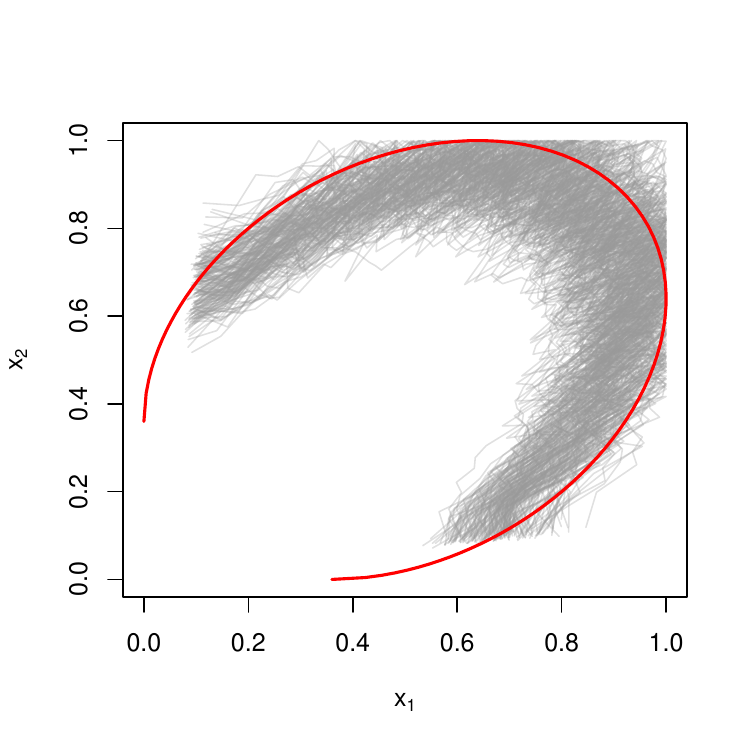}
	\includegraphics[width=0.3\textwidth]{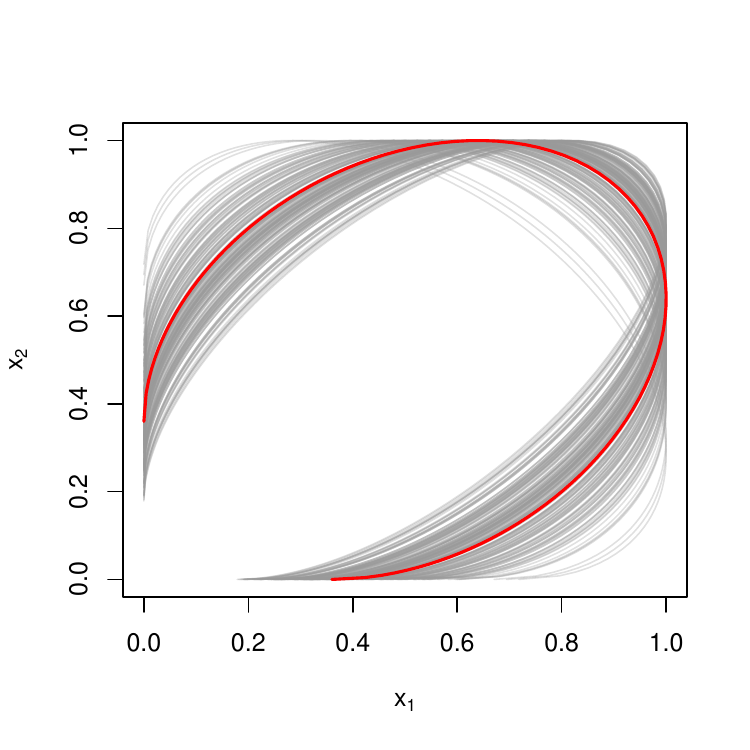}
	\includegraphics[width=0.3\textwidth]{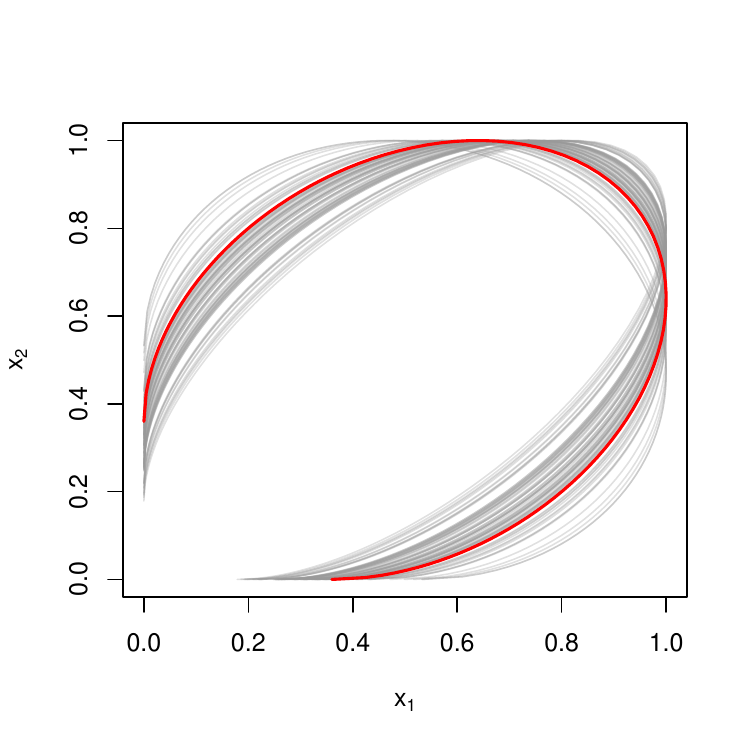}\\
	\includegraphics[width=0.3\textwidth]{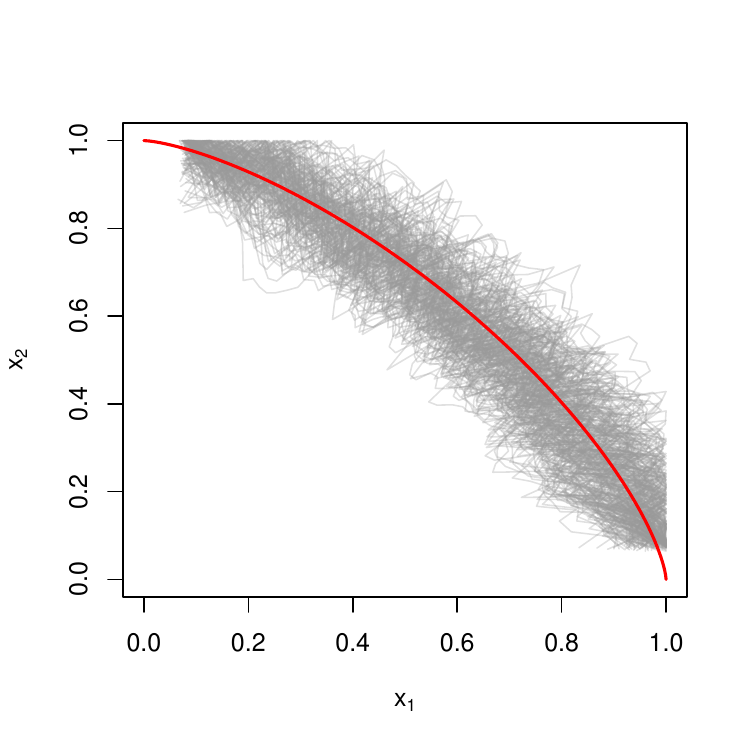}
	\includegraphics[width=0.3\textwidth]{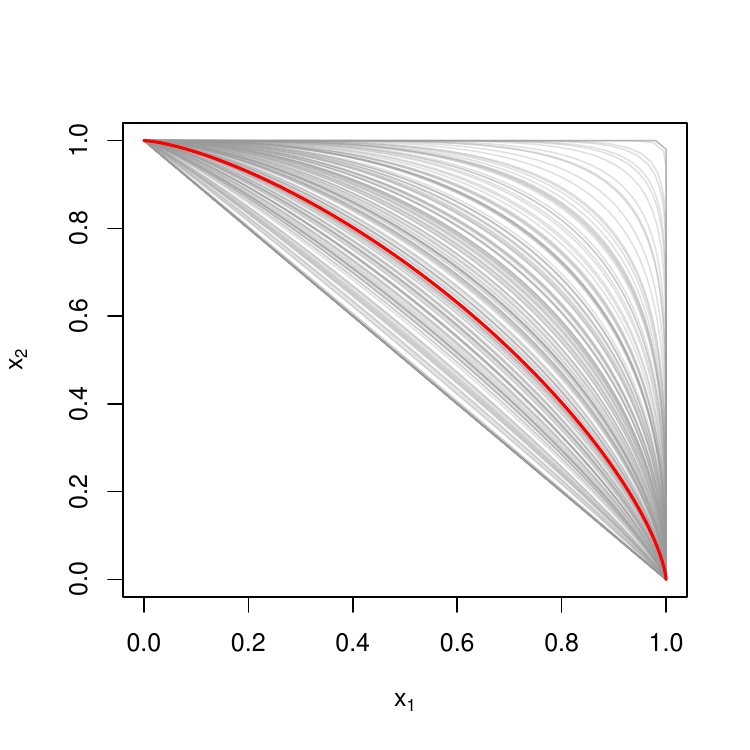}
	\includegraphics[width=0.3\textwidth]{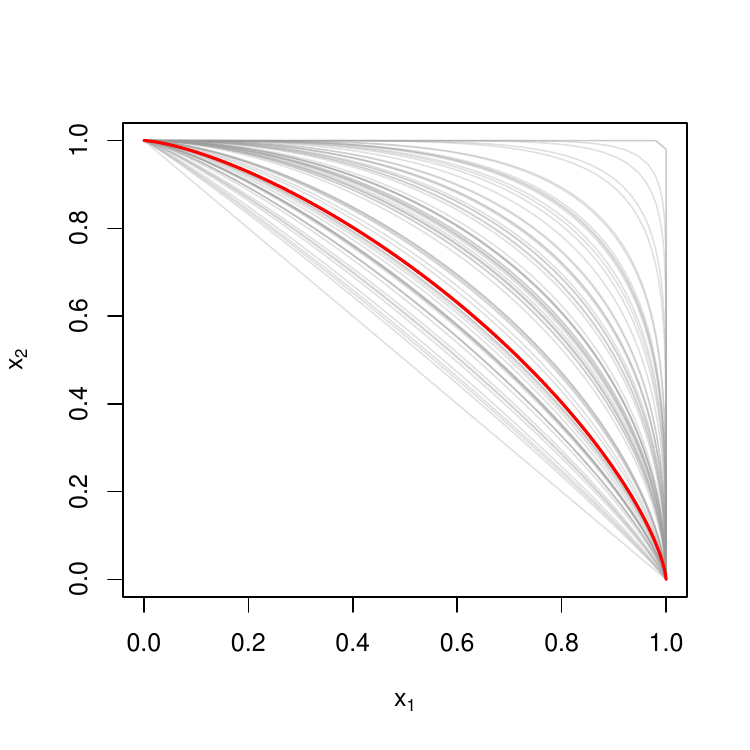}\\
	\includegraphics[width=0.3\textwidth]{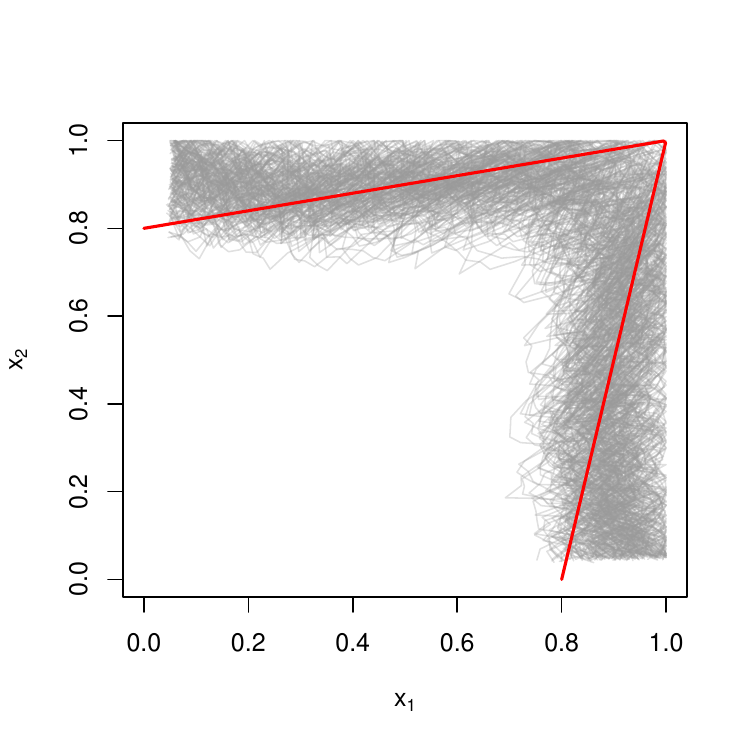}
	\includegraphics[width=0.3\textwidth]{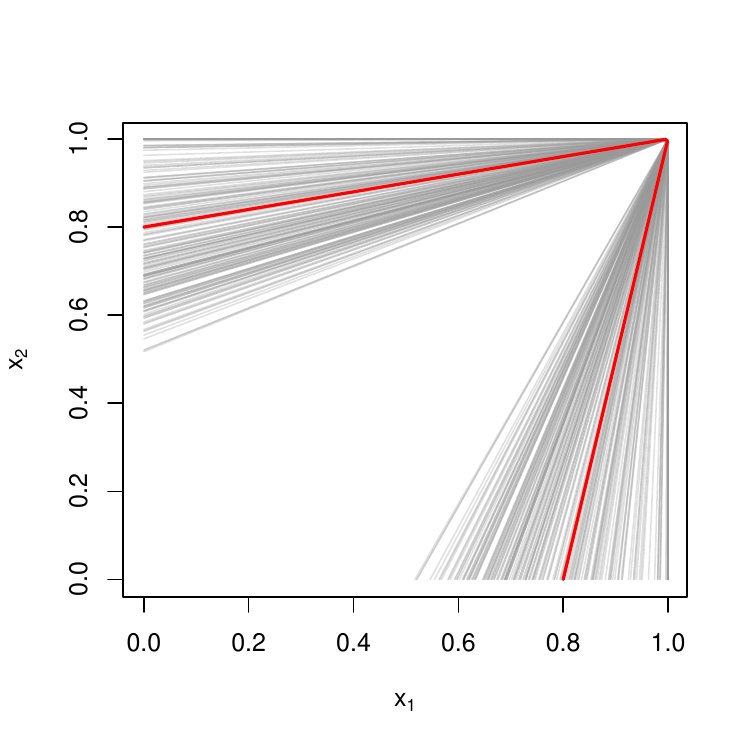}
	\includegraphics[width=0.3\textwidth]{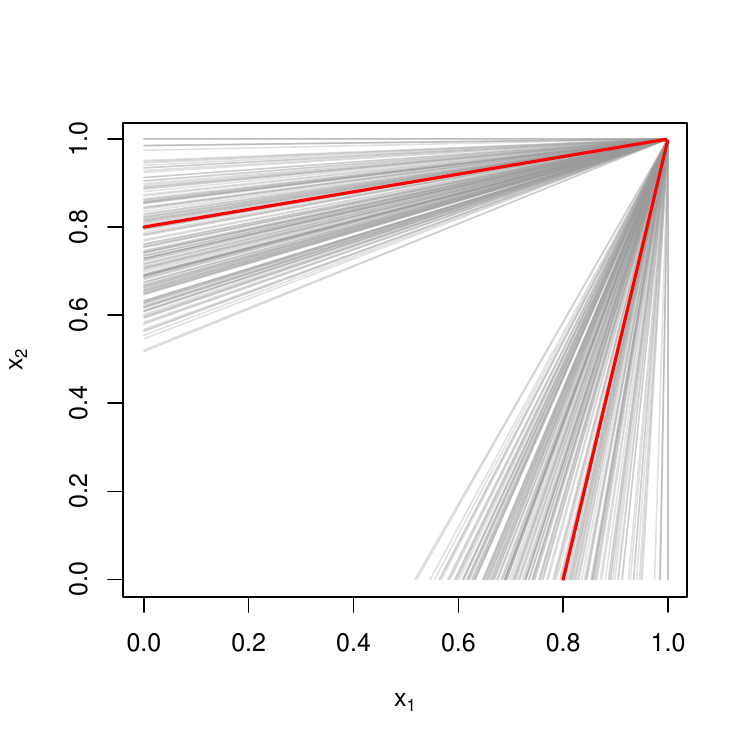}
	\caption{Non-parametric (left column) and parametric estimates (centre and right columns) of unit level sets of $g$. The unit level sets of the true $g$ are shown in red. \edit{Top-bottom: distributions (I), (II), (III) and (IV), respectively}. The centre column includes all parametric estimates from the correct gauge function; the right column includes only those parametric estimates where the true model had the lowest AIC.}
	\label{fig:estimated_gauges}
\end{figure}

Figure~\ref{fig:estimated_gauges} displays estimates of the limit set shape via non-parametric estimation of $g$ using rolling-windows quantiles, as described in Section~\ref{sec:threshold} of the main paper, and parametric estimation from the \edit{maximum likelihood estimates} of the gauge function parameters. We display parametric estimates using both all fits from the correct gauge function, and only the fits where the correct gauge function returned the minimum AIC. For distributions \edit{(I), (II), (III) and (IV), this is 82.5\%, 41.5\%, 39.5\% and 82\%, respectively}. The non-parametric estimates do not quite join to the axes because we ascribe the rolling-windows estimate of $r_0(w)$ to the centre of each window for $w$. \edit{For distribution (III) we note that the non-parametric estimates display lower variability than the parametric ones. In spite of this, the performance of the method for probability estimation, which relies both on non-parametric estimation of $r_0(\bm{w})$ and parametric estimation of $g$, appears quite reasonable.}

\subsection{$d=3$}
\begin{figure}[htpb]
	\centering
	\includegraphics[width=0.32\textwidth]{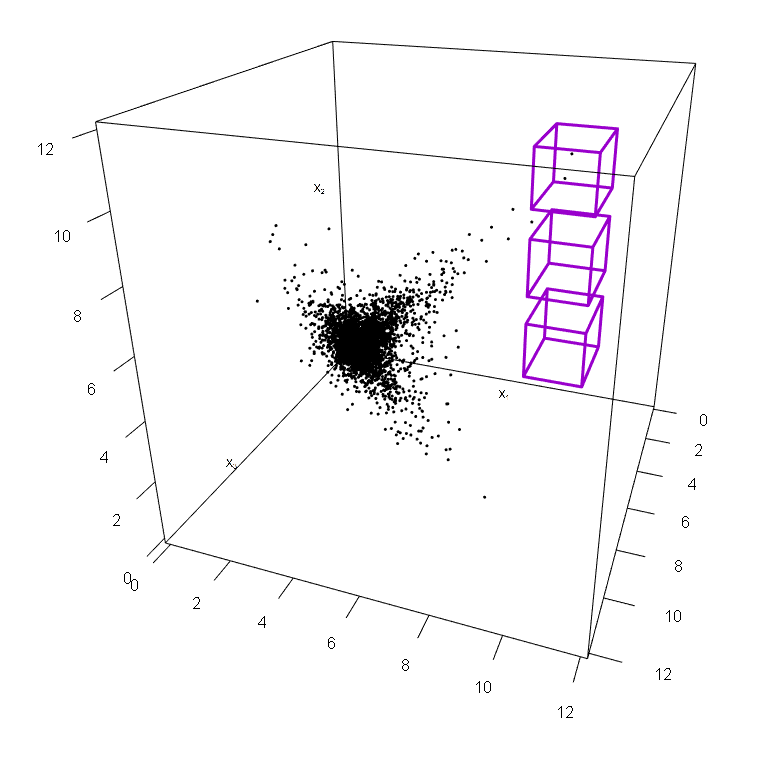}
	\includegraphics[width=0.32\textwidth]{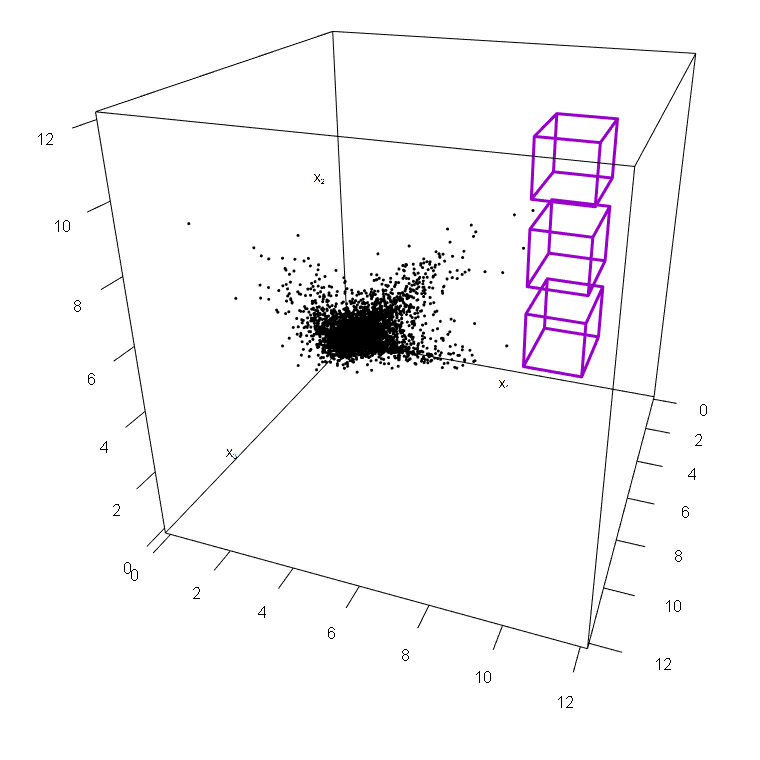}
	\includegraphics[width=0.32\textwidth]{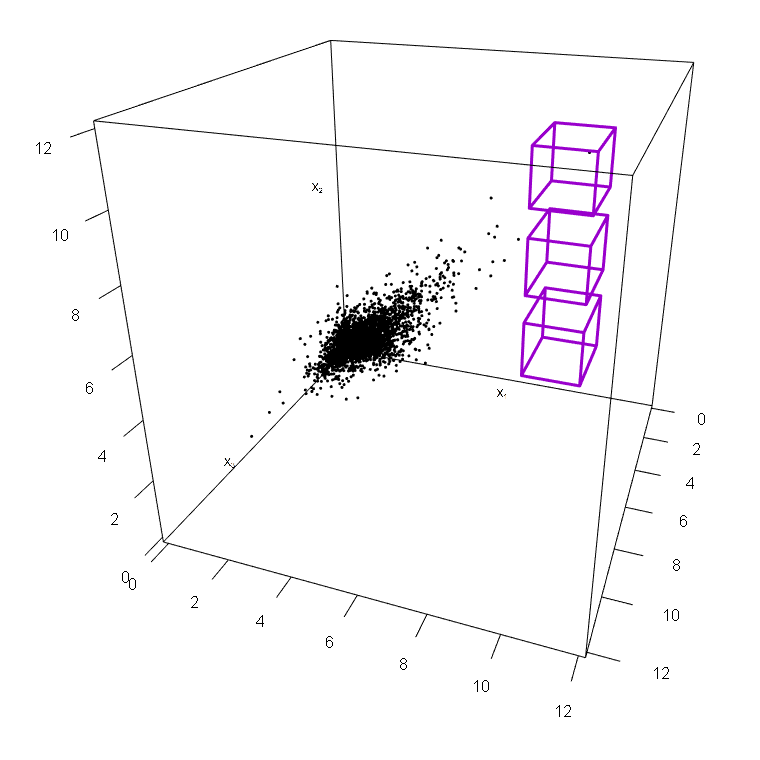}
	\caption{\edit{$d=3$}: Example data from distributions (I), (II) and (III) (left to right), and illustration of sets $B_1,B_2,B_3$ (purple boxes).}
	\label{fig:3ddatasets}
\end{figure}

Figure~\ref{fig:3ddatasets} depicts examples of the three datasets for the $d=3$ simulation study, along with sets of interest $B_1 = (8,10)\times(8,10)\times(0.01,3)$, $B_2=(8,10) \times (5,7) \times (0.01,3)$ and $(8,10)\times (2,4) \times (0.01,3)$.

\begin{figure}[h]
	\centering
	\includegraphics[width=0.3\textwidth]{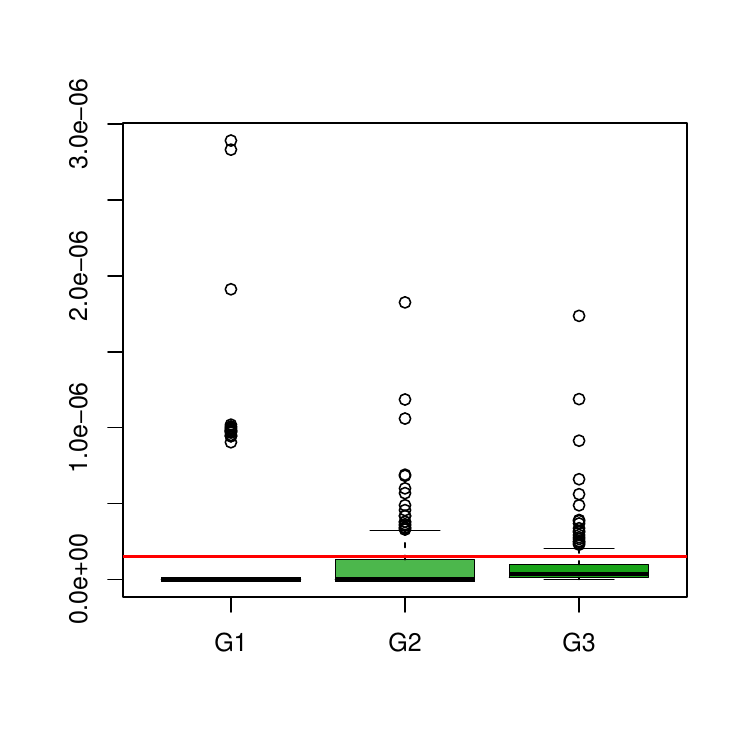}
	\includegraphics[width=0.3\textwidth]{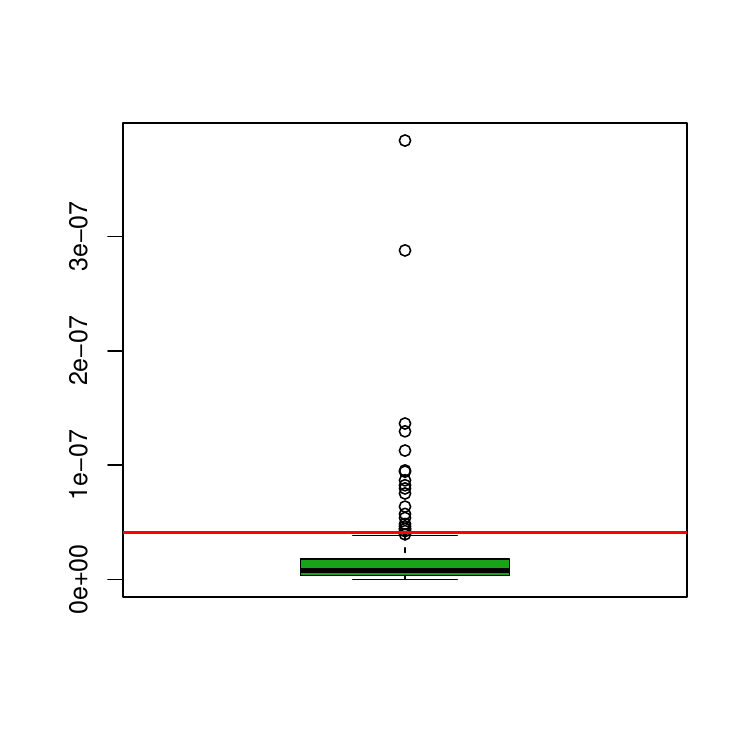}
	\caption{Boxplots of probability estimates for $\Pr(\bm{X} \in B_3)$ \edit{for $d=3$} under the geometric approach. Left: for distribution (I); right: for distribution (III). For clarity only a single boxplot, corresponding to that labelled G3 in the main manuscript, is included in the right panel.}
	\label{fig:d3extraboxplots}
\end{figure}

Figure~\ref{fig:d3extraboxplots} displays boxplots relating only to the geometric approach for $d=3$ from Figure~\ref{fig:3dboxplots} of the main manuscript, with a clearer vertical scale. Although there is a downwards bias in the estimation, the geometric approach still provides reasonable estimates in these difficult cases.

\newpage

\section{Additional figures for river flow analysis}
\label{sec:dataextra}

\begin{figure}[htbp]
	\centering
	\includegraphics[width=0.32\textwidth]{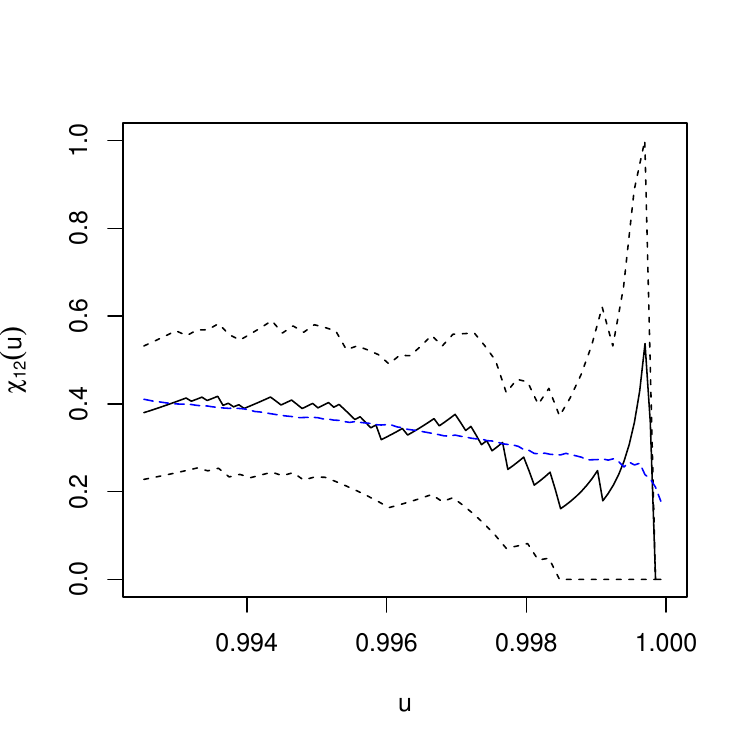}
	\includegraphics[width=0.32\textwidth]{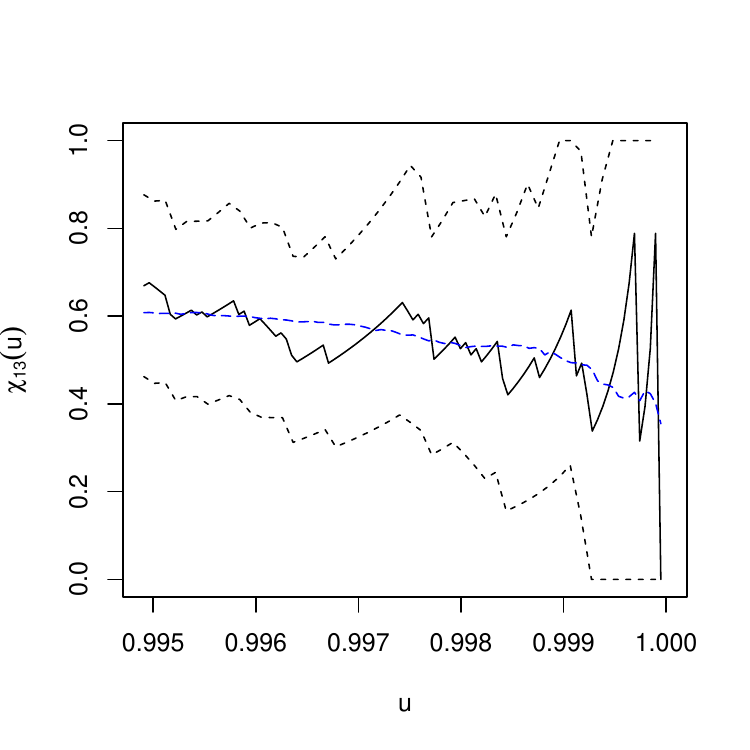}
	\includegraphics[width=0.32\textwidth]{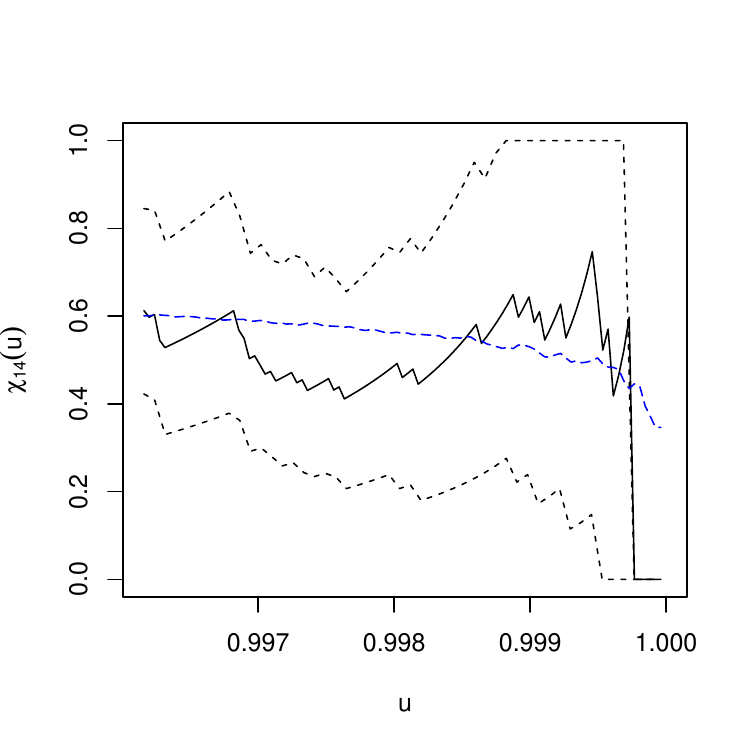}\\
	\includegraphics[width=0.32\textwidth]{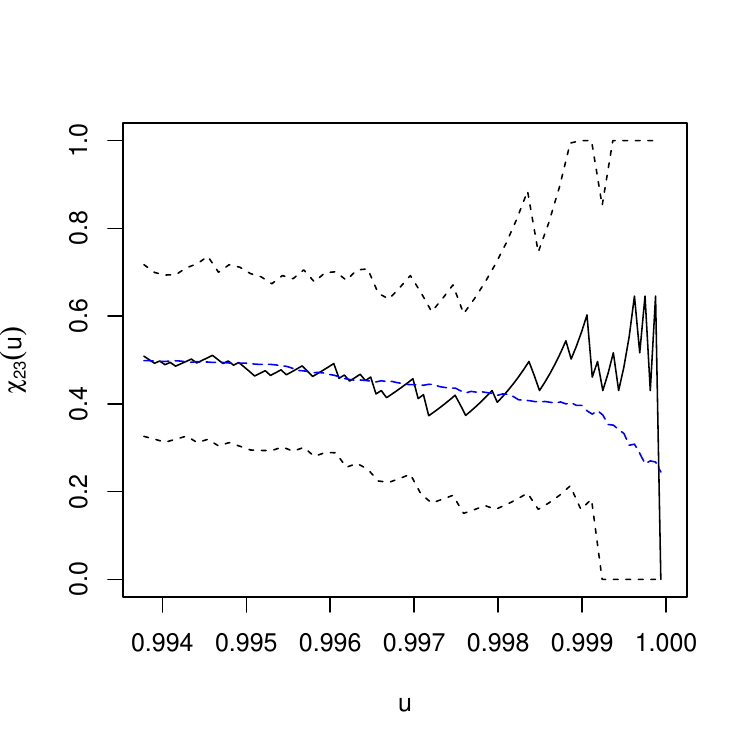}
	\includegraphics[width=0.32\textwidth]{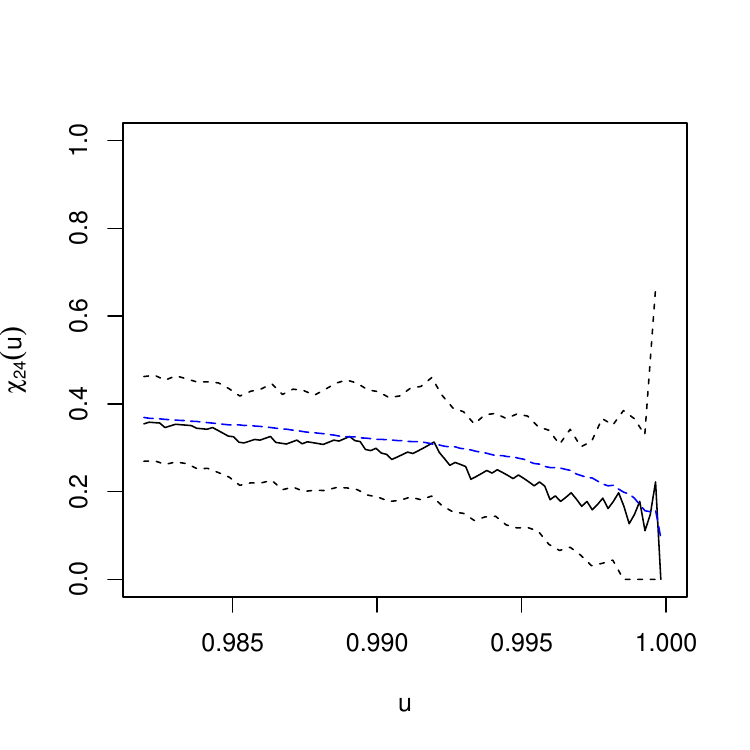}
	\includegraphics[width=0.32\textwidth]{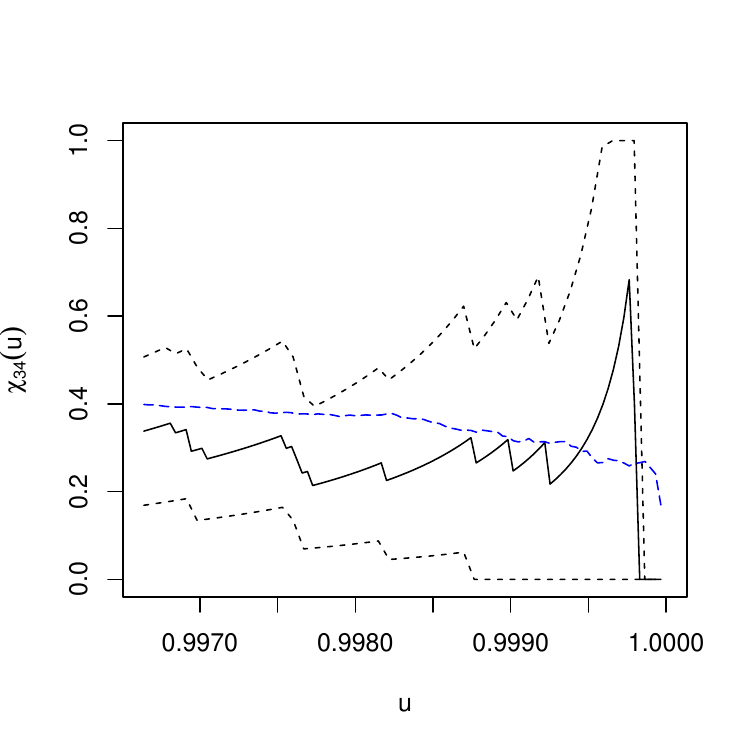}
	\caption{Empirical (black) estimates of $\chi_C(u)$ with 95\% confidence interval (dashed), and model-based estimate (blue) for groups $C= \{1,2\}$, $\{1,3\}$, $\{1,4\}$, $\{2,3\}$, $\{2,4\}$, and $\{3,4\}$ (top left - bottom right).}
	\label{fig:river-chi-extra}
\end{figure}

\begin{figure}[htbp]
	\centering
	\includegraphics[width=0.4\textwidth]{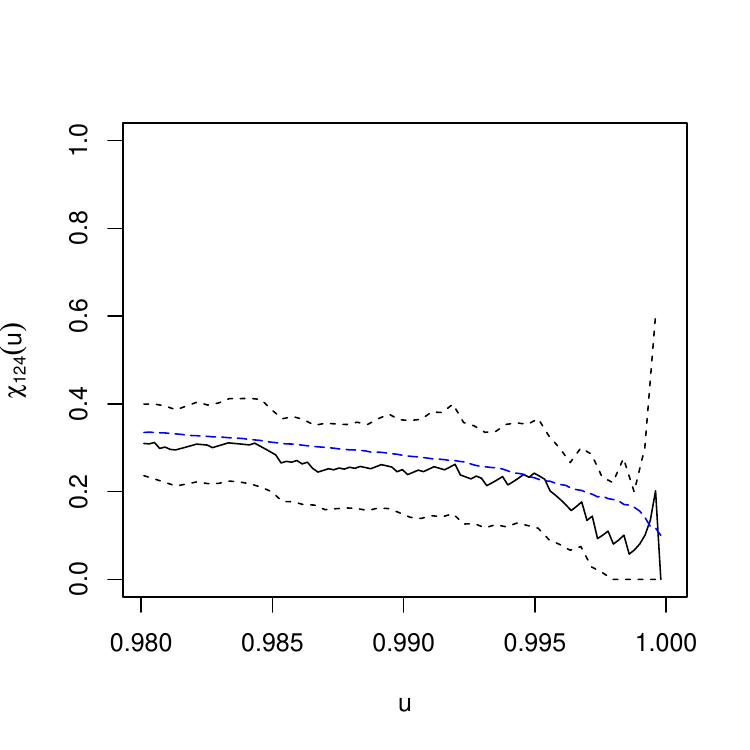}
	\includegraphics[width=0.4\textwidth]{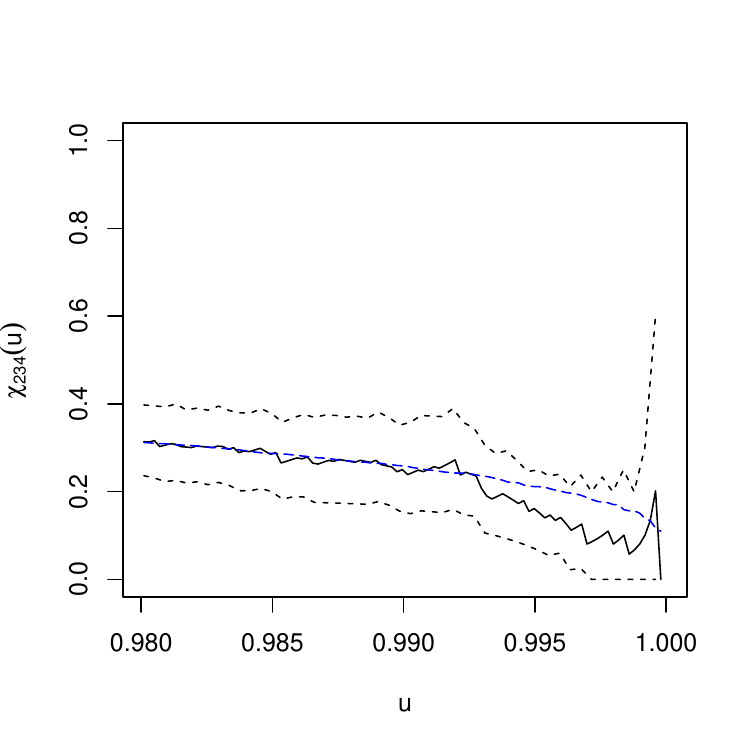}\\
	\includegraphics[width=0.4\textwidth]{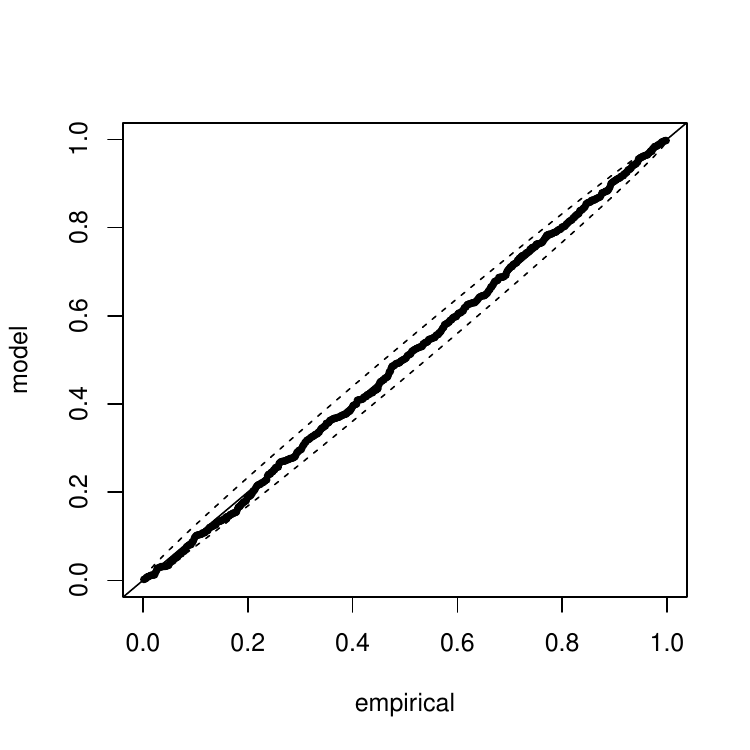}
	\includegraphics[width=0.4\textwidth]{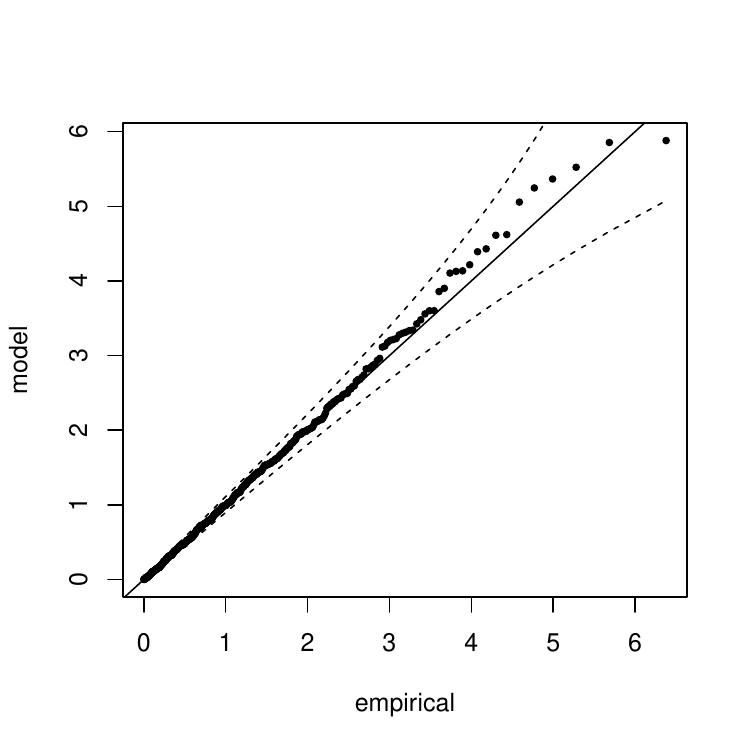}
	\caption{Top row: Empirical (black) estimates of $\chi_C(u)$ with 95\% confidence interval (dashed), and model-based estimate (blue) for groups $C=\{1,2,4\}$ and $\{2,3,4\}$. Bottom row: PP and exponential QQ plots for the fitted geometric model with Gaussian gauge function.}
	\label{fig:river-chi-pp-extra}
\end{figure}

Figures~\ref{fig:river-chi-extra} and~\ref{fig:river-chi-pp-extra} display plots of $\chi_C(u)$ for the remaining groups of variables. The bottom row of Figure~\ref{fig:river-chi-pp-extra} displays the PP and exponential QQ plots for the fit of model~\eqref{eq:rwgamma}.

\newpage
\bibliographystyle{apalike}
\bibliography{GeometricBib}

\end{document}